\documentclass[11pt,a4paper]{article}

\usepackage[T1]{fontenc}
\usepackage[utf8]{inputenc}
\usepackage{amsmath, amsfonts, amssymb, graphicx, authblk, geometry, color, xcolor, url, ascmac, bm, tensor, microtype} 
\usepackage{jheppub}
\usepackage{cleveref}
\usepackage{caption}
\usepackage{subcaption}
\usepackage{float}
\usepackage{soul}
\usepackage{subcaption}
\usepackage{slashed}
\usepackage{cancel}
\usepackage{tabularx} 
\usepackage{comment}
\usepackage{enumitem}

\usepackage{booktabs}
\usepackage{longtable}

\newcommand{\mpl}{M_{\text{pl}}}

\newcommand{\bra}{\langle\,}
\newcommand{\ket}{\rangle\,}
\newcommand{\paren}[1]{\left( #1 \right)}
\newcommand{\sqbracket}[1]{\left[ #1 \right]}
\newcommand{\subt}[1]{_{\rm #1}} 

\newcommand{\fig}[1]{figure~\ref{#1}}
\newcommand{\eq}[1]{Eq.~(\ref{#1})}
\newcommand{\eqs}[2]{Eqs.~(\ref{#1}) and (\ref{#2})}
\newcommand{\Sec}[1]{section~\ref{#1}}
\newcommand{\Secs}[2]{sections~\ref{#1} and \ref{#2}}
\newcommand{\App}[1]{appendix~\ref{#1}}

\newcommand{\hinf}{H_{\rm inf}}

\newcommand{\f}[1]{f_{\rm #1}}

\newcommand{\afactor}[1]{a_{\rm #1}} 

\newcommand{\figs}[1]{figures~\ref{#1}}

\newcommand{\vev}[1]{ \left< {#1} \right> }
\newcommand{\dd}{\mathrm{d}}

\newcommand{\GW}{\text{GW} }

\newcommand{\tot}{\text{tot} }
\newcommand{\ee}{\text{e}}

\newcommand{\kin}{\text{kin}}

\newcommand{\RM}{\text{RM}}
\newcommand{\MK}{\text{MK}}
\newcommand{\KR}{\text{KR}}

\newcommand{\osc}{\text{osc}}
\newcommand{\NL}{\text{NL}}

\newcommand{\equ}{\text{eq}}
\newcommand{\kr}{\text{KR}}
\newcommand{\Mpc}{\text{Mpc}}

\newcommand{\Hz}{\text{Hz}}
\newcommand{\ls}{\text{ls}}

\newcommand{\bfk}{\mathbf{k}}

\newcommand{\app}{\text{app}}

\newcommand{\hh}{\text{h}}

\newcommand{\rr}{\text{rad}}

\newcommand{\kd}{\text{KD}}
\newcommand{\emd}{\text{eMD}}

\newcommand{\cc}{\text{c}}

\definecolor{lightpink}{rgb}{1.0, 0.55, 0.75}

\definecolor{purple}{rgb}{0.4 ,0, 0.85}

\title{
Anisotropic Gravitational Waves from Anisotropic Axion Rotation
}

\author[a,b,c]{Arushi Bodas,}
\author[a,b,d,e]{Keisuke Harigaya,}
\author[f]{Keisuke Inomata,}
\author[g]{Takahiro Terada,}
\author[a,b,d]{and Lian-Tao Wang}

\affiliation[a]{Department of Physics, University of Chicago, Chicago, IL 60637, USA}
\affiliation[b]{Enrico Fermi Institute and Leinweber Institute for Theoretical Physics, University of Chicago, Chicago, IL 60637, USA}
\affiliation[c]{Particle Theory Department, Fermilab, Batavia, Illinois 60510, USA}
\affiliation[d]{Kavli Institute for Cosmological Physics, University of Chicago, IL 60637, USA}
\affiliation[e]{Kavli Institute for the Physics and Mathematics of the Universe (WPI),\\
        The University of Tokyo Institutes for Advanced Study,\\
        The University of Tokyo, Kashiwa, Chiba 277-8583, Japan}
\affiliation[f]{William H. Miller III Department of Physics \& Astronomy, Johns Hopkins University, 3400 N. Charles St., Baltimore, MD 21218, USA}
\affiliation[g]{Kobayashi-Maskawa Institute for the Origin of Particles and the Universe, Nagoya University, Furo-cho Chikusa-ku, Nagoya 464-8602, Japan}

\emailAdd{arushib@uchicago.edu}
\emailAdd{kharigaya@uchicago.edu}
\emailAdd{kinomat1@jhu.edu}
\emailAdd{terada@eken.phys.nagoya-u.ac.jp}
\emailAdd{liantaow@uchicago.edu}

\abstract{
Gravitational waves (GWs) provide a powerful probe of the early universe due to their ability to free-stream across cosmic history. We study GW production in a compelling scenario where a rotating axion(-like) field becomes relevant for a brief period in the early universe before transitioning into a kination fluid and rapidly dissipating its energy through cosmic expansion. During this short epoch, the curvature perturbation can be predominantly sourced by the rotating axion and may significantly exceed 
the adiabatic component. Moreover, axion field perturbations grow on superhorizon scales during this phase. These effects can generate a strong stochastic background of induced GWs. This GW background also exhibits a pronounced large-scale anisotropy inherited from the axion fluctuations, serving as a distinctive signature of the scenario. Importantly, the transient nature of axion relevance enables this scenario to evade stringent bounds on large-scale perturbations. We analyze various observational constraints and find that both the amplitude and anisotropy of the resulting GW signal could be accessible to future detectors.
} 

\keywords{Induced gravitational waves, rotating axion, kination, anisotropy}

\begin{document}

\setcounter{tocdepth}{2}
\maketitle
\flushbottom


\section{Introduction}
\label{sec:intro}

The early universe, with its high energy, sets the stage for novel phenomena stemming from fundamental theories that are beyond the reach of terrestrial experiments. Stochastic gravitational waves (GWs) associated with these phenomena can be detected by current and future GW observatories~\cite{LISA:2017pwj, Harry:2006fi, Kawamura:2011zz,Sesana:2019vho,NANOGrav:2023gor, Braun:2015zta,Hall:2022dik,Punturo:2010zz}. The signals from these waves, including their strength, spectral shape, and anisotropy on large scales, carry invaluable information about new-physics processes that occurred in the early universe. (See Ref.~\cite{Roshan:2024qnv} for a recent review).

One intriguing possibility for new physics is the presence of axion(-like) fields (henceforth, axions) in the early universe that arise from spontaneous symmetry breaking of approximate global symmetry.
Recent research has highlighted the complex and rich dynamics that these axions could exhibit. Field-theoretical axions are the angular directions of complex scalar fields and may rotate in the field space, initiated by the Affleck-Dine mechanism~\cite{Affleck:1984fy}. 
The angular momentum in the rotation may be partially transferred into baryon charges to explain the observed matter-antimatter asymmetry of the universe~\cite{Co:2019wyp,Domcke:2020kcp,Co:2020xlh,Co:2020jtv}.
Such dynamics has also been invoked to explain the observed dark matter abundance~\cite{Co:2019jts,Co:2020dya,Eroncel:2022vjg,Eroncel:2022efc,Eroncel:2025qlk,Bodas:2025eca,Fasiello:2025ptb}.

If the radial component of the complex scalar has a nearly quadratic potential, the energy stored in the rotation initially behaves like matter and later transitions to kination.\footnote{Here, kination refers to the dominance of kinetic energy over potential and gradient energies in a scalar field. When this fluid dominates the total energy density, we refer to the era as the kination-dominated era.} 
During the matter phase, the fraction of the energy density in the complex field increases, and later decreases during the kination phase.
This matter-to-kination (MK) behavior allows the rotating complex field to temporarily constitute a significant, or even dominant, fraction of the total energy density of the universe~\cite{Co:2019wyp,Co:2021lkc,Gouttenoire:2021wzu}.

In this work, we study the GW signal sourced by the fluctuations of such a rotating axion field.
These fluctuations are generated during inflation, similarly to the inflaton.
However, they can be of very different from, and
in particular, much larger than the adiabatic ones sourced by the inflaton.
If the complex scalar occupies a large portion of the total energy density near MK transition, its fluctuations can become the main source of curvature perturbations around that time, enhancing their amplitude for a short duration.
In addition, superhorizon modes of the axion field fluctuations experience significant growth during this time, leading to large velocity perturbations. 
The density and velocity perturbations with a non-zero sound speed oscillate upon horizon re-entry and can generate GWs through second-order perturbations, known as \textit{induced} GWs~\cite{Tomita:1967wkp,Ananda:2006af, Baumann:2007zm, Kohri:2018awv}.\footnote{They are also known as secondary gravitational waves.}
The GWs induced from the standard adiabatic perturbations are typically weak, with an exception studied in Refs.~\cite{Inomata:2019ivs,Inomata:2020lmk,Harigaya:2023mhl}.
In our scenario, however, the transient enhancement of curvature perturbations and the super-horizon growth of axion fluctuations can produce a strong and potentially observable induced gravitational wave background (GWB) over a wide range of frequencies.

Induced GWB also exhibits anisotropy on large scales~\cite{Bartolo:2019oiq, Bartolo:2019yeu, Dimastrogiovanni:2022eir, Chen:2022qec, Li:2023qua, Li:2023xtl, Wang:2023ost, Yu:2023jrs, Ruiz:2024weh, Li:2025met}.
Importantly, in our model, these are inherited directly from the axion fluctuations.
As a result, the GWB anisotropy can differ significantly from the adiabatic perturbations, and, notably, it can be larger.
This makes them not only more accessible to future experiments~\cite{LISACosmologyWorkingGroup:2022kbp, Kudoh:2005as, Mentasti:2023gmg, Depta:2024ykq, Cusin:2025xle}, but also provides us with a novel observable to probe the spectator axion field and its interactions~\cite{Kumar:2021ffi, Bodas:2022zca}.
While large anisotropies in GWBs from phase transitions have been studied before \cite{Geller:2018mwu, Bodas:2022urf}, this is the first example where such anisotropies occur within induced GWBs.
The combination of detecting the induced GWs and observing their anisotropic features could provide compelling evidence for the axion dynamics in our scenario.

We would like to emphasize the advantage of the transient relevance of the axion field in our scenario. 
When aiming to produce a highly anisotropic GWB, one typically encounters a trade-off: a strong isotropic GW signal requires the source sector to constitute a substantial fraction of the total energy density at the time of production.
However, large perturbations in such a sector lead to long-wavelength curvature perturbations that exceed observed values if the source sector eventually thermalizes with the Standard Model (SM) radiation, or imprint significant dark radiation isocurvature in the CMB if it remains decoupled. 
Avoiding these requires the sector to be sufficiently subdominant, which hurts the GW signal strength.
This trade-off can be softened if the model features a mechanism by which the fractional energy density of the GW-generating sector decreases significantly between the time of GW production and the CMB epoch. 
Our model naturally achieves this; during the early matter-like phase, the axion energy density becomes substantial, boosting GW production, while it dilutes faster than radiation during the subsequent kination phase, thereby evading CMB (iso)curvature constraints on large scales.

The rest of the paper is organized as follows. 
In \Sec{sec:axion_dynamics_fluctuations}, we describe the dynamics and inflationary fluctuations of the complex scalar field. 
Section~\ref{sec:induced-GW} presents the computation of induced GWs, and \Sec{sec:anisotropy} analyzes their anisotropy. 
The axion dynamics leaves behind remnants such as primordial black holes (PBHs) and axion radiation. In \Sec{sec:constraints}, we consider observational constraints on these remnants, which in turn limit the amplitude of axion fluctuations, and therefore the strength of the induced GWB.
Detection prospects for both GWB amplitude and anisotropy produced in our model are compared to future experimental projections in \Sec{sec:detectability}. 
We conclude in \Sec{sec:conclusion}.
The appendices contain supplementary material relevant to our analysis.
Appendix~\ref{app:notation} summarizes the notation used throughout the paper.
Appendix~\ref{app:thermalization} reviews the thermalization process that leads to the circularization of the axion rotation. A supersymmetric realization of the axion dynamics is presented in \App{app:susy_realization}.
A detailed calculation of the GWs induced before the axion rotation completely transitions to kination fluid is included in \App{app:induced-GW}.
The final \App{app:PBH_finiteMD} contains the derivation for PBH formation probability during a short matter dominance.


\section{Classical dynamics and fluctuations}
\label{sec:axion_dynamics_fluctuations}
In this section, we describe the model of the rotating axion field, focusing on its post-inflationary dynamics and the fluctuations that play a crucial role in generating a large gravitational-wave signal.

\subsection{Axion rotation and post-inflationary cosmology}
\label{sec:scalar_rotations}

Consider a complex scalar field $S$, 
\begin{equation}\label{eq:S_def}
    S = \frac{1}{\sqrt{2}} r e^{i \theta},
\end{equation}
where $r$ and  $\theta$ are the radial and angular field directions, respectively.
The field has a global $U(1)$ symmetry, which is taken to be spontaneously broken at $r =f_a$,  where $f_a$ is called the axion decay constant.
The angular direction is the axion-like field $\chi$ (henceforth referred to as the ``axion''):
\begin{equation}
    \label{eq:theta_chi}
    \chi = r \, \theta.
\end{equation}

\begin{figure}[tb!]
	\centering	
    \subfloat[]{\includegraphics[width=0.4\linewidth, trim={0cm 0cm 0cm 0cm},clip]{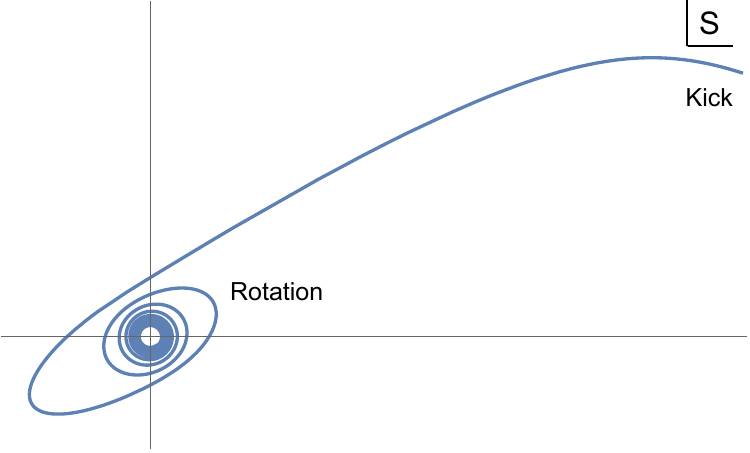}}
    \hspace{1.8em}
    \subfloat[]{
    \includegraphics[width=0.5\columnwidth,trim={3.5cm 3.8cm 4.2cm 5.3cm},clip]{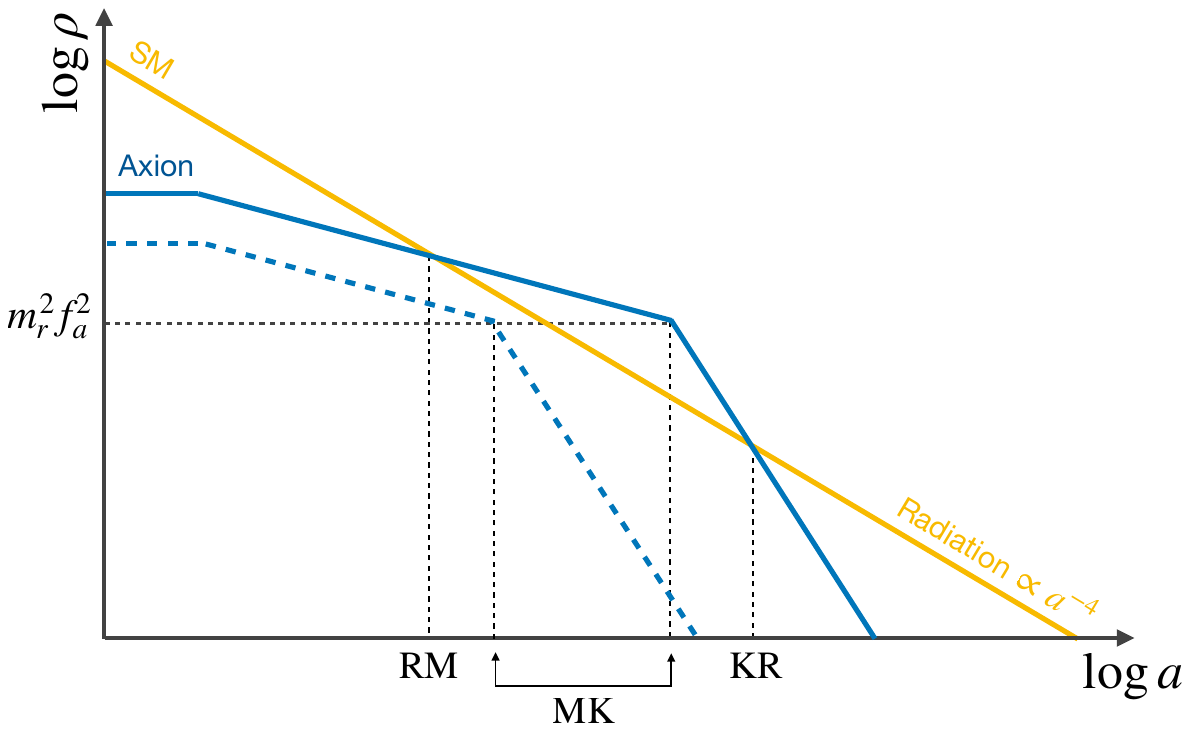}}
	\caption{(a) A schematic showing the initiation of axion rotations and the subsequent thermalization that makes the rotation circular. (b) Possible post-inflationary cosmologies with rotation dynamics.
    The rotating field may either dominate briefly (solid blue) or remain subdominant 
    (dashed blue line).} 
    \label{fig:S_class_dynamics}
\end{figure}

We consider an initial displacement of $r$ far from its vacuum expectation value.
As long as the Hubble parameter $H$ exceeds the radial mass $m_r$, the field remains effectively frozen near its initial value.
Once $H \sim m_r$, the field begins to roll. 
If the global $U(1)$ symmetry is also explicitly broken, a gradient develops in the angular direction, which imparts angular velocity to the field. This effect, labeled “Kick” in \fig{fig:S_class_dynamics}(a), initiates the classical rotational dynamics of $S$.
Such explicit symmetry breaking may arise from higher-dimensional operators of the form $ V_{\cancel{U(1)}} \sim \frac{S^{n}}{\Lambda^{n-4}} + \text{h.c.}$, which are important at large field values. 
The resulting angular kick puts the field in an elliptical orbit.

As $r$ decreases, the $U(1)$ breaking terms become negligible and the symmetry is approximately preserved. 
The $U(1)$ charge $n_\theta = i( S\dot{S}^{*}-S^{*}\dot{S}) = \dot{\theta} \, r^{2}$ induced from the initial kick is thereafter conserved up to dilution by cosmic expansion.
Here, `$ \,  \dot{} \,$' $  \equiv \partial_t$ denotes partial derivative with respect to proper time.
It is convenient to define the yield, $Y_\theta$, that remains constant:
\begin{equation}\label{eq:Y_def}
    Y_{\theta} \equiv \frac{n_{\theta}}{s} = \frac{\dot{\theta} r^{2}}{s},
\end{equation}
where $s$ is the entropy density.

The initial motion of the field is typically elliptical due to gradients in both the radial and angular directions at the time of the kick. However, this motion can later become circular through thermalization of $S$ with light species in the thermal bath. These interactions dissipate the radial oscillations and redistribute the $U(1)$ charge between coherent rotation and particle-like excitations of $S$. If the charge density at the time of thermalization exceeds $m_r T^2$, most of the charge remains stored in the coherent motion~\cite{Co:2019wyp,Domcke:2022wpb}, which can be interpreted as a Bose-Einstein condensate of $U(1)$ charges. As a result, the elliptical motion transitions into a circular orbit, as illustrated in \fig{fig:S_class_dynamics}(a). Eventually, the rotation settles at $r = f_a$,  
and the energy density in $S$ is dominated by rotational energy of the axion, leading to a kination phase, as we will discuss shortly.
In the absence of thermalization, the energy density in the radial oscillatory mode remains comparable to or dominant over the axion rotation energy, preventing the onset of the kination phase. We therefore assume that thermalization takes place sometime after the initiation of rotation. Constraints on the parameter space required for efficient thermalization are discussed in Appendix~\ref{app:thermalization}.

Let us now examine the equation of state of $S$ across different stages.
After the initial kick and thermalization, the equations of motion impose $\dot{\theta}^{2} = \partial_r V(r)/r $, where $V(r)$ is the potential of the radial direction~\cite{Co:2019wyp,Harigaya:2023mhl}. For an approximately quadratic potential $V(r)\sim m_r^2 r^2$, the angular velocity evaluates to $\dot{\theta} \approx m_r$. 
The potential energy $m_r^2 r^2$ and the rotational kinetic energy $\dot{\theta}^2 r^2$ are then comparable.
Given the conservation of $n_{\theta} = \dot{\theta} r^2 \approx m_r r^2 \propto a^{-3}$, where $a$ is the scale factor, the radius 
decreases as $r \propto a^{-3/2}$, leading to a matter-like dilution of the energy density $\rho_{S} \sim m_r^2 r^2 \propto a^{-3}$.
Once the field settles into the minimum at $r= f_a$, 
the conservation of charge $n_{\theta} = \dot{\theta}f_a^2 \propto a^{-3}$ implies $\dot{\theta} \propto a^{-3}$.
The kinetic energy of rotation, now dominant, dilutes as $\rho_{S} \propto \dot{\theta}^2 f_a^2 \propto a^{-6}$, and the field transitions into a kination phase.
We refer to this matter-to-kination transition point as MK.
In summary, 
\begin{equation}
 \rho_{S} \propto
    \begin{cases}
        a^{-3} \quad({\rm matter\text{-}like}) , & r > f_a \\
        a^{-6} \quad({\rm kination\text{-}like}) , & r \sim f_a. 
    \end{cases}
    \label{eq:rho_s_propto}
\end{equation}
A supersymmetric (SUSY) model realizing this dynamics is presented in \App{app:susy_realization}.%
\footnote{
Before the completion of the thermalization when the rotation is elliptic, fluctuations around the rotating background can grow by parametric resonance~\cite{Shtanov:1994ce,Kofman:1994rk,Kofman:1997yn,Co:2020dya,Co:2020jtv,Fedderke:2025sic}. 
For nearly quadratic potentials, the resonance is ineffective at large $r$~\cite{Co:2020jtv}, and as we discuss in Appendix~\ref{app:thermalization}, thermalization should occur at large $r$ to avoid non-Gaussianity constraints. Therefore, it is expected that the rotation remains nearly homogeneous. Even if fluctuations grow, subsequent thermalization will remove fluctuations except for long cosmic strings (if produced by non-thermal symmetry restoration~\cite{Tkachev:1995md,Kasuya:1996ns,Kasuya:1997ha,Kasuya:1998td,Tkachev:1998dc,Kasuya:1999hy,Co:2020dya}) that are protected by topology and phonon modes whose thermalization rate is suppressed by the smallness of the momentum of the fluctuations. The energy density of long cosmic strings is negligible unless the decay constant is near the Planck scale. That of phonon modes scales as radiation and is negligible. Note that $U(1)$ symmetry is preserved and the $U(1)$ charge is not dissipated by parametric resonance.
}

During inflation, $S$ is taken to be a subdominant spectator field.
The inflaton $\phi$ reheats the Standard-Model (SM) plasma, initiating radiation dominance with $\rho_\rr \propto a^{-4}$. 
As the $S$-field evolves, its fractional energy density $ \Omega_S = \rho_S/\rho_{\rm tot}$ evolves differently from that of radiation. 
During the matter-like phase, $\Omega_S$ grows and may even dominate, leading to an early matter-dominated (eMD) epoch. 
The transition from radiation dominance to eMD is denoted by RM. 
After MK, the universe enters a kination-dominated phase, during which $\Omega_S$ reduces rapidly, eventually restoring radiation domination. We denote this transition by KR. 
The possible scenarios are illustrated in \fig{fig:S_class_dynamics}(b).

When $S$ temporarily dominates the energy budget, the duration of eMD and kination dominance are related as
\begin{equation}
    \frac{\afactor{KR}}{\afactor{MK}} \approx \paren{\frac{\afactor{MK}}{\afactor{RM}}}^{1/2} .
\end{equation}
The relevant phenomenological parameters are 
the duration of early matter dominance, $(\afactor{MK}/\afactor{RM})$ and the temperature of the SM radiation at MK transition,  $T_{\rm MK}$.


\subsection{Axion fluctuations and enhanced curvature perturbation}
\label{sec:enhanced_curv_pert}

During inflation, the subdominant  $S$ field develops quantum fluctuations, similarly to the inflaton. 
As mentioned in the previous section, the field is at a large radius $r_{\rm inf}$ during inflation. To be concrete, we take the effective mass of the radial direction at this field value to be comparable to or larger than the Hubble parameter during inflation, $\hinf$, which can be achieved by a negative Hubble induced mass term $- H^2 |S|^2$ balanced by an up-lifting potential. 
In this case, quantum fluctuations along the radial direction are suppressed.
However, if the explicit $U(1)$ symmetry breaking is small, the angular direction can be effectively massless, and $S$ develops an approximately scale-invariant spectrum along the angular direction,%
\footnote{We will also consider a case where the spectrum of $\delta \theta$ is (slightly) blue-tilted.
This is useful in the part of the parameter space where the constraints on long-wavelength fluctuations are stronger than those on the short-wavelength modes (see \Sec{sec:constraints}).
Such a blue-tilt can be achieved if $r$ has not settled into $r_{\rm inf}$ when the long-wavelength modes exit the horizon.}
\begin{align}\label{eq:P_def} 
    \bra \delta \theta({\mathbf k}) \delta \theta({\mathbf k}')\ket = (2\pi)^3 \delta({\mathbf k} + {\mathbf k}') \frac{2\pi^2} {k^3} \mathcal P_\theta, \qquad {\rm with} \,\, \sqrt{\mathcal{P}_\theta} \approx \frac{\hinf}{2 \pi r_{\rm inf}}.
\end{align}

Due to the primordial fluctuations in the initial angle $\theta_{\rm init}$, the strength of the kick, and hence $n_{\theta}$, is modulated~\cite{Enqvist:1998pf,Kasuya:2008xp,Co:2022qpr}. 
The two can be related as
\begin{align}\label{eq:deltaNTheta_deltaTheta}
    \frac{\delta n_{\theta}}{n_{\theta}} = \frac{(\partial n_{\theta}/\partial \theta)}{n_{\theta}} \delta \theta = \left. \frac{(\partial n_{\theta}/\partial \theta)}{n_{\theta}} \right|_{\theta_{\rm init}}\paren{\frac{\hinf}{2\pi r_{\rm inf}}}.
\end{align}
The evolution of $n_{\theta}$ depends on the explicit $U(1)$-breaking term and is therefore model-dependent,
\begin{equation}\label{eq:n_theta_dynamic}
    \dot{n}_{\theta} + 3H n_{\theta} = -\frac{\partial V_{\cancel{U(1)}}}{\partial \theta}.
\end{equation}
We can generically parametrize the symmetry breaking potential as 
\begin{align}
     V_{\cancel{U(1)}} \sim \bar{m}^4 (r) \cos[n \theta], 
\end{align}
where $\bar{m}(r)$ and $n$ depend on the exact form of the symmetry breaking operator. 
For instance, in the SUSY model in \App{app:susy_realization}, $\bar{m}^4 = \alpha m_r \frac{r^{n}}{\Lambda^{n-3}}$, where $\alpha$ is an $O(1)$ constant and $\Lambda$ is the cutoff scale in the  higher dimensional symmetry-breaking operator.
Since the kick is strongest in a short interval around $H_{\rm kick} \sim m_r$,
the expression for $n_{\theta}$ simplifies to
\begin{align}\label{eq:n_theta}
    n_{\theta} \approx n  \frac{\bar{m}^4 (t_{\rm kick})}{H_{\rm kick}} \sin[n\theta_{\rm init}].
\end{align}
If $\delta\theta \ll 1$ and $\theta_{\rm init}$ is not close to the extrema of the potential, the fluctuation in the number density is
\begin{equation}\label{eq:delta_nTheta}
    \delta_S \equiv \frac{\delta n_{\theta}}{n_{\theta}} \approx n \, \frac{\cos[n \theta_{\rm init}]}{\sin[n \theta_{\rm init}]} \delta\theta.
\end{equation}
Since thermalization approximately conserves $n_{\theta}$, its fluctuations are also preserved.  
The boundedness of $n_{\theta}$ from Eq.~\eqref{eq:n_theta} implies that $\delta n_{\theta}$ is not strictly Gaussian.
While this does not impact most of our results, it becomes important when estimating primordial black hole (PBH) abundance, as discussed in \Sec{sec:constraint_pbh}.  

Note that the axion has independent quantum fluctuations  different from the adiabatic perturbations sourced by $\phi$. In particular, the regime of $\mathcal{P}_{\delta_S} \gg \mathcal{P}_{\phi} (\approx 2.1 \times 10^{-9})$  
has interesting consequences as we will see shortly, and we will therefore restrict to such a case in the rest of the paper.
The gauge-invariant axion perturbation, $\zeta_S$, is 
\begin{align}
    \zeta_S = -\Phi + \frac{1}{3(1+w_S)}  \frac{\delta\rho_S}{\rho_S},
\end{align}
where $\Phi$ is the gravitational potential and $w_S$ is the equation of state of $S$. 
Consider some early time when $S$ is thermalized, subdominant, and dilutes like matter. Then $\Phi \sim 10^{-5} \ll (\delta \rho_S /\rho_S)$, giving $\zeta_S \approx (\delta \rho_S /\rho_S) /3$.
The energy density in $S$ during this time is $\rho_S = \dot{\theta} n_{\theta} = m_r n_{\theta}$.
Then $\delta \rho_S /\rho_S = \delta n_{\theta}/n_\theta =\delta_S$, giving
\begin{align}\label{eq:zeta_S}
    \zeta_S \approx \frac{\delta_S}{3}.
\end{align}
$\zeta_S$, which is conserved on superhorizon scales, can thus be related to the primordial fluctuation of $\theta$ using \eqs{eq:P_def}{eq:delta_nTheta}.

In a universe with multiple non-interacting fluids, the total curvature perturbation is a weighted sum of the individual perturbations,
\begin{align}
    \zeta_{\rm tot} = \sum_{i=\phi, S} \tilde\Omega_i \zeta_i ,
    \label{eq:zeta_tot}
\end{align}
where $\tilde\Omega_i\equiv \frac{\rho_i + P_i}{(\rho+P)_\text{total}}$ 
and $\zeta_\phi$ denotes the curvature perturbations of the SM plasma inherited from the inflaton.
From CMB observations, $\sqrt{\mathcal{P}_{\text{tot}}(t_{\rm CMB})} \approx 4.5\times 10^{-5}$, where $t_\text{CMB}$ is the time of the CMB epoch.
In our scenario, since we are interested in $\sqrt{\mathcal{P}_{\zeta_S}} \gg 4.5 \times 10^{-5}$ on all scales including those relevant for the CMB, we require $\tilde \Omega_{S}(t_{\rm CMB}) \ll 1$ to be compatible with the CMB observations.
However, $\tilde \Omega_{S}$ evolves over time, as illustrated in \fig{fig:S_class_dynamics}, 
and can be large enough around $t_{\rm MK}$ such that the total curvature perturbation is significantly enhanced for a short duration: $\mathcal{P}_{\text{tot}} (t_{\rm MK}) \approx \tilde{\Omega}_S^2(t_{\rm MK})\, \mathcal{P}_{\zeta_S}\gg 2 \times 10^{-9}$.

This temporary enhancement of $\tilde{\Omega}_{S}$ and the curvature perturbation around MK has two important consequences.
First, the modes that re-enter the horizon during this period produce a significant amount of induced GWs at second order, with the characteristic frequency determined by the horizon scale at that time.
Additionally, the angular (axion) fluctuation grows during this period on superhorizon scales. 
When these modes ($k < k_{\rm MK} (= a_\MK H_\MK)$) re-enter the horizon later during the radiation era, they source large velocity perturbations that again source induced GWs.
Both these effects can lead to an induced GWB with observable strength over a wide frequency range.
The second consequence is that the induced GWB also inherits the long-wavelength fluctuations from the axion, making it highly anisotropic on large scales.
This is in sharp contrast to the adiabatic nature of the CMB and LSS perturbations.
Therefore, a single (approximately) scale-invariant spectrum of $\zeta_S$, combined with the classical dynamics of $S$, can lead to two key effects: 1) a large \emph{isotropic} background of induced GWs sourced by short wavelength modes, $\zeta_S(k_{s})$, and 2) enhanced \emph{anisotropies} within the GWB on large scales from long wavelength fluctuations, $\zeta_S(k_l)$.
We elaborate on these two aspects in \Secs{sec:induced-GW}{sec:anisotropy}, respectively.


\section{Induced GWB} 
\label{sec:induced-GW}

In this section, we summarize the key formulas for the induced GWs in the case with and without a rotation dominated era.
In the following, we take the conformal Newtonian gauge
\begin{align}
	\dd s^2 = a^2 \left[-(1+2\Phi) \dd \eta^2 +\left( (1-2\Phi) \delta_{ij} + \frac{h_{ij}}{2} \right) \dd x^i \dd x^j\right],
\end{align}
where $\Phi$ is the first-order scalar perturbation and $h_{ij}$ is the second-order tensor perturbation.
We have neglected the first-order vector and the tensor perturbations to focus on the GWs induced by the first-order scalar perturbations. 
Also, we have neglected the first-order anisotropic stress to equate one of the Bardeen potentials with (the negative of) the other. Indeed, the leading-order anisotropic stress $\pi_{ij}$ originating from the scalar field $S$ is at the second order, $\pi_{ij} \sim \partial_i S \partial_j S$, which is the source of the induced GWs, while we neglect any second-order anisotropic stress in the radiation.

In the following, we consider the two cases: 1) the axion rotation dominates the Universe at some point  and 2) it never dominates the Universe. 
Notably, for both cases, a scale-invariant GW spectrum can be produced by the perturbations that enter the horizon after the energy fraction of the axion rotation is negligible in the late time, which we focus on in section~\ref{subsec:late_gw}.

\subsection{Early-time GW production}
\label{subsec:early_gw}
 
In this subsection, we focus on the GWs induced when the energy fraction of the axion rotation is non-negligible, i.e., around the MK transition.
During this phase, the axion rotation induces GWs through both the gravitational potential and the fluid velocity field.

\subsubsection{
Induced GWs with rotation domination 
}\label{subsec:iGW_KD}

We begin with the case with a rotation domination era.
In this case, the amount of induced GWs generally depends on the timescale of the transition from the eMD era to the subsequent cosmological 
era~\cite{Inomata:2019zqy,Inomata:2019ivs,Inomata:2020lmk, Pearce:2023kxp}. 
If the transition occurs suddenly, the induced GWs are enhanced through the poltergeist mechanism~\cite{Inomata:2019ivs, Inomata:2020lmk}, as studied in detail in Ref.~\cite{Harigaya:2023mhl} in the context of the MK transition.
On the other hand, in this paper, we propose a novel case for strong GW production from large isocurvature (axion) perturbations. 
To focus on this new point, we consider the case where the transition from the eMD era to the kination era occurs gradually.
Such a gradual transition is readily achieved with a log-potential model described in \App{app:induced-GW}.

\begin{figure}
        \centering \includegraphics[width=0.75\columnwidth]{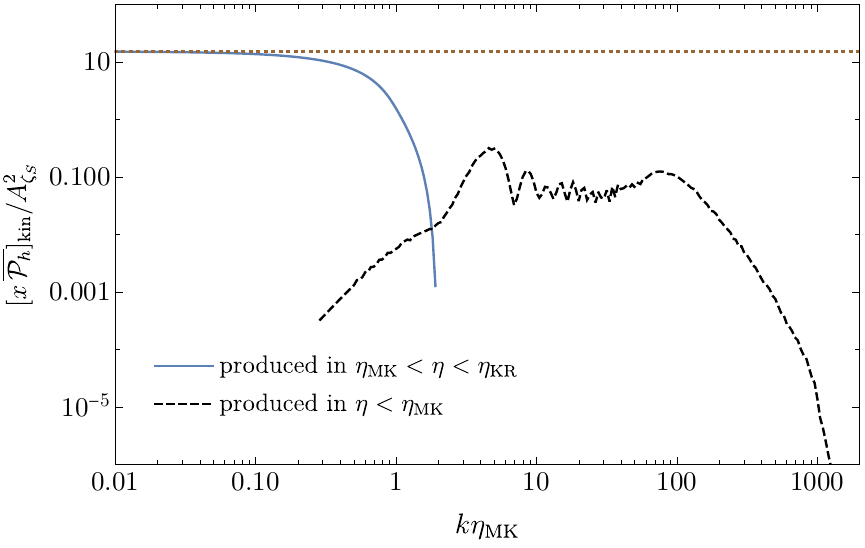}
        \caption{ 
        The blue solid line shows the power spectrum of induced GWs with the curvature power spectrum given by Eq.~(\ref{eq:p_zeta}).
        The brown dotted line is the guideline for $15.28$, which indicates that the value of the flat region in $k \eta_\MK \ll 1$ is $[x \overline{\mathcal P_h}]_\kin/A_{\zeta_S}^2 \simeq 15.28$.
        The black dashed line shows the rough estimate of the GWs induced by the perturbations that enter the horizon when $w,c_s \ll 1$ (before the kination era).
        See Appendix~\ref{app:induced-GW} for details.
	}
        \label{fig:ph}
\end{figure}

As we will see below, the scalar perturbations start to oscillate even when $w \ll 1$ due to a nonzero sound speed $c_s$. 
Because of this, GWs are also produced before the kination era begins when $w=c_s^2 =1$.\footnote{Strictly speaking, in the gradual transition, there is no strict boundary between the kination era and the eMD era. In this paper, we just call an era with $w, c_s \ll 1$ an eMD era and an era with $w \simeq c_s \simeq 1$ a kination era regardless of the perturbation behavior.}
However, the oscillation timescale during the transition is given by $\sim 1/(k c_s)$ and is long because $w < 1$ and $c_s^2 < 1$ during that period.
From this, we can naively expect that the GWs induced before the kination era are not much larger than those induced during the kination era.
We will see that this expectation is true in \fig{fig:ph} and  Appendix~\ref{app:induced-GW}.
The precise calculation of the GWs induced before the kination era requires model-dependent analyses and high computational costs (see Appendix~\ref{app:induced-GW} for details). 
Because of this, we only focus on the GWs induced during the kination era in the main text, which leads to conservative results in terms of detectability.

The power spectrum of GWs induced during the kination era is given by
\begin{align}
        \overline{\mathcal P_h(\eta, k)} = 4 \int^\infty_0 \dd v \int^{1+v}_{|1-v|} \dd u \left( \frac{4v^2 - (1+v^2-u^2)^2}{4 u v} \right)^2 \overline {I^2(u,v,x)} \mathcal P_\zeta(k u) \mathcal P_\zeta(k v),
        \label{eq:p_h_uv}
\end{align}
where $x = k \eta$ and the overline denotes the time average over oscillations of the tensor modes. 
The analytic expression of $I$ in the late-time limit $x\gg 1$ is given by~\cite{Domenech:2019quo}
\begin{align}
  \overline{I^2(u,v,x(\gg 1))} \simeq \frac{9}{16 \pi u^4 v^4 x} \left\{ \frac{(3(u^2 + v^2 -1)^2 - 4 u^2 v^2)^2}{4 u^2 v^2 - (u^2 + v^2-1)^2} + 9(u^2 + v^2-1)^2 \right\}.
\end{align}
We note that $x \overline{\mathcal P_h(\eta, k)}$ approaches a constant value during the kination era because of the redshift of the induced GWs and the decay of the source terms. 
For convenience, we here define the constant value as $[x \overline{\mathcal P_h(\eta, k)}]_\kin$.
After the axion rotation (kination) energy density becomes much smaller than the radiation energy density, the energy density parameter of the induced GWs during the radiation-dominated (RD) era can be expressed as~\cite{Harigaya:2023pmw}
\begin{align}
         \Omega_{\GW}(\eta_\cc,k) 
         &= \frac{k}{6\sqrt{2} \mathcal H_\kr} [x \overline{\mathcal P_h(\eta, k)}]_\kin,
        \label{eq:gw_r_ratio}
\end{align}
where $\eta_\cc \gg \eta_\kr$, $\mathcal H \equiv a H$, and the subscript `$\kr$' denotes the value at the equality time for the kination and the radiation energy density, $\rho_S = \rho_\rr$.

Then, the current $\Omega_\GW$ is given by~\cite{Inomata:2020lmk,Harigaya:2023pmw}
\begin{align}
        \Omega_\GW(\eta_0, k) h^2 \simeq 0.39 \left( \frac{g_{\rho,\hh}}{106.75} \right) \left( \frac{g_{s,\hh}}{106.75} \right)^{-4/3} \Omega_{\rr,0} h^2 \,  \Omega_\GW(\eta_\cc,k),
        \label{eq:omega_gw0}
\end{align}
where the prefactors come from the effect of the later (dark) matter-dominated era, and $g_{\rho,\hh}$ and $g_{s,\hh}$ are the  effective relativistic degrees of freedom for the energy and entropy density at the horizon crossing of the GWs with the scale of $k$. 
$\Omega_{\rr,0} h^2  \simeq 4.2 \times 10^{-5}$ is the current radiation energy density parameter with the normalized Hubble parameter $h = H_0/(100 \, \mathrm{km} \, \mathrm{s}^{-1} \, \mathrm{Mpc}^{-1})$.

To focus on the GWs induced by the scale-invariant power spectrum of curvature perturbations that enter the horizon during the kination era as described above,
let us parametrize the curvature power spectrum
as 
\begin{align}
	\mathcal P_\zeta (k) = A_{\zeta_S} \, \Theta(1/\eta_\MK- k),
    \label{eq:p_zeta}
\end{align}
where the
modes that enter the horizon before the kination era are dropped.
See Appendix~\ref{app:induced-GW} for the estimation of the neglected contributions.

Figure~\ref{fig:ph} shows the late-time limit ($\eta \gg 1/k$) of the GW power spectrum produced during the kination era, where we have numerically calculated Eq.~(\ref{eq:p_h_uv}).
In particular, in the limit of $k \ll 1/\eta_\MK$, we numerically obtain 
\begin{align}
	[x \overline{\mathcal P_h(\eta, k)}]_\kin \simeq 15.28\, A_{\zeta_S}^2.
        \label{eq:xP_h|kin}
\end{align}
In the figure, we also show a rough estimate of GWs induced by the perturbations that enter the horizon when $w,c_s \ll 1$ (before the kination era). 
This rough estimate is calculated with a concrete potential of the radial direction in Eq.~\eqref{eq:pot_vs} and can change in different potentials.
See Appendix~\ref{app:induced-GW} for details.

\begin{figure}
        \centering \includegraphics[width=0.75\columnwidth]{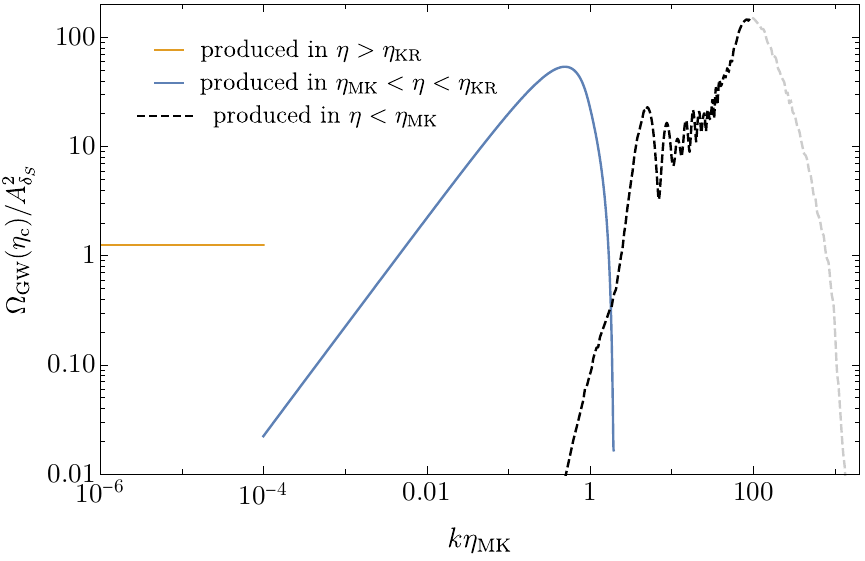}
        \caption{ 
        The energy density parameter of the induced GWs. 
        The blue solid and black dashed lines in this figure are converted from those in \fig{fig:ph} using Eq.~(\ref{eq:gw_r_ratio}).
        The orange line is the GWs induced after the KR transition (during $\eta > \eta_\KR$), calculated with \eqs{eq:delta_chi_ini}{eq:omega_gw_si}.
        We take $1/(\eta_\MK \mathcal H_\KR) = 10^4$ for all lines. 
        The small-scale cut-off of the orange line corresponds to the horizon scale at $\eta_\KR$.
        In the regime $k \eta_\MK \gtrsim 100$, the perturbations enter the horizon before the eMD era (during $\eta < \eta_\RM$), and therefore the dashed line is unreliable, though we still plot the dashed line in gray for reference.
        We normalize $\Omega_\GW$ by $A_{\delta_S}^2$ instead of $A_{\zeta_S}^2$, which are related as $A^2_{\zeta_S} = A^2_{\delta_S}/81$ through \eq{eq:zeta_S}.
        \label{fig:omegw}
        }
\end{figure}
Substituting Eq.~\eqref{eq:xP_h|kin} into Eq.~(\ref{eq:gw_r_ratio}), we obtain 
\begin{align}
	\Omega_{\GW}(\eta_\cc,k) \simeq \frac{k}{6\sqrt{2} \mathcal H_{\kr}} \times 15.28\, A_{\zeta_S}^2.
    \label{eq:omega_gw_a}
\end{align}
Figure~\ref{fig:omegw} shows the normalized $\Omega_\GW(\eta_\cc, k)$, including the above contribution shown as blue line.
We can see that for the perturbations that enter the horizon during the kination era ($ 1/\eta_{\rm KR} < k < 1/\eta_\MK$), a stronger GW signal is produced by the smaller-scale perturbations i.e., by $k_{\rm MK}$.  
This is because, once the GWs are produced, their contribution to the total energy density grows proportionally to $a^2$ (recall $\rho \propto a^{-6}$ for the kination fluid).
Figure~\ref{fig:omegw} also shows that the peak height of the GWs induced before the kination era (shown as black dashed line) can be comparable to the peak of those induced during the kination era.
However, since these are rough estimates and depend on the radial potential, we neglect these contributions in the following analysis for simplicity and give general and conservative discussion.
The precise analysis of the neglected contribution is beyond the scope of this paper.

The conformal Hubble at the end of the kination era can be expressed as~\cite{Harigaya:2023pmw}
\begin{align}
    \mathcal H_\kr = 2.5\times 10^{10}\,\Mpc^{-1} \left(\frac{g_{s,\kr}}{106.75}\right)^{-1/3} \left(\frac{g_{\rho,\kr}}{106.75}\right)^{1/2} \left(\frac{T_\kr}{\text{TeV}} \right), \label{eq:H_KR}
\end{align}
where we have normalized the scale factor as $a_0=1$ at present time.
The corresponding GW frequency is given by 
\begin{align}
\label{eq:nu_KR}
    \nu_\kr = \frac{\mathcal H_\kr}{2\pi} = 3.9\times 10^{-5}\,\Hz \left(\frac{g_{s,\kr}}{106.75}\right)^{-1/3} \left(\frac{g_{\rho,\kr}}{106.75}\right)^{1/2} \left(\frac{T_\kr}{\text{TeV}} \right).
\end{align}
Using this and \eqs{eq:omega_gw0}{eq:omega_gw_a}, we obtain 
\begin{equation}
    \Omega_\GW(\eta_0,\nu) h^2 \simeq 2.9\times 10^{-5} \, A_{\zeta_S}^2 \left(\frac{\nu}{3.9\times 10^{-5}\Hz} \right) \left(\frac{T_\kr}{\text{TeV}} \right)^{-1},
\end{equation}
where we have substituted $g_s = g_\rho = 106.75$.

\subsubsection{
Induced GWs without rotation domination 
}
\label{subsubsec:gw_wo_ax_dom}

Next, we discuss induced GWs in the case without the rotation domination era.
This situation is similar to the case for GWs induced by CDM isocurvature perturbations, which are studied in Refs.~\cite{Domenech:2021and,Domenech:2023jve}.
In particular, we focus on the GWs induced before the MK transition (during $\eta < \eta_\MK$) in this subsection, while we will discuss those induced later during $\eta > \eta_\text{MK}$ in section~\ref{subsec:late_gw}.
For simplicity, in the following, we assume $c_s^2 = w = 0$ for the axion rotation (not for the radiation background) during its matter-like phase ($\eta < \eta_{\rm MK}$). 
See Appendix~\ref{app:induced-GW} for the GWs produced in $\eta < \eta_\MK$ with nonzero $c_s$ in a concrete potential.
Note that, even in $c_s^2 = w=0$ for the axion fluid during the RD era, the evolution of the radiation component is affected by the axion isocurvature perturbation through the gravitational potential, and the radiation perturbations oscillate after their horizon entry. This oscillation of radiation perturbations produces GWs, which we focus on in the following.

We here define the power spectrum of the superhorizon isocurvature perturbation as 
\begin{align}
    \vev{\delta_{S}(\bfk) \delta_{S}(\bfk')} = (2\pi)^3 \delta(\bfk + \bfk') \frac{2\pi^2}{k^3}\mathcal P_{\delta_S}(k). \label{eq:P_iso_def}
\end{align}
Then, from Ref.~\cite{Domenech:2023jve}, the energy density parameter of the induced GWs in this situation is given by 
\begin{align}
	\Omega_{\GW}(\eta_\cc, k) = \frac{2}{3} \int^\infty_0 \dd v \int^{1+v}_{|1-v|} \dd u \left(\frac{4 v^2 - (1-u^2+v^2)^2}{4 uv} \right)^2 \overline{J^2(x_\cc,k,u,v)} \mathcal P_{\delta_S}(ku) \mathcal P_{\delta_S}(kv),
	\label{eq:omega_gw_wo_ax_dom}
\end{align}
where $x_\cc = k \eta_\cc$ with $\eta_\cc$ the time when the GW spectrum becomes constant after the source becomes negligible. 
The time average of the kernel in the late-time limit can be expressed as 
\begin{align}
	\overline{J^2(x \to \infty,k,u,v)} \simeq \frac{1}{2}(J^2_{c,\infty}(k,u,v) + J^2_{s,\infty}(k,u,v)),
    \label{eq:j_lim}
\end{align}
where 
\begin{align}
	J_{c,\infty}(k,u,v) &= \frac{9}{32 u^4 v^4 \kappa^2} \left\{ -3 u^2 v^2 + (-3 + u^2)(-3 + u^2 + 2v^2) \ln \left|1 - \frac{u^2}{3} \right| \right. \nonumber \\
	&\qquad \left. + (-3 + v^2)(-3 + v^2 + 2 u^2) \ln \left|1 - \frac{v^2}{3} \right| \right. \nonumber \\
	&\qquad \left. - \frac{1}{2}(-3 + v^2 + u^2)^2 \ln \left[ \left|1 - \frac{(u+v)^2}{3} \right| \left|1 - \frac{(u-v)^2}{3} \right| \right] \right\}, \\
	J_{s,\infty}(k,u,v) &= \frac{9 \pi}{32 u^4 v^4 \kappa^2} \left\{ 9 - 6v^2 - 6u^2 + 2u^2 v^2 + (3-u^2)(-3 + u^2 + 2v^2) \Theta\left(1 - \frac{u}{\sqrt{3}}\right) \right. \nonumber \\
	&\qquad \left. + (3- v^2)(-3 + v^2 +2 u^2) \Theta\left(1 - \frac{v}{\sqrt{3}}\right)  \right. \nonumber \\
	&\qquad \left. + \frac{1}{2}(-3 + v^2 + u^2)^2 \left[ \Theta\left(1 - \frac{u+v}{\sqrt{3}} \right) + \Theta\left( 1 + \frac{u-v}{\sqrt{3}} \right) \right] \right\},
 \label{eq:Is}
\end{align}
where $\kappa = k/k_{\overline{\RM}}$ with $k_{\overline{\RM}}$ being the scale that would have corresponded to the horizon scale at the beginning of the MD era, if the axion rotation had continued to behave like matter and come to dominate the Universe (though it does not actually dominate in the scenario considered in this subsection). 
Strictly speaking, the late-time limit is not a good approximation for the perturbations that enter the horizon around $\eta_\text{MK}$ because the above expressions are valid only when the perturbations are non-relativistic matter isocurvature perturbations. 
In the next subsection~\ref{subsec:late_gw}, we will discuss the GWs induced by the perturbations that enter the horizon after the MK transition (during $\eta > \eta_\MK$).

\begin{figure}[t]
        \centering \includegraphics[width=0.7\columnwidth]{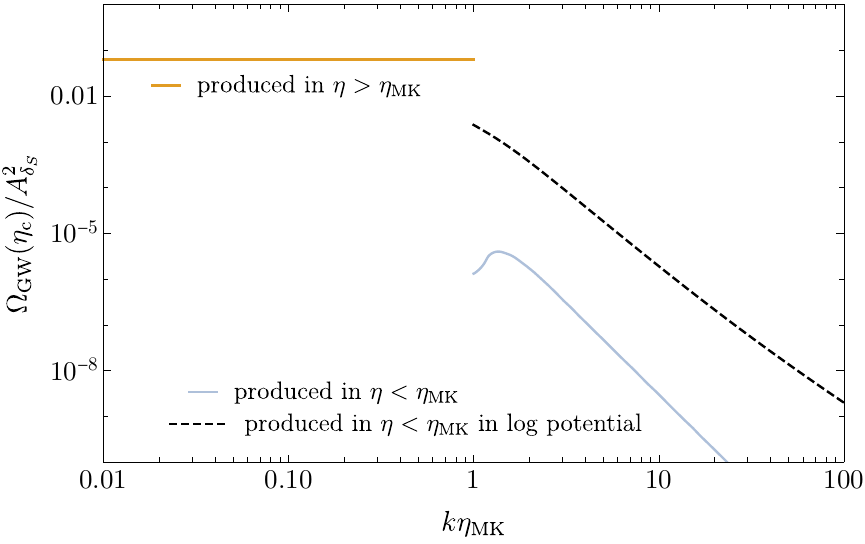}
        \caption{ 
        $\Omega_\GW(\eta_\cc)$ in the case without the rotation domination. 
        The (light) blue line is the GWs produced in $\eta < \eta_\MK$, calculated with Eqs.~(\ref{eq:omega_gw_wo_ax_dom}) and (\ref{eq:p_s_ansatz}).
        The orange line is the GWs produced in $\eta > \eta_\MK$, calculated with Eqs.~(\ref{eq:delta_chi_ini}) and (\ref{eq:omega_gw_si}).
        The black dashed line is the GWs produced in $\eta < \eta_\MK$ in the case of the log potential, discussed in Appendix~\ref{app:induced-GW} (Eq.~(\ref{eq:omega_gw_log_pot})).
        We take $F_S = 0.1$ for all lines.
		}
        \label{fig:gw_wo}
\end{figure}

For convenience, let us here connect $\kappa$ with other quantities.  
We introduce the ratio of $\rho_S/\rho_\rr$ at $\eta_\MK$ as
\begin{equation}
    \label{eq:F_s_def}
    F_S \equiv \frac{\rho_S(\eta_\MK)}{\rho_\rr(\eta_\MK)}. 
\end{equation}
We can easily see $F_S = k_{\overline{\RM}} \eta_\text{MK}$. 
Then, we can express $\kappa = k\eta_\text{MK}/F_S$.
Since we are focusing on the GWs induced by the scalar perturbations that enter the horizon at $\eta < \eta_\text{MK}$, let us consider the following power spectrum,
\begin{align}
	\mathcal P_{\delta_S}(k) = A_{\delta_S} \Theta(k - 1/\eta_\text{MK}).
    \label{eq:p_s_ansatz}
\end{align}
The blue line in Figure~\ref{fig:gw_wo} shows $\Omega_\GW$ with $x_c \to \infty$ in Eq.~(\ref{eq:omega_gw_wo_ax_dom}) for $k>1/\eta_\MK$. 
We do not show the blue line in $k < 1/\eta_\MK$ because the approximation of Eq.~(\ref{eq:j_lim}) is invalid for the perturbations that enter the horizon after $\eta_\MK$, as mentioned above.
See below Eq.~(\ref{eq:Is}) for the caveat.

The UV tail of the blue-line in \fig{fig:gw_wo} obeys the $k^{-4}$ power law.  This is because the isocurvature perturbations are not yet substantially converted to the curvature perturbations when high $k$ modes enter the horizon. This suppression involves $\kappa^{-2}$~\cite{Domenech:2021and,Domenech:2023jve}, and the induced GWs are produced by the second-order effect, while the assumed $\mathcal{P}_{\delta_S}(k)$ for $k > 1/\eta_\text{MK}$ does not involve another power of $k$. 
This explains the $k^{-4}$ suppression on the high-$k$ side.  
In \fig{fig:gw_wo}, we also show the GW spectrum in the log potential model by the dashed black line for comparison, discussed in Appendix~\ref{app:induced-GW}. 
In the log potential model, $c_s$ is non-negligible even in $\eta < \eta_\MK$, which leads to $k^{-2}$ UV tail.
The difference of the magnitude of the GWs between blue and black-dashed lines can be understood as follows.
The blue line in \fig{fig:gw_wo} mainly comes from the density perturbations, which give $\Omega_{\rm GW} \propto F_S^4$. On the other hand, the black-dashed line for the log potential is dominated by the velocity perturbation, which becomes larger for $c_s\neq 0$,  giving $\Omega_{\rm GW} \propto F_S^2$.

\subsection{Late-time GW production}
\label{subsec:late_gw}

After the KR transition, the contribution of the axion to the total energy density, and thus to the gravitational potential, becomes negligible.
However, significant GW production occurs even after $\eta_\text{KR}$ from the fluid velocity perturbation of the axion. Intuitively, this large contribution can be understood as follows.

The angular velocity imparted during the kick depends on the initial angle, and therefore, the regions with different inflationary fluctuation $\delta\theta_{\rm inf}$ also acquire a differential angular velocity $\delta\dot\theta$.
As a result, $\delta\theta$ grows from initial fluctuation even when the mode is superhorizon.
The total angular fluctuation at horizon re-entry ($a_k$) of comoving mode $k$ is then
$\delta\theta(k, a_k) = \delta\theta_{\rm inf}(k) + \delta(\Delta\theta)$,
where $\Delta\theta$ is the accumulated angular displacement from the kick until re-entry. The growth can be estimated as
\begin{align}\label{eq:delta_theta_estimate}
    \delta (\Delta\theta) \sim \int \delta \dot{\theta} \, dt \sim  \delta_S \, \left.\frac{\dot{\theta}}{H} \right|_{\rm MK} \sim \delta_S \, \Omega_{S,\rm MK}^{1/2} \frac{\mpl}{f_a} .
\end{align}
Since $(\dot{\theta}/H)_{\rm MK}\gg  1$ (or $\mpl/f_a \gg 1$), this superhorizon evolution can significantly enhance $\delta\theta(k, a_k)$ relative to the inflationary seed.
As the angular direction remains very light, the resulting $\delta\theta$ behaves like radiation after horizon re-entry~\cite{Eroncel:2025bcb,Eroncel:2025qlk}.
Physically, this corresponds to the phonon mode (sound waves) of the rotating condensate \cite{ Bodas:2025eca}.

$\delta \theta$ produces velocity perturbations, which can be sizable due to the superhorizon growth of $\delta\theta$.
These velocity perturbations then source GWs at horizon re-entry.
Interestingly, in the case without rotation domination, or $\Omega_{S,\rm MK} \approx F_S \ll1$, this contribution gives $h_{ij} \propto \delta\theta^2 \propto F_{S} \, \delta_S^2$, unlike the more suppressed contribution from the gravitational potential, $h_{ij} \propto \zeta_{\rm tot}^2 \sim F_{S}^2 \, \delta_S^2$.  
Therefore, the amplitude of the GW signal goes as $\Omega_{\rm GW} \propto F_S^2 \, \delta_S^4$, larger than expected from only considering curvature perturbation. 
From \eq{eq:delta_theta_estimate}, we see that the growth of $\delta \theta$ occurs dominantly around MK for all modes, irrespective of the duration between the kick and their horizon re-entry. 
Since the growth is the same for all superhorizon modes, the GWs produced exhibit a 
distinctive flat frequency spectrum (seen as orange lines in \figs{fig:omegw} and \ref{fig:gw_wo}).

In the following, we will perform a more detailed calculation of the evolution of the angular-direction field fluctuations $\delta \chi \equiv r \delta \theta$ and GW production from the velocity perturbations of the axion.

\subsubsection{
Growth of axion field fluctuation}
\label{subsec:angle_fluc}

We here estimate $\delta \chi$ in $\eta \gg \eta_\KR$ on superhorizon scales.
To this end, let us first obtain the total displacement of $\theta$ ($\Delta \theta$) due to the angular velocity ($\dot{\theta}$) in terms of $Y_\theta$:
\begin{align}
    \Delta \theta = \int^\infty_0 \dot \theta \dd t 
    = -\int^\infty_0 \frac{n_\theta}{r^2} \frac{\dd t}{\dd s} \dd s 
    = \frac{Y_\theta}{3} \int \frac{\dd s}{r^2 H},
    \label{eq:Delta_theta}
\end{align}
where $s$ is the entropy density.
The spatial fluctuation of $\theta $ at a later time is then $\delta (\Delta \theta)$, which can be related to $\delta Y_\theta$ using the expression of $\Delta \theta$. 
For simplicity, we parametrize $\rho_\rr = b_\rho T^4$ and $s = b_s T^3$ for shorthand notation of $b_\rho = \pi^2 g_\rho(T)/30$ and $b_s = 2 \pi^2 g_s(T) / 45$. 
With this, we can express the Hubble parameter as 
\begin{align}
    \label{eq:hubble_s_rho}
    H = \frac{(c s^{4/3} + \rho_S)^{1/2}}{\sqrt{3}\mpl},
\end{align}
where $c \equiv b_\rho b_s^{-4/3} \simeq 0.99 g_\rho(T)g_s(T)^{-4/3}$.
We here approximate the $s$ dependence of $\rho_S$ as 
\begin{align}
    \label{eq:rho_s}
    \rho_S \simeq \cfrac{r^2 \dot \theta^2}{2} = \begin{cases}  
    \cfrac{m_r n_\theta}{2} = \cfrac{m_r Y_\theta s}{2}  & (\eta < \eta_\MK) \\
    \cfrac{\dot \theta^2 f_a^2}{2} = \cfrac{n_\theta^2}{2 f_a^2} = \cfrac{Y_\theta^2 s^2}{2 f_a^2}  & (\eta > \eta_\MK)
    \end{cases},
\end{align}
where we have used $n_\theta = r^2 \dot \theta$.
In the following, we separately calculate $\Delta \theta$ induced before and after $\eta_\MK$ by assuming an instantaneous matter-kination transition for simplicity.
Note that, strictly speaking, $\rho_S$ should be $r^2 \dot \theta^2$ in $\eta \ll \eta_\MK$, which is different from Eq.~(\ref{eq:rho_s}) by a factor $2$. 
However, we stick to Eq.~(\ref{eq:rho_s}) because it simplifies calculation thanks to the continuity at $\eta_\MK$ and the factor difference does not change the order of $\delta (\Delta \theta)$. 
With Eq.~(\ref{eq:rho_s}), we can reexpress $F_S$ as 
\begin{equation}
   \label{eq:F_s_def2}
    F_S = \frac{1}{2c} \left( \frac{m_r Y_\theta^2}{f_a} \right)^{2/3}. 
\end{equation}

\noindent
{\bf 1) In $\eta < \eta_\MK$}

\noindent
For $\Delta \theta$ at $\eta_\MK$, we can express Eq.~(\ref{eq:Delta_theta}) as
\begin{align}
    \Delta \theta|_{<\eta_\MK} &= \frac{Y_\theta \mpl}{\sqrt{3}} \int^\infty_{s_\MK} \frac{\dd s}{r^2 (c s^{4/3} + \frac{m_r Y_\theta s}{2})^{1/2}} \nonumber \\
    &= \frac{m_r \mpl}{\sqrt{3}} \int^\infty_{s_\MK} \frac{\dd s}{s (c s^{4/3} + \frac{m_r Y_\theta s}{2})^{1/2}},
\end{align}
where we have used $Y_\theta s = r^2 \dot \theta \simeq r^2 m_r$ in $\eta < \eta_\MK$ and $s_\MK (= m_r f_a^2/Y_\theta)$ is the entropy density at $\eta_\MK$.
Changing the variable with $x \equiv s (mY_\theta/(2c))^{-3}$, we can re-express this as 
\begin{align}
    \Delta \theta|_{<\eta_\MK} 
    &= \frac{\mpl}{\sqrt{3}} \frac{4c^{3/2}}{m_r Y_\theta^2} \int^\infty_{x_\MK} \frac{\dd x}{x (x^{4/3} + x)^{1/2}} \nonumber \\
    &= \frac{\mpl}{\sqrt{3}} \frac{4c^{3/2}}{m_r Y_\theta^2} \left[ \frac{2(-1 + 2 x^{1/3}) \sqrt{x + x^{4/3}}}{x}\right]^\infty_{x_\MK}.
\end{align}
Note that $x_\MK = F_S^{-3}$.
Therefore, $x_\MK > 1$ corresponds to the case of rotation non-dominance.  
We can approximate $\Delta \theta$ as 
\begin{align}\label{eq:d_theta_1}
    &\Delta \theta|_{<\eta_\MK} \nonumber \\ 
    &\simeq \begin{cases}
     \cfrac{\mpl}{\sqrt{3}} \cfrac{4c^{3/2}}{m_r Y_\theta^2} \cfrac{2}{\sqrt{x_\MK}} \left( 1- \cfrac{3}{2}x_\MK^{1/3}\right)= 2\sqrt{\cfrac{2}{3}} \cfrac{\mpl}{f_a}\left(1 - 3c \left( \cfrac{f_a}{m_r Y_\theta^2} \right)^{2/3}  \right) & ( F_S \gg 1  ) \\[2em]
     \cfrac{\mpl}{\sqrt{3}} \cfrac{4c^{3/2}}{m_r Y_\theta^2} \left(\cfrac{3}{2} x_\MK^{-2/3}\right) = \cfrac{\sqrt{3}\mpl}{2\sqrt{c}} \left( \cfrac{m_r Y^2_\theta}{f_a^4}\right)^{1/3} &( F_S \ll 1)
    \end{cases}.
\end{align}
From this, we can see that 
\begin{align}
    \label{eq:d_th_before_mk}
    \delta (\Delta \theta)|_{<\eta_\MK} \simeq \begin{cases}
    8 \sqrt{\cfrac{2}{3}} \cfrac{c \mpl}{f_a}\left( \cfrac{f_a}{m_r Y_\theta^2} \right)^{2/3}  \cfrac{\delta Y_\theta}{Y_\theta}  & (x_\MK \ll 1 \,,\, F_S \gg 1) \\
    \cfrac{\sqrt{3}\mpl}{2\sqrt{c}} \left( \cfrac{m_r Y^2_\theta}{f_a^4}\right)^{1/3} \cfrac{2}{3} \cfrac{\delta Y_\theta}{Y_\theta}  & (x_\MK \gg 1 \,,\, F_S \ll 1)
    \end{cases}.
\end{align}

\noindent
{\bf 2) In $\eta > \eta_\MK$}

\noindent
Next, we consider the contribution in $\eta > \eta_\MK$:
\begin{align}
    \Delta \theta|_{>\eta_\MK} &\simeq \frac{Y_\theta \mpl}{\sqrt{3} f_a^2} \int^{s_\MK}_0 \frac{\dd s}{\left( c s^{4/3} +  \frac{Y_\theta^2 s^2}{2 f_a^2} \right)^{1/2}},
\end{align}
where we have used $r \simeq f_a$ in $\eta > \eta_\MK$.
We here change the variable with $y \equiv s (2cf_a^2/Y_\theta^2)^{-3/2}$:
\begin{align}
    \Delta \theta|_{>\eta_\MK} &\simeq 
    \sqrt{\frac{2}{3}}\frac{\mpl}{f_a} \int^{y_\MK}_0 \frac{\dd y}{\left( y^{4/3} +  y^2 \right)^{1/2}} \nonumber \\
    &= \sqrt{\frac{2}{3}}\frac{\mpl}{f_a} \left[ -2 \log y + 3 \log\left[ y + \sqrt{y^{4/3} + y^2} \right] \right]^{y_\MK}_0.
\end{align}
We note $y_\MK = x^{-1/2}_\MK = F_S^{3/2}$.
This means that the axion rotation never dominates if $y_\MK < 1$.
Then, we can approximate this as 
\begin{align}\label{eq:d_theta_2}
    \Delta \theta|_{>\eta_\MK} \simeq \begin{cases}
      \sqrt{\cfrac{2}{3}}\cfrac{\mpl}{f_a} \log (8 y_\MK) =  \sqrt{\cfrac{2}{3}}\cfrac{\mpl}{f_a} \log \left( 2\sqrt{2} \cfrac{m_r Y^2_\theta}{c^{3/2} f_a} \right) & (y_\MK \gg 1 \,,\, F_S \gg 1)\\
      \sqrt{\cfrac{2}{3}}\cfrac{\mpl}{f_a} (3 y_\MK^{1/3}) = \cfrac{\sqrt{3} \mpl}{\sqrt{c}} \left( \cfrac{m_r Y_\theta^2}{f_a^4} \right)^{1/3}  & (y_\MK \ll 1 \,,\, F_S \ll 1)      
    \end{cases}.
\end{align}
From these, we can obtain 
\begin{align}
    \label{eq:d_th_after_mk}
    \delta (\Delta \theta)|_{>\eta_\MK} \simeq \begin{cases}
    2\sqrt{\cfrac{2}{3}}\cfrac{\mpl}{f_a} \cfrac{\delta Y_\theta}{Y_\theta}  & (y_\MK \gg 1 \,,\, F_S \gg 1) \\    
     \cfrac{\sqrt{3} \mpl}{\sqrt{c}} \left( \cfrac{m_r Y_\theta^2}{f_a^4} \right)^{1/3} \cfrac{2}{3} \cfrac{\delta Y_\theta}{Y_\theta}  & (y_\MK \ll 1 \,,\, F_S \ll 1) 
    \end{cases}.
\end{align}

Combining \eqs{eq:d_th_before_mk}{eq:d_th_after_mk} and using Eq.~(\ref{eq:F_s_def2}), we finally obtain the expression of $\delta \chi$ on superhorizon in $\eta \gg \eta_\MK$ as 
\begin{align}
\label{eq:delta_chi_ini}
   \frac{\delta \chi_i}{\mpl} \simeq \frac{f_a \delta \theta_i}{\mpl} \simeq 
   \begin{cases}
    2\sqrt{\cfrac{2}{3}} \delta_{S,i}  & (F_S \gg 1) \\    
    \sqrt{6} F_S^{1/2} \delta_{S,i} & (F_S \ll 1)
    \end{cases},
\end{align}
where we have used $\delta Y_\theta/Y_\theta = \delta_S$ and put the subscript ``$i$'' to explicitly mean the value when the perturbations are on superhorizon scales in $\eta \gg \eta_\MK$.
This equation determines the initial condition of $\delta \chi$ on superhorizon scales.
We see that it matches our naive estimate in \eq{eq:delta_theta_estimate}.
In Appendix~\ref{subsubsec:perturbations}, we check this estimate by numerically solving the equation of motion for the perturbations with the fluid picture (see \fig{fig:pertb_ax_dom_k}).

$\delta \chi$ finally enters the horizon.
Since the angular-direction field does not have a potential, $\delta \chi$ follows
\begin{equation}
	\delta \chi_\bfk'' + 2 \mathcal H \delta \chi_\bfk' + k^2 \delta \chi_\bfk = 0.
\end{equation}
For convenience, we define the transfer function as 
\begin{align}
    \delta \chi_\bfk(\eta) = \delta \chi_{i,\bfk} T(k\eta).  
    \label{eq:chi_trans}
\end{align}
In the following, we omit the subscript ``$i$'' for brevity.
For example, we express the power spectrum of $\delta \chi_i$ just as $\mathcal P_{\delta \chi}$.

\subsubsection{
Induced GWs
}
\label{sssec:late_gw}

Let us calculate the induced GWs.
Since we can neglect the gravitational potential in $\eta \gg \eta_\KR$ or in the case of $F_S \ll 1$, we can express the equation of motion for tensor perturbation in those cases as 
\begin{align}
    h_{ij}'' + 2 \mathcal H h_{ij}' + k^2 h_{ij} \simeq \frac{4}{\mpl^2} \mathcal T_{ij}^{\ \ lm} \partial_l \delta \chi \partial_m \delta \chi,
\end{align}
where $\mathcal T$ is the traceless-transverse projection operator.
Note that the source term corresponds to the velocity field contribution ($\propto \mathcal T_{ij}^{\ \ lm}v_l v_m$) in the fluid picture.
By solving this equation of motion with the Green function, we finally obtain~\cite{Inomata:2021zel}
\begin{align}
\label{eq:ph_express_v_u}
\mathcal P_h(k,\eta) = \frac{4}{\mpl^4}&
\int^\infty_0 \dd v \int^{|1+v|}_{|1-v|} \dd u \left[ \frac{4v^2 - (1 + v^2 - u^2 )^2}{4uv} \right]^2 I^2(u, v, k, \eta) \mathcal P_{\delta \chi}(u k) \mathcal P_{\delta \chi} ( v k ),
\end{align}
where $I(u,v,k,\eta)$ is defined as
\begin{align}
  \label{eq:i_vux_def}
  I(u,v,k,\eta) \equiv k^2 \int^\eta_0 \dd \bar \eta \, g_k(\eta; \bar \eta) T(uk\bar\eta) T(vk\bar\eta).
\end{align}
$g_k(\eta;\bar\eta)$ is the Green function that follows 
\begin{align}
    g''_k(\eta;\bar \eta) + 2 \mathcal H g'_k(\eta;\bar \eta) + k^2  g_k(\eta;\bar \eta) = \delta(\eta- \bar \eta),
\end{align}
where the prime denotes the derivative with respect to $\eta$ (not $\bar \eta$).

\noindent
{\bf 1) GWs induced during RD era}

We can express the energy density parameter of GWs induced by the perturbations that enter the horizon in the late time (during RD era) as 
\begin{align}
    \label{eq:omega_gw_rad_late}
 \Omega_\GW(\eta, k) = \frac{(k \eta)^2}{6\mpl^4}&
\int^\infty_0 \dd v \int^{|1+v|}_{|1-v|} \dd u \left[ \frac{4v^2 - (1 + v^2 - u^2 )^2}{4uv} \right]^2 \overline{I^2(u, v, k, \eta)} \mathcal P_{\delta \chi}(u k) \mathcal P_{\delta \chi} ( v k ).
\end{align}
For the perturbations that enter the horizon during the RD era in $\eta \gg \eta_\MK$, the transfer function becomes
\begin{equation}
\label{eq:trans_rd}
    T(x) = \frac{\sin x}{x} \ \ \text{(during RD era)}.
\end{equation}
In addition, the Green function during a RD era is given by 
\begin{align}
    k g_k(\eta;\bar \eta) = \frac{\bar x}{x} \sin(x - \bar x) \ \ \text{(during RD era)}.
\end{align}
Substituting these into Eq.~(\ref{eq:i_vux_def}) and taking the oscillation average in the subhorizon limit ($k\eta \to \infty$), we obtain 
\begin{align}
         \overline{I^2(u,v,x(\gg 1))} \simeq \frac{\pi}{32 u^2 v^2 x^2} \left[ \pi^2 + \left(\log \left[ \frac{1-(u-v)^2}{(u+v)^2 -1} \right]\right)^2 \right],
\end{align}
where $x \equiv k\eta$.
For simplicity, let us assume a scale-invariant power spectrum:
\begin{align}
   \frac{\mathcal P_{\delta \chi}(k)}{\mpl^2} = A_{\delta \chi}.
\end{align}
Substituting this into Eq.~(\ref{eq:omega_gw_rad_late}) and performing the $u$ and $v$ integrals, we find $\Omega_\GW$ asymptotes to 
\begin{equation}
    \label{eq:omega_gw_si}
    \Omega_\GW(\eta_c, k) \simeq 0.175 A_{\delta \chi}^2.
\end{equation}
Note that the GW spectrum in the above expression is flat. This is because the growth of the source perturbations ($\delta \chi$) is independent of the scale, as can be seen from \eq{eq:delta_chi_ini}. Also, in the case of rotation non-dominance, $\Omega_{\rm GW} \propto F_S^2\, \delta_S^4$ as expected.

We use \eq{eq:omega_gw_si} in $k<1/\eta_\KR$ of figure~\ref{fig:omegw} and $k<1/\eta_\MK$ of figure~\ref{fig:gw_wo}.
The frequency corresponding to $1/\eta_\KR$ is given by Eq.~(\ref{eq:nu_KR}). 
Also, the frequency corresponding to $1/\eta_\MK$ in the absence of the axion domination can be calculated with Eq.~(\ref{eq:nu_KR}) by replacing $\KR \to \MK$ in the expression.
We note that the discontinuous jump of the GW spectrum at $k = 1/\eta_\KR$ in figure~\ref{fig:omegw} and $k = 1/\eta_\MK$ in figure~\ref{fig:gw_wo} (the difference between orange and blue lines) is artificial.
To obtain the precise GW spectrum around those scales, we need to take into account the fact that, around those scales, 1) $\delta \chi$ deviates from Eq.~(\ref{eq:delta_chi_ini}), and 2) the contribution from the gravitational potential is non-negligible. 
We leave the analysis on these modifications for future work.

\noindent
{\bf 2) GW spectrum for CMB B-mode}

Finally, let us roughly estimate the GW spectrum on the scales constrained through the CMB B-mode, which corresponds to the snapshot of the tensor perturbations at the CMB recombination. 
For simplicity, we hereafter focus on the superhorizon tensor perturbations at the recombination, which mainly contribute to the large-scale (small-$\ell$) B-mode.

For superhorizon tensor perturbations, we can approximately obtain the Green function as 
\begin{equation}
	\tilde g''(\eta;\bar\eta) + 2 \mathcal H \tilde g'(\eta;\bar\eta) = \delta(\eta- \bar\eta).
\end{equation}
To derive $\tilde g$, we first solve the following differential equation:
\begin{align}
	v'' + 2 \mathcal H v' = 0. 
\end{align}
One solution is $v_1 = C$ with $C$ some constant. 
The other solution is
\begin{align}
	&\frac{v''_2}{v'_2} = -2 \frac{a'}{a} \nonumber \\
	\Rightarrow \ 	
	& v_2(\eta) = D \int^\eta \dd \eta' a^{-2}(\eta'),
\end{align}
where $D$ is some constant of integration.
The retarded Green function is then given by~\cite{Baumann:2007zm}
\begin{align}
	\tilde g(\eta;\bar\eta) &= \Theta(\eta-\bar\eta) \frac{v_1(\eta) v_2(\bar\eta) - v_1(\bar \eta) v_2(\eta)}{v_1'(\bar \eta) v_2(\bar \eta) - v_1(\bar \eta) v_2'(\bar \eta)} \nonumber \\
	&= \Theta(\eta-\bar\eta) a^2(\bar \eta) \int^\eta_{\bar \eta} \dd \eta' a^{-2}(\eta').
	\label{eq:green}
\end{align}
The scale factor around the last-scattering time can be expressed as~\cite{Mukhanov:991646} 
\begin{align}
  a(\eta) = a_\equ \left( \left(\frac{\eta}{\eta_*}\right)^2 + 2 \left(\frac{\eta}{\eta_*}\right) \right),
  \label{eq:a_r_m}
\end{align}
where $\eta_* \equiv \eta_\equ/(\sqrt{2}-1)$ with $\eta_\equ$ the conformal time at the matter-radiation equality.
The Planck TT,TE,EE+lowE+lensing+BAO result gives~\cite{Planck:2018vyg}
\begin{align}
  \label{eq:keq}
  \frac{k_\equ}{\Mpc^{-1}} = 0.010339 \pm 0.000063,
\end{align}
where $k_\text{eq} \equiv a_\equ H_\equ = 2(2-\sqrt{2})/\eta_\equ$.
This leads to 
\begin{align}
  \frac{\eta_\equ}{\Mpc} = 113.32 \pm 0.69.
\end{align}
Then, we can obtain the primitive function of $a^{-2}$ as
\begin{align}
	F(\eta) \equiv \int^\eta \dd \eta' a^{-2}(\eta') = -\frac{\eta_*}{4a_\equ^2} \left[ \frac{2(y+1)}{y(y+2)} + \ln \left(\frac{y}{y+2}\right) \right],
\end{align}
where $y \equiv \eta/\eta_*$.
Using this, we can reexpress Eq.~(\ref{eq:green}) as 
\begin{align}
	\tilde g(\eta;\bar \eta) &= \Theta(\eta-\bar\eta) a^2(\bar \eta) \left( F(\eta) - F(\bar \eta) \right).
\end{align}

Once $\delta \chi$ enters the horizon, it starts to decay and finally stops inducing GWs. 
This roughly sets the UV cutoff of $\delta \chi$ at each time. 
Given this, we here simplify the calculation by focusing on the tensor perturbations and angular field fluctuations on superhorizon scales at the last scattering. 
Specifically, we consider the following power spectrum:
\begin{align}
	\frac{\mathcal P_{\delta \chi}(k)}{\mpl^2} = A_{\delta \chi} \Theta(1/\eta_\ls - k).
	\label{eq:ps}
\end{align}
We further set $T(x) \simeq 1$ for simplicity because the angular perturbations with $k < 1/\eta_\ls$ are outside the horizon at $\eta_\ls$. 
Substituting $g_k = \tilde g$, we can analytically solve Eq.~(\ref{eq:i_vux_def}) as 
\begin{align}
	I(u,v,k,\eta) &\simeq k^2 \int^\eta_{0} \dd \bar \eta \, \tilde g(\eta; \bar \eta) \nonumber \\
	&= (k\eta_*)^2 \left[ \frac{y(8+3y(4+y))}{30(2+y)} - \frac{4}{15} \ln \left(\frac{2+y}{2}\right) \right].
	\label{eq:i_uv_y}
\end{align}

\begin{figure}  
\centering \includegraphics[width=0.6\columnwidth]{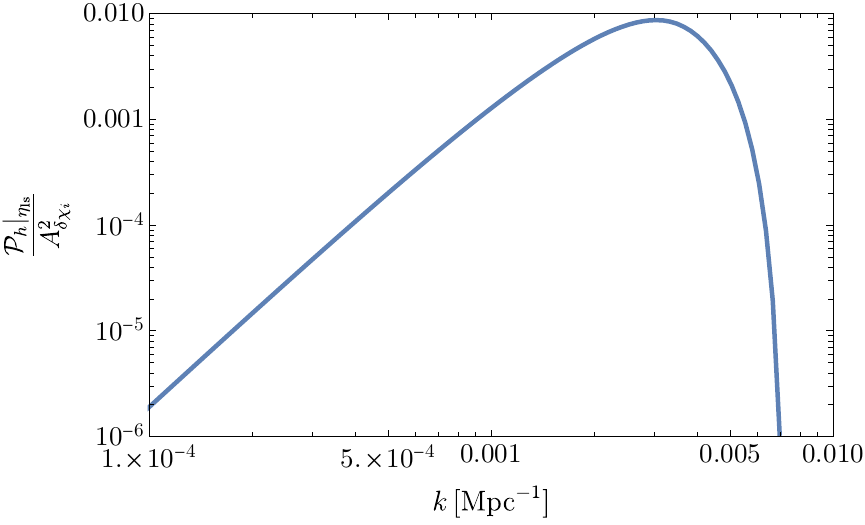}
\caption{ GW power spectrum at the last scattering surface, which is constrained by CMB B-mode measurements. 
We assume the power spectrum of $\delta \chi$ given by Eq.~(\ref{eq:ps}).
}
\label{fig:gw_sh}
\end{figure}

Let us substitute concrete values. 
The last scattering occurs around $z_\ls \sim 1100$~\cite{Dodelson:1282338}.
The redshift at the equality time is $z_\equ \simeq 3400$~\cite{Planck:2018vyg}.
From these, we can obtain $\eta_\ls/\eta_*$ by solving 
\begin{align}
	&\frac{1+z_\equ}{1+z_\ls} = \frac{a_\ls}{a_\equ} = \left(\frac{\eta_\ls}{\eta_*}\right)^2 + 2 \left(\frac{\eta_\ls}{\eta_*}\right) \nonumber \\ 
	\Rightarrow\  &
	\frac{\eta_\ls}{\eta_*} \simeq 1.0, \nonumber \\
	\Rightarrow\ &
	\eta_\ls \simeq 274\, \Mpc.
\end{align}
Substituting this into Eq.~(\ref{eq:i_uv_y}), we obtain 
\begin{align}
	I(u,v,k,\eta_\ls) &\simeq 0.15 \times (k\eta_\ls)^2.
	\label{eq:i_uv_y2}
\end{align}
With this and \eqs{eq:ph_express_v_u}{eq:ps}, we obtain \fig{fig:gw_sh}.
This figure shows $\mathcal P_h \propto k^3$ on large scales, which are superhorizon scales at the last scattering.
The $k^3$ slope is related to the causality~\cite{Cai:2019cdl}, which means that the angular field fluctuations around $k \sim 1/\eta_\ls$ mainly induce the large-scale tensor perturbations. 
The sharp UV cutoff in the GW spectrum in the figure is due to the UV cutoff in the power spectrum given by Eq.~(\ref{eq:ps}), which is introduced just for simplicity. 
If we consider a scale-invariant power spectrum without the UV cutoff, the UV cutoff in the GW spectrum will disappear and the GW spectrum should be smoothly connected to those induced during the RD era (Eq.~(\ref{eq:omega_gw_si})).
To see the precise spectrum between the large-scale limit spectrum, calculated here, and the spectrum of the GW induced during the RD era, we need to carefully take into account the transition from the RD to MD era, which is beyond the scope of this paper. 

Furthermore, when we compare the tensor power spectrum in \fig{fig:gw_sh} to the CMB B-mode constraints, we need to be careful about the fact that these tensor perturbations come from the isocurvature perturbations, which are different from radiation.
These induced GWs lead to a quadrupole moment of radiation through gravitational interaction, which is the origin of the B-mode. 
The detailed calculation of the B-mode in this case is left for future work.


\section{Anisotropy in the Induced GWB}
\label{sec:anisotropy}
In the previous section, we evaluated the production of the induced GWB.
Here, we compute its large-scale anisotropies. 
Because our scenario involves multiple cosmological components, the resulting anisotropy can exceed the level predicted by the consistency condition for the single-component case~\cite{Rey:2024giu}.

{\bf (1) Anisotropy in GWs produced at MK transition:} 

The strength of the GW signal produced at MK transition (from comoving mode $k_{\rm MK}$) is
\begin{align}
    \rho_{\rm GW, MK}(k_{\rm MK}) &= \mathcal{C} \left[\paren{\tilde{\Omega}_S \zeta_S(k_{\rm MK})}^4 \rho_{\rm tot} \right]_{\rm MK}  \\ \notag 
    &= \mathcal{C} \left[ \paren{\frac{3\rho_S}{3 \rho_S + 2 \rho_{\rm rad}} }^4 (\rho_S+\rho_{\rm rad}) \right]_{\rm MK} \paren{\frac{\delta Y_{\theta} (k_{\rm MK})}{Y_{\theta}}}^4,
\end{align}
where $\mathcal{C}$ collects constant prefactors that are irrelevant for the analysis in this section, 
$\rho_{\rm rad}$ is the energy density in the SM radiation, and we have taken $w_{S} \approx 1$. 
The fractional energy density in GWB today corresponding to the frequency $\nu_\MK$ (the mode $k_{\rm MK}$), is
\begin{align}\label{eq:GW_MK}
    \Omega_{\rm GW,0}(\nu_{\rm MK}) &= \Omega_{\rr,0} 
    \left. \frac{\rho_{\rm GW} (k_{\rm MK})}{\rho_{\rr}} \right|_{\rm today} \approx \Omega_{\rr,0} \left. \frac{\rho_{\rm GW} (k_{\rm MK})}{\rho_{\rr}} \right|_{\rm MK} \nonumber \\
     &= \Omega_{\rr,0}  \,{\cal C} \,\left[ \paren{\frac{3\rho_S}{3 \rho_S + 2 \rho_{\rm rad}} }^4 \, \frac{\rho_S+\rho_{\rm rad}}{\rho_{\rm rad}} \right]_{\rm MK} \paren{\frac{\delta Y_{\theta} (k_{\rm MK})}{Y_{\theta}}}^4.
\end{align}

At MK transition, the axion energy density $\rho_{S ,\rm MK} = m_r^2 f_a^2$ is fixed.
However, $\rho_{\rr} $ at MK varies due to the long-wavelength fluctuations in the field $S$, which shift the timing of the transition.
In addition, the ratio $\delta Y_{\theta}(k_\MK)/Y_\theta$ also exhibits long-wavelength modulation.
Note that we are considering two distinct wavelength modes here. For convenience, we denote the short-wavelength modes (responsible for GW production) by $k_s$ and the long-wavelength ones (responsible for large-scale anisotropy) by $k_l$.
Since $Y_\theta$ is conserved, the short-wavelength fluctuation $\delta Y_\theta(k_s)$ can be evaluated at an earlier time, for example, at the moment of the angular kick. On constant-entropy slices, $\delta Y_\theta(k_s) \propto \delta n_\theta(k_s) \propto \delta\theta(k_s)$, and since the inflationary $\delta\theta$ fluctuation is stochastic, $\delta Y_\theta(k_s)$ does not exhibit long-wavelength modulation, provided $Y_\theta$ depends linearly on the initial $\theta$.\footnote{In principle,  Eq.\eqref{eq:n_theta} implies a nonlinear dependence: $\delta Y_{\theta} (k) \propto \delta n_\theta (k) \propto \cos(n\theta) \,\delta \theta(k)$, 
which induces a long-wavelength modulation in $\delta Y_\theta(k_s)$. Specifically,
\begin{align}
    \frac{\delta [\delta Y_\theta (k_s)] (k_l)}{\delta Y_\theta (k_s)} \sim - n \frac{\sin(n \theta_{\rm init})}{\cos(n \theta_{\rm init} )} \delta \theta(k_l) = -\frac{\sin^2(n \theta_{\rm init})}{\cos^2(n \theta_{\rm init})} \delta_S (k_l). \notag
\end{align}
The size of this modulation depends on the initial angle
$\theta_{\rm init}$. We take $\tan(n \theta_{\rm init})<1$, such that this contribution to large-scale anisotropy is subdominant. Nevertheless, this effect can be readily incorporated into the analysis if needed.}
However, in the ratio $\delta Y_\theta(k_s)/Y_\theta$, the denominator $Y_\theta$ is actually the \emph{local average} over a region larger than the relevant GW production scale.
On long wavelengths, using separate universe approach, we can see that this local average is modulated:
\begin{align}\label{eq:aniso_delta_Y_short}
    \frac{\delta Y_\theta (k_s)}{Y_\theta} \approx \frac{\delta Y_\theta (k_s)}{\overline{Y}_\theta} \paren{1-\frac{\delta Y_\theta (k_l)}{\overline{Y}_\theta} },
\end{align}
where we have decomposed the local average into the global part and long-wavelength fluctuation: $Y_\theta \approx \overline{Y}_\theta + \delta Y_{\theta}(k_l)$. This modulation is a generic effect for any quantity sourced by small-scale fluctuations, including the induced GWs and the axion radiation discussed later in section~\ref{sec:constraint_axions}.

To get the long-wavelength fluctuation in $\rho_{\rr, \rm MK}$, let us first relate it to $Y_{\theta}$:
\begin{align}
    s = \frac{n_{\theta}}{Y_{\theta}} \xrightarrow{\rm at\, MK} \frac{m_r f_a^2}{Y_{\theta}}  \qquad \rightarrow \qquad \rho_{\rr ,\rm MK}\simeq  c\paren{\frac{m_r f_a^2}{Y_{\theta}}}^{4/3} .
\end{align}
Then, the long-wavelength fluctuation in $\rho_{\rr, \rm MK}$ is
\begin{align}\label{eq:deltaRad_to_deltaY}
    \left. \frac{\delta \rho_{\rr}(k_l)}{\rho_{\rr}}\right|_{\rm MK} =  -\frac{4}{3} \frac{\delta Y_{\theta} (k_l)}{Y_{\theta}} = -4 \zeta_S (k_l).
\end{align}
Therefore, using Eqs.~(\ref{eq:GW_MK}), (\ref{eq:aniso_delta_Y_short}), and (\ref{eq:deltaRad_to_deltaY}), the large-scale anisotropy in GWB today at the frequency corresponding to $k_{\rm MK}$ is
\begin{align}\label{eq:deltaGW_MK}
    \delta_{\rm GW}(k_l; \nu_{\rm MK}) \equiv \frac{\delta [\Omega_{\rm GW,0}(\nu_{\rm MK})] (k_l)}{\Omega_{\rm GW,0}(\nu_{\rm MK})} = 4 \paren{-11+\Omega_{S, \rm MK}+\frac{24}{2+\Omega_{S, \rm MK}}  } \zeta_S (k_l).
\end{align}
In the case of rotation dominance, $\Omega_{S, \rm MK} \approx 1$, and the anisotropy in GWB is $\delta_{\rm GW} = -8 \zeta_{S}$.
On the other hand, if the rotation remains subdominant such that $\Omega_{S, \rm MK} <1$ but $\tilde{\Omega}_{S,\rm MK} \zeta_{S} > \zeta_\phi$, then $\delta_{\rm GW} \approx 4 \zeta_{S}$.

Interestingly, the GWB frequency $\nu_{\rm MK}$ also exhibits large-scale fluctuations,
\begin{align}
    \nu_{\rm MK} = \frac{H_{\rm MK}}{2 \pi} \frac{a_{\rm MK}}{a_0} \approx \frac{H_{\rm MK}}{2 \pi} \paren{\frac{T_0}{T_{\rm MK}}} \approx \frac{H_{\rm MK} }{2 \pi} T_0 \,b_{\rho,\MK}^{1/4}\,\rho_{\rr,\rm MK}^{-1/4},
\end{align}
giving  
\begin{equation}
    \frac{\delta \nu_\text{MK} (k_l)}{\nu_\text{MK}} = \frac{\delta H_\text{MK} (k_l)}{H_\text{MK}} - \frac{1}{4} \left. \frac{\delta \rho_{\rr} (k_l)}{\rho_{\rr}} \right|_{\rm MK}.
\end{equation}
Using $3 H^2_{\rm MK} \mpl^2 = \rho_{\rm tot, MK}$, 
\begin{align}
    \frac{\delta H_\text{MK} (k_l)}{H_\text{MK}} = \frac{1}{2} \Omega_\text{rad,MK} \left. \frac{\delta \rho_\text{rad} (k_l)}{\rho_\text{rad}} \right|_{\rm MK}.
\end{align}
Then together with \eq{eq:deltaRad_to_deltaY}, we obtain
\begin{align}\label{eq:deltaNu_MK}
    \frac{\delta \nu_\text{MK}(k_l)}{\nu_\text{MK}} = (2 \Omega_{S,\rm MK} - 1) \zeta_S (k_l) .
\end{align}
Note that $\nu_{\rm MK}$ can be identified by looking for a distinct feature in the spectrum. 
In the case of rotation dominance, there is a local peak in the GW spectrum close to $\nu_{\rm MK}$ as seen in \fig{fig:omegw}.\footnote{The contributions from the modes re-entering before MK in a specific model (for example, the contribution in the black dashed line in \fig{fig:omegw}) might shift the peak towards a higher frequency. However, we still expect the shifted peak to also exhibit a similar long-wavelength modulation.}
In the case of rotation non-dominance, $\nu_{\rm MK}$ can be identified by 
the change in the slope of the GW spectrum from flat for $\nu <\nu_{\rm MK}$ to damped for $\nu >\nu_{\rm MK}$, as seen in \fig{fig:gw_wo}.

In the case with rotation domination, we expect another dip feature around $\nu_{\rm KR}$, as the spectrum transitions between the blue and the orange line in \fig{fig:omegw}.
The anisotropy in the amplitude of the GWB at $\nu_{\rm KR}$ can be calculated using the same method as above, but instead using the energy densities at KR in \eq{eq:GW_MK}, which is characterized by the condition $\rho_S = \rho_{\rm rad}$. Then 
\begin{align}\label{eq:deltaGW_KR}
     \delta_{\rm GW}(k_l; \nu_{\rm KR}) \equiv \frac{\delta [\Omega_{\rm GW,0}(\nu_{\rm KR})] (k_l)}{\Omega_{\rm GW,0}(\nu_{\rm KR})} = -12 \zeta_S (k_l).
\end{align}
The location of the dip, i.e., $\nu_{\rm KR}$, also modulates. From Eq.~\eqref{eq:nu_KR} and using $T_{\rm KR} \propto Y_{\theta}^{-1}$, which can be obtained from Eq.~(\ref{eq:rho_s}) and $\rho_{S,\KR} = \rho_{\rr,\KR}$,
we obtain  $\nu_{\rm KR} \propto Y_\theta^{-1}$.
Therefore, its large-scale fluctuation is 
\begin{equation}\label{eq:deltaNu_KR}
    \frac{\delta \nu_\kr(k_l)}{\nu_\kr} = -\frac{\delta Y_{\theta} (k_l)}{Y_{\theta}} = -3 \zeta_S (k_l).
\end{equation}

Since both the amplitude ($\Omega_{\rm GW}$) and the locations of the features ($\nu_{\rm MK}$ and $\nu_{\rm KR}$) exhibit large-scale fluctuations, their simultaneous detection could allow for independent extraction of $\zeta_S(k_l)$ and $\Omega_{S,\rm MK}$, provided both fluctuations are measurable. 
The expressions in \eqs{eq:deltaGW_MK}{eq:deltaNu_MK} may be modified in scenarios with non-instantaneous matter-to-kination transitions, or \eqs{eq:deltaGW_KR}{eq:deltaNu_KR} could be slightly different after a more detailed treatment of the KR transition. Nevertheless, the presence of two distinct observables offers a promising opportunity to identify the axion rotation model as the origin of the induced GWB.

{\bf (2) Anisotropy in GWs produced during late radiation domination:} 
We denote by $k_{\rm rd}$ the comoving momenta of modes that re-enter the horizon during the later radiation domination (when the kination fluid is subdominant). These modes give rise to the approximately flat portion of the GW spectrum shown in \figs{fig:omegw} and \ref{fig:gw_wo}:
\begin{align}
    \rho_{\rm GW,rd}(k\subt{rd}) \sim \rho\subt{tot}(a\subt{rd}) \paren{\frac{\delta \chi (k\subt{rd})}{\mpl}}^4.
\end{align}
The phase fluctuation $\delta\chi$ for these modes grows predominantly during the MK epoch (as discussed at the beginning of \Sec{subsec:late_gw}), and thus inherits implicit dependence on parameter evaluated at MK:
\begin{align}
   \frac{\delta\chi(k\subt{rd})}{\mpl} =\frac{f_a \delta \theta(k\subt{rd})}{\mpl}  \sim \frac{\delta Y_\theta(k\subt{rd})}{Y_{\theta} } \Omega_{S,\rm MK}^{1/2}.
\end{align}
Following the same method as in the previous subsection,
the anisotropy in this flat region of the GWB is given by 
\begin{align}
    \delta\subt{GW}( k_l; \nu\subt{rd}) =  \sqbracket{(1+2 \Omega_{S})\frac{\delta\rho\subt{rad}}{\rho\subt{rad}}}_{\rm MK} = -4(1+2 \Omega_{S,\rm MK}) \, \zeta_S(k_l).
\end{align}
Remarkably, GWs sourced by modes that re-enter long after MK, when the field $S$ is already highly subdominant, still inherit large-scale anisotropies imprinted during the MK epoch. Since this flat portion of the spectrum is typically stronger and spans a broader frequency range (see  \figs{fig:omegw} and \ref{fig:gw_wo}), it presents a more promising target for detecting isocurvature GW anisotropies.


\section{Observational constraints}
\label{sec:constraints}

In this section, we discuss the observational constraints on our scenarios, which are from PBH production, axion radiation, isocurvature perturbations, CMB B-mode, and non-Gaussianity.

\subsection{PBH production}
\label{sec:constraint_pbh}

The production of induced GWB and PBHs is closely connected, as both arise from the same small-scale perturbations~\cite{Saito:2008jc, Saito:2009jt}. In scenarios where the scalar field $S$ dominates the energy density, PBH formation is enhanced during the eMD era due to a larger total curvature perturbation $\zeta_{\rm tot} \approx \zeta_S$ and the linear growth of matter overdensities, $\delta_{\rm matter} \propto a(t)$. This enhancement can lead to PBH overproduction, thereby constraining the amplitude of initial overdensities, $\delta_S$, and consequently limiting the strength of the resulting GWB.

The fraction of sub-volumes on a given comoving length scale ($1/k$) that collapse into PBHs is denoted by $\beta(M(k)) = \rho_{\rm PBH}(M(k))/\rho_{\rm total}(a_k)$, where $M(k)$ is approximately the horizon mass at the time when the mode $k$ re-enters the horizon.
The PBH formation during eMD is affected by the following factors:
\begin{enumerate}
    \item {\bf Spherical asymmetry within overdensity:} 
    In the matter-dominated era, PBH formation typically occurs well after the overdense region re-enters the horizon.
    During this time, even a small initial asymmetry in the overdensity can grow significantly. 
    These highly asymmetric regions are more likely to form virialized halos rather than PBHs. This effect is accounted for in the PBH formation probability as $\beta_{\rm asym} \sim {\sigma}_{S}^{5}$, where $\sigma_S = \sqrt{{\cal P}_{\delta_S}}$.
    \cite{khlopov1980primordial, Polnarev:1985btg, Harada:2016mhb, Harada:2017fjm, Kokubu:2018fxy, deJong:2021bbo, Harada:2022xjp}.
    This expression assumes a long duration of matter domination, while the eMD era in our model can be short.
    For a short eMD, we obtain (see \App{app:PBH_finiteMD} for derivation)

    \begin{align}\label{eq:betaPBH_asym}
    \beta_{\rm asym} \approx 
    \begin{cases}
        \dfrac{\delta_{\rm end}}{\delta_c} \exp\left[-\dfrac{\delta_{c}^{2}}{2 \delta_{\rm end}^{2}}\right]\quad  & (\sigma_S^{1/2}  \geq \dfrac{\delta_{\rm end}}{\delta_c}) \vspace{0.8em} \\
        \dfrac{\sigma_S^{5}}{(\delta_{\rm end}/\delta_c)^9} \exp\left[-\dfrac{\delta_{c}^{2}}{2 \delta_{\rm end}^{2}}\right]\quad  & (\sigma_S^{1/2}  < \dfrac{\delta_{\rm end}}{\delta_c}),
    \end{cases}\, 
    \end{align}
    where $\delta_c$ is the critical overdensity for collapse, and
    $\delta_{\rm end}(k)$ is the maximum overdensity reached, which is determined 
    either by the end of eMD or by the Jeans length consideration. For us, the latter is more important, as explained in point 3 below. 

    \item {\bf Inhomogeneity within overdensity:} 
    Even if the overdensity is spherically symmetric, significant radial inhomogeneity can suppress PBH formation. During the infall, 
    the velocity dispersion can become sufficiently large prior to the formation of an apparent horizon such that the overdensity disperses rather than form a PBH. 
    A lower bound on the probability of PBH formation 
    in a spherically symmetric but radially inhomogeneous overdensity was estimated in Refs.~\cite{khlopov1980primordial,Kokubu:2018fxy} to be $\beta_{\rm inhom} \sim \sigma_S^{3/2}$.

    \item {\bf Non-zero sound speed:}
    In this paper, we are considering a gradual transition from the matter to kination phase of $S$. One way of achieving this is to consider a log potential given in \eq{eq:pot_vs}.
    This results in a small but nonzero time-dependent sound speed $c_s$ around the end of the matter-like phase of $S$ (see for example \fig{fig:phi2}). This introduces a Jeans length ($1/k_J$), below which overdensity growth is halted. 
    Given the time dependence of $c_s$ in \eq{eq:cs_s}, the growth ceases before the end of eMD, determined by the criterion
    $k_{\rm phy} \approx k_J \rightarrow (a_{\rm end}/a_{k}) = 1/c_{s}^{2}(a_{\rm end})$, where $k_\text{phy}$ is the physical momentum of the perturbation, $a_k$ and $a_\text{end}$ are the scale factors when a comoving mode $k$ re-enters the horizon and when its growth stops, respectively.
    The maximum growth is then $\delta_{\rm end}(k) =\sigma_S\cdot (a_{\rm end}(k)/a_{k})$, which is then used in \eq{eq:betaPBH_eMD}. 
    The critical overdensity at collapse is $\delta_c = \frac{3(1+w_{\rm end})}{5+3w_{\rm end}} \sin\paren{\frac{\sqrt{\pi w_{\rm end}}}{1+3w_{\rm end}}}$~\cite{Harada:2013epa}, where $w_{\rm end} = c_s^2 (a_{\rm end})$.
\end{enumerate}
It is relatively straightforward to combine the effects of anisotropy and finite sound speed by substituting the critical density $\delta_c(c_s^2)$ into $\beta_\text{asym}$. On the other hand, it is nontrivial how to combine the effect of inhomogeneity.  In Ref.~\cite{Kokubu:2018fxy}, the effects of inhomogeneity and anisotropy were combined by simply multiplying $\beta_\text{inhom}$ and $\beta_\text{asym}$. We follow this prescription for simplicity. Taking into account the effects of asymmetry, inhomogeneity, and finite sound speed, 
the total PBH collapse fraction during the eMD era is given by\footnote{During kination dominance (KD), overdensities also grow linearly, $\delta \propto a$, albeit with possible oscillatory suppression~\cite{Harigaya:2023mhl, Eroncel:2025bcb}. 
However, since the Jeans length during KD is comparable to the horizon scale, subhorizon overdensities cannot grow efficiently. As a result, PBH formation is limited to horizon re-entry and is less efficient than during eMD. Even with large overdensities, $\delta_S \sim 10^{-2}$, the PBH collapse fraction remains lower during KD. Therefore, we focus on PBH production during the eMD era.} 
\begin{align}\label{eq:betaPBH_eMD}
    \beta_{\rm eMD} \approx \beta_{\rm asym}\times \beta_{\rm inhom} \approx
    \begin{cases}
        \frac{\sigma_S^{3/2} \delta_{\rm end}}{\delta_c} \exp\left[-\frac{\delta_{c}^{2}}{2 \delta_{\rm end}^{2}}\right]\quad  & (\sigma_S^{1/2} \geq \frac{\delta_{\rm end}}{\delta_c}) \vspace{0.8em} \\
        \frac{\sigma_S^{13/2}}{(\delta_{\rm end}/\delta_c)^9} \exp\left[-\frac{\delta_{c}^{2}}{2 \delta_{\rm end}^{2}}\right]\quad  & ( \sigma_S^{1/2} < \frac{\delta_{\rm end}}{\delta_c}).
    \end{cases}
\end{align} 

The fraction of DM energy density in PBH today, $f_{\rm PBH}$, can be written as 
\begin{align}
    \f{PBH} \equiv \left.\frac{\rho_{\rm PBH}}{\rho_{\rm DM}}\right|_{\rm today} & \simeq \int d 
    (\log M) \, \beta_{\rm MD}(M) \paren{\frac{a_{\rm KR}}{a_{\rm MK}}}^{3}  \paren{\frac{a_{\rm eq}}{a_{\rm KR}}}.
\end{align}
The scale factors account for the relative enhancement of PBH abundance during kination and radiation dominance.
The PBH constraints are often given for a monochromatic distribution $\int dM f_{\rm mono}(M_c)\,\delta(M-M_c)$. 
However, they can also be extended to a general mass function using the method in Ref.~\cite{Carr:2017jsz}, giving
\begin{align}\label{eq:fPBH}
    \int d(\log M) \frac{f_{\rm PBH}(M)}{f_{\rm mono}^{\rm max}(M)} \leq 1 ,
\end{align}
where $f_{\rm mono}^{\rm max}$ is the maximum allowed fraction considering observational constraints (taken from Ref.~\cite{carr2021constraints}). 

The mass of the PBH associated with the collapse of a comoving length scale $k^{-1}$ is approximately the horizon mass at the time of horizon re-entry of the mode $k$,
\begin{align}
    M(k) \approx \frac{4 \pi}{3} \rho_{\rm tot}(a_{k}) H_{k}^{-3} = 4 \sqrt{3} \pi \frac{\mpl^3}{\rho_{\rm tot}^{1/2}(a_k)},
\end{align}
where we have used $\rho_{\tot} = 3 H^2 \mpl^2$ in the second expression.
In the rotation dominance scenario (i.e. $F_S >1$), the modes re-entering during eMD, $k_{\rm MK} <k < k_{\rm RM}$, are most relevant for PBH production. For these modes, we find
\begin{align}
    \rho_{\rm tot}(a_k) \approx \paren{\frac{a_{\rm MK}}{a_k}}^3 \paren{\frac{a_{\rm KR}}{a_{\rm MK}}}^6  \rho_{\rm rad}(a_{\rm KR}).
\end{align}
Then using $\mpl \approx 2.4 \times 10^{18} \, {\rm GeV}$ and $M_\odot \approx 1.1\times 10^{57}\, {\rm GeV}$, we obtain
\begin{align}
    M(k) \approx 0.01\, \paren{\frac{g_{\rho}(a_k)}{106.75}}^{-1/2} \paren{\frac{a_{k}}{a_{\rm MK}}}^{3/2} \paren{\frac{a_{\rm MK}}{a_{\rm KR}}}^{3}  \paren{\frac{{\rm GeV}}{T_{\rm KR}}}^{2} M_{\odot}.
\end{align}

The large-scale distribution of PBHs is expected to reflect the long-wavelength fluctuations in $S$ (i.e. $\zeta_S(k_l)$), analogous to the case of induced GWs. Since PBHs contribute to the dark matter density, their isocurvature fluctuations are constrained by the CMB observations~\cite{Planck:2018jri}, leading to the bound 
\begin{align}\label{eq:fPBH_constraint}
    3 \f{PBH} \, \sqrt{\mathcal{P}_{\zeta_S}(k_{l})} < 5\times 10^{-6}.
\end{align}
\eqs{eq:fPBH_constraint}{eq:fPBH} together yield a stringent upper bound on $\delta_S (k_s)$ in the rotation-dominated scenario.

The situation is different for the case without rotation dominance.
Since $n_\theta$ is a bounded function of $\theta$ (as shown in \eq{eq:n_theta}), the distribution of $\delta n_\theta$ is approximately Gaussian near the mean but becomes skewed toward the tails, with hard cutoffs at both ends.%
\footnote{It may seem that if the radial direction is light during inflation (see appendix~\ref{app:susy_realization}), $\delta n_\theta$ would not have a cutoff due to fluctuations along $r$.
However, a light radial direction during inflation requires a negative Hubble-induced mass after inflation to drive the field to a large field value. Then the radial fluctuation is dampened, making $\delta \theta$ the dominant source of $\delta n_\theta$.}
In the rotation-dominant case, overdensities grow during eMD, and the upper cutoff can exceed the critical value $\delta_c$ across most of the $\theta_{\rm init}$ range. 
Therefore, PBH production can be reliably estimated assuming a Gaussian $\delta_S$ distribution. 
In contrast, when $S$ does not dominate, there is no growth, and the upper cutoff may lie below $\delta_c$ for a wide range of $\theta_{\rm init}$. 
In this regime, we expect the PBH production to be negligibly small compared to the Gaussian case.
Accordingly, we neglect PBH constraints in the non-dominant case.


\subsection{Axion radiation }
\label{sec:constraint_axions}
 
Fluctuations in the rotation of axion condensate contain a phonon mode, corresponding to the sound waves in the condensate~\cite{Bodas:2025eca}. 
These are the same phase fluctuations that are responsible for GW production discussed in \Sec{subsec:late_gw}.
The energy density in these sound waves ($\rho_{\rm sw}$) dilutes like radiation~\cite{Eroncel:2025bcb}, in contrast to the axion condensate, which dilutes like kination at the bottom of the potential.
Consequently, this radiation component does not dilute relative to the SM radiation, and must therefore be sufficiently small at the time of production.

Using \eq{eq:delta_chi_ini}, the energy density in sound waves for a mode $k$ at its horizon re-entry is given by 
\begin{equation}\label{eq:rho_sw_tk}
    \rho_{{\rm sw},t_k}(k)=  k_{\rm phy}^2 \delta\chi^2(k) \approx H_k^2 \mpl^2  
    \begin{cases}
        \paren{  2\sqrt{\frac{2}{3}} \,\epsilon(k)\,   \frac{\delta Y_\theta(k)}{Y_{\theta}}}^2 &\qquad (F_S \gg 1) \vspace{
        0.3em} \\
         \paren{ \sqrt{6}\,\epsilon(k)\, \Omega_{S,\rm MK}^{1/2} \,  \frac{\delta Y_\theta(k)}{Y_{\theta}}}^2 &\qquad(F_S \ll 1)
    \end{cases},
\end{equation}
where $H_k$ is the Hubble parameter at the horizon re-entry of the $k$ mode and $\epsilon(k)$ accounts for the deviation in $\delta \chi$ from \eq{eq:delta_chi_ini} for modes that re-enter the horizon close to the MK transition, as can be seen from \fig{fig:pertb_ax_dom_k}.
The energy density $\rho_{\rm sw}$ constitutes a form of dark radiation and contributes to both $\Delta N_{\rm eff}$ and isocurvature fluctuations, with the latter typically imposing a stronger constraint.  
We first consider the energy density in axion sound waves for the mode $k_{\rm MK}$:
\begin{align}
    \Omega_{\rm sw,ls}(k_{\rm MK}) \sim 
    \frac{\rho_{\rm sw,ls}(k_{\rm MK})}{\rho_{\rm rad,ls}} \sim \frac{\rho_{\rm sw,MK}(k_{\rm MK})}{\rho_{\rm rad,MK}} \sim \frac{10^{-1} \rho_{S,\rm MK} }{\rho_{\rm rad,MK}} \paren{\frac{\delta Y_\theta(k_{\MK})}{Y_{\theta}}}^2,
\end{align}
where the subscript ``ls'' denotes the time of last scattering and we have set $\epsilon(k_{\rm MK})\approx 0.3$ from \fig{fig:pertb_ax_dom_k}.
Following the same method as in \Sec{sec:anisotropy}, the long-wavelength ($k_l$) fluctuations in $\rho_{\rm sw}(k\subt{MK})$  are 
\begin{equation}\label{eq:delta_sw_MK}
    \delta_{{\rm sw}}(k_l; k\subt{MK}) \equiv\frac{\delta [\Omega_{\rm sw,ls}(k_{\rm MK})] (k_l)}{\Omega_{\rm sw,ls}(k_{\rm MK})} = -2 \, \zeta_S(k_l).
\end{equation}

For the modes (denoted by $k\subt{rd}$) that re-enter after KR in the case of rotation dominance or sufficiently after MK in the case of rotation non-dominance,
\begin{align}
    \Omega_{\rm sw,ls}(k_{\rm rd}) \sim \frac{\rho_{\rm sw,rd}(k_{\rm rd})}{\rho_{\rm rad,rd}} \sim \paren{\frac{\rho_{S,\rm MK}}{\rho_{S,\rm MK}+\rho_{\rm rad, MK}}} \paren{\frac{\delta Y_\theta(k)}{Y_{\theta}}}^2.
\end{align}
Then the long-wavelength anisotropy in these modes is
\begin{align}\label{eq:delta_sw_rd}
    \delta_{{\rm sw}}(k_l; k\subt{rd})=  -4\paren{\frac{1}{2}+\Omega_{S,\rm MK}} \, \zeta_S(k_l).
\end{align}
Axion sound waves act like dark radiation, which is constrained by neutrino isocurvature bound, $\sqrt{\mathcal{P}_{3(\zeta_\nu -\zeta_\gamma)}} \approx 3 \frac{\rho_{\rm sw}}{\rho_{\nu}} \sqrt{\mathcal{P}_{\zeta_{\rm sw}}} < 6 \times 10^{-6}$~\cite{Planck:2018jri}, where $\zeta_{\rm sw}$ is the gauge-invariant fluctuation in axion radiation. Then using $\zeta_{\rm sw} = \delta_{\rm sw}/4$ and $\rho_{\gamma,0}/\rho_{\nu,0} \approx 1.5$, we get a constraint 
\begin{align}\label{eq:sw_isocurv_brief}
    \frac{15}{8} \paren{\frac{\rho_{\rm sw,ls}}{\rho_{\rm rad,ls}}} \sqrt{\mathcal{P}_{\delta_{\rm sw}}} < 6\times 10^{-6}.
\end{align}
To get the total isocurvature contribution, we must add contributions from sound waves on all short-wavelength modes, i.e., all modes re-entering between MK and the epoch of last scattering.
More precisely, the fraction of energy density in mode $k_{\rm rd}$ at last scattering is
\begin{align}
    \frac{\rho_{\rm sw, ls}}{\rho_{\rr, \rm ls}} = \paren{\frac{g_{\rho, \rm rd}}{g_{\rho, \rm ls}}} \paren{\frac{g_{s,  \rm rd}}{g_{s,\rm ls}}}^{-4/3}  \frac{\rho_{{\rm sw},  \rm rd}}{\rho_{\rr,  \rm rd}},
\end{align}
where factors of $g_\rho$ and $g_s$ account for the change in the degrees of freedom in the SM plasma. 
To simplify the calculation, we take $g_{\rho} = g_s = 106.75$ for all times before the QCD transition at $T \sim 100$ MeV, and $g_{\rho} \approx g_s = 3.36$ for all times after.
Using Eqs.~(\ref{eq:rho_sw_tk}), (\ref{eq:delta_sw_MK}), (\ref{eq:delta_sw_rd}), and (\ref{eq:sw_isocurv_brief}), and summing up contributions from all modes, we get the expression for the isocurvature constraint as:
\begin{align} 
        \frac{10}{3}\left[\frac{g_{\rho, \rm MK} \, g_{s,\rm MK}^{-4/3}}{ g_{\rho, \rm ls} \, g_{s,\rm ls}^{-4/3}}\paren{0.1  \frac{a^2_{\rm KR}}{a^2_{\rm MK}} + 3 \log\paren{\frac{a_{\rm QCD}}{a_{\rm KR}}}} + 3 \log\paren{\frac{a_{\rm ls}}{a_{\rm QCD}}} \right]\mathcal{P}_{\delta_S}(k_{s})\mathcal{P}_{\zeta_S}^{1/2}(k_l)\leq 6 \times 10^{-6}& \nonumber\\
        \cdots (S\, {\rm dom.} / F_S \gg 1),& \nonumber \\ 
        \frac{15}{2} \, \Omega_{S,\rm MK} \left[ \frac{g_{\rho, \rm MK} \, g_{s,\rm MK}^{-4/3}}{ g_{\rho, \rm ls} \, g_{s,\rm ls}^{-4/3}}\log\paren{\frac{a_{\rm QCD}}{a_{\rm MK}}} + \log\paren{\frac{a_{\rm ls}}{a_{\rm QCD}}} \right] \mathcal{P}_{\delta_S}(k_{s}) \mathcal{P}_{\zeta_S}^{1/2}(k_l) \leq 6 \times 10^{-6}& \nonumber \\
        \cdots (S\, {\rm non\text{-}dom.}/ F_S \ll 1).
\end{align} 
The logarithmic enhancement accounts for the contribution from all modes re-entering during the late radiation era.
We have also accounted for the sharp drop in the relativistic degrees of freedom ($g_{\rho}$) around the QCD crossover.
This constraint becomes even stronger if the axion acquires a mass, which may also lead to domain wall formation.  
To avoid these complications, we assume $m_a \ll H_{\rm CMB}$.

There is also a constraint from $\Delta N_\text{eff}$. 
From the Planck+BAO data, $\Delta N_\text{eff} < 0.28$, which translates to $ \frac{\rho_{\rm sw}}{\rho_{\gamma}} < 0.063$.
The isocurvature constraint requires 
$(\frac{9}{8}\frac{\rho_{\rm sw}}{\rho_{\gamma}}) \sqrt{\mathcal{P}_{\delta_{\rm sw}}(k_l)} < 6\times 10^{-6}$.
Then using \eq{eq:delta_sw_rd}, we can quickly see that for $\sqrt{\mathcal{P}_{\zeta_S}} \gtrsim 5\times 10^{-5}$, $\frac{\rho_{\rm sw}}{\rho_{\gamma}} <0.06$, and therefore satisfying the isocurvature constraint automatically satisfies the $\Delta N_\text{eff}$ bound in our regime of interest.

\subsection{Isocurvature from late kination-to-radiation transition}
\label{sec:constraint_Tkr}
In the case of rotation dominance, the transition from kination to radiation is constrained by Big Bang nucleosynthesis (BBN), as the observed primordial abundances of light elements require BBN to occur during an RD era.
Additionally, the kination fluid must be sufficiently subdominant at the onset of BBN, since its presence can alter the freeze-out ratio of protons to neutrons.
This imposes the following lower bound on the temperature at kination–radiation equality \cite{Co:2021lkc} 
\begin{equation}\label{eq:Tkr_min_BBN}
    T_{\rm KR} \geq 2.5 \, {\rm MeV}. 
\end{equation}

Even if the kination fluid becomes subdominant in energy density by the time of BBN, the curvature perturbation may remain larger than the typical amplitude of $\sim 10^{-5}$.
A larger curvature perturbation leads to two effects.
Firstly, the proton-neutron ratio, and hence the helium-to-hydrogen abundance, can be modulated at the level of $\sim \Omega_{S, \rm MeV} \, \delta_{S}(k_l)$.
However, direct observational constraints from primordial helium measurements are weak\footnote{Since the binding energies of helium and hydrogen are different, the modulated helium-to-hydrogen abundance also leads to modulated matter density. However, this effect is suppressed by the ratio of the binding energy difference to nucleon mass, i.e., MeV/GeV $=10^{-3}$, making its contribution to matter isocurvature small.}, and indirect effects on the CMB, such as modifications to the damping scale, are second order and subdominant.
The second and more important effect arises from neutrino decoupling, which occurs around $T \sim 2 \, {\rm MeV}$, and will therefore also be modulated.
Below $T \lesssim 0.5 \, {\rm MeV}$, most $e^+ e^-$ pairs annihilate to photons. 
However, because neutrino decoupling is not instantaneous, a small fraction of these pairs instead annihilate into neutrinos.
This partial transfer of energy to neutrinos is one of the main reasons the SM predicts $N_{\rm eff} \approx 3.044$ instead of exactly 3 \cite{Akita:2022hlx}.
The fractional contribution of these annihilations to the total neutrino energy density is approximately $(3.044-3)/3.044 \approx 0.014$.
If neutrino decoupling is modulated, this small neutrino energy contribution from $e^+ e^-$ annihilations will also be modulated, thereby generating neutrino isocurvature in this small fraction.
From Planck 2018~\cite{Planck:2018jri}, we obtain the following constraint: 
\begin{align}\label{eq:Tkr_min_isocurv}
    \sqrt{\mathcal{P}_{3(\zeta_\nu - \zeta_\gamma) }}
    &\approx 0.014 \paren{ 3\,\Omega_S \sqrt{\mathcal{P}_{\zeta_{S}}}}_{0.5 {\rm MeV}} \notag \\ 
    &= 
    \begin{cases}
         0.042 \paren{\frac{T_{\rm KR}}{0.5 {\rm MeV}}}^{-2} \sqrt{\mathcal{P}_{\zeta_{S}}(k_l)} < 6\times 10^{-6} & \,{(F_S \gg 1)} \\ 
         0.042 \,\Omega_{S,\rm MK}\paren{\frac{T_{\rm MK}}{0.5 {\rm MeV}}}^{-2} \sqrt{\mathcal{P}_{\zeta_{S}} (k_l)} < 6\times 10^{-6} & \,{(F_S \ll 1)}
    \end{cases}.
\end{align}


\subsection{CMB B-modes}
\label{sec:constraint_Bmode}

The GWs produced from the phase fluctuations have been evaluated in section~\ref{sssec:late_gw}.
Modes that re-enter the horizon near the epoch of recombination can generate B-mode polarization in the CMB.
Combined observations from Planck and BICEP2/Keck constrain the tensor-to-scalar ratio to $r_{0.002} < 0.064$
at scale $k = 0.002\, {\rm Mpc}^{-1}$~\cite{Planck:2018jri}. 
This sets an upper bound on the long-wavelength fluctuations of the axion.
Using \fig{fig:gw_sh}, we obtain the following constraint:
\begin{align}
    & 10^{-2} A^{2}_{\delta \chi}(k_l) < 1.5 \times 10^{-10} \notag \\ 
    \rightarrow \quad& \frac{\sqrt{\mathcal{P}_{\delta\chi} (k_{l})}}{\mpl}=
    \left.
    \begin{cases}
    2\sqrt{\cfrac{2}{3}} \, \sqrt{\mathcal{P}_{\delta_{S}}(k_{l})}  & (F_S \gg 1) \\    
    \sqrt{6} F_S^{1/2} \, \sqrt{\mathcal{P}_{\delta_{S}}(k_{l})} & (F_S \ll 1)
    \end{cases}
    \right\}
    \, <1.1 \times 10^{-2},
\end{align}
where we have used $A_{\delta \chi}
= \sqrt{\mathcal{P}_{\delta\chi}}/\mpl$ and $\delta \chi/\mpl$ is given by \eq{eq:delta_chi_ini}.

\subsection{Non-Gaussianity in curvature perturbation}
\label{sec:constraint_NG}

In this subsection, we discuss the CMB non-Gaussianity induced by the phase fluctuations.

\subsubsection{$k < 1/\eta_\ls$}
We first focus on the modes that enter the horizon after the CMB last scattering ($k < 1/\eta_\ls$).
For such modes, the phase fluctuations can change CMB anisotropies through the integrated Sachs-Wolfe effect, which comes from the time-dependence of the gravitational potential. 
Unlike the B-mode, the temperature fluctuations are affected by the phase fluctuations even after the last scattering.
This contribution leads to non-Gaussianities in the CMB anisotropies.
Given that the phase fluctuations decay after their horizon entry, we here estimate the order of magnitude of the non-Gaussianities by focusing on the gravitational potential around their horizon entry.
Note that the gravitational potential becomes maximal at their horizon entry, which happens after the last scattering for $k < 1/\eta_\ls$ (see also footnote~\ref{ft:phi_evo}).
This means that, for $k < 1/\eta_\ls$, the integrated Sachs-Wolfe effect is dominant over the normal Sachs-Wolfe effect, since the latter is determined by the gravitational potential at the last scattering.

We can express the energy density of the phase fluctuations as
\begin{align}
	\rho_{\chi} = \frac{1}{2a^2} \left[ (\delta \chi')^2 + (\partial_i \delta \chi)^2 \right].
\end{align}
Its first-order energy density fluctuation is given by 
\begin{align}
	\delta \rho_{\chi} = \frac{1}{2a^2} \left[ (\delta \chi')^2 + (\partial_i \delta \chi)^2 - \vev{(\delta \chi')^2 + (\partial_i \delta \chi)^2} \right].
\end{align} 
We can approximate this as
\begin{equation}
	\delta \rho_{\chi} \sim  \nabla^2 (\delta \chi^2 - \vev{\delta \chi^2}).
    \label{eq:delta_rho_d_chi}
\end{equation}
We here use Poisson's equation at the horizon entry: 
\begin{align}
	\nabla^2 \Phi_{\chi} \simeq \frac{\delta \rho_{\chi}}{2\mpl^2},
    \label{eq:poisson_eq}
\end{align}
where we have neglected the 
 cosmic expansion for simplicity.\footnote{ 
 \label{ft:phi_evo}When the modes are outside the horizon, the cosmic expansion effect is dominant and the left hand side of Eq.~(\ref{eq:poisson_eq}) becomes $\sim \mathcal H^2 \Phi_\chi$. 
On the other hand, the approximation Eq.~(\ref{eq:delta_rho_d_chi}) remains the same.
From the fact that $\delta \chi$ is constant on superhorizon scales, this leads to $\Phi \propto (k\eta)^2$ on superhorizon scales.
}
Using Eqs.~(\ref{eq:delta_rho_d_chi}) and (\ref{eq:poisson_eq}), we obtain
\begin{equation}
	\Phi_{\chi} \sim \frac{\delta \chi^2 - \vev{\delta \chi^2}}{2\mpl^2}. \label{eq:Phi_in_terms_of_delta_chi}
\end{equation}

To estimate the order of magnitude, we use the approximation $\zeta_{\chi} \sim \Phi_{\chi}$ with $\zeta_{\chi}$ being the axion component in the total curvature perturbations. 
The integrated Sachs-Wolfe effect with $\Phi_\chi$ induces temperature fluctuations,
which are of $\mathcal O(\zeta_\chi) (\sim \mathcal O(\zeta_S^2))$ and inherit the non-Gaussianity of $\Phi_\chi$.
This enables us to compare the stochastic properties of $\Phi_\chi (\sim \zeta_\chi)$ with the observational constraints on the primordial non-Gaussianity.
We can express the power spectrum as
\begin{equation}
    \vev{\zeta_{\chi} \zeta_{\chi}} \sim \vev{ \left(\frac{\delta \chi^2 - \vev{\delta \chi^2}}{2\mpl^2}  \right)^2} \nonumber = \frac{\vev{\delta \chi^2}^2}{2\mpl^4},
    \label{eq:zeta_a_2}
\end{equation}
where we have assumed the Gaussian distribution for $\delta \chi$.
Similarly, the bispectrum is 
\begin{align}
    \vev{\zeta_{\chi} \zeta_{\chi} \zeta_{\chi}} = \vev{ \left(\frac{\delta \chi^2 - \vev{\delta \chi^2}}{2\mpl^2}  \right)^3} = \frac{\vev{\delta \chi^2}^3}{\mpl^6} \sim \vev{\zeta_{\chi}\zeta_{\chi}}^{3/2}.
    \label{eq:zeta_a_3}    
\end{align}
The local type non-Gaussianity is parameterized by~\cite{Komatsu:2001rj}
\begin{align}
    \zeta = \zeta_g + \frac{3}{5}f_\text{NL} (\zeta_g^2 - \vev{\zeta_g^2}),
\end{align}
where the Planck results put the constraint $f_\NL = -0.9 \pm 5.1$~\cite{Planck:2019kim}.
With this parametrization, we can express the bispectrum as 
\begin{align}
    &\vev{\zeta\zeta \zeta} \sim \frac{18}{5} f_\NL \vev{\zeta_g\zeta_g}^2 \nonumber \\
    \Rightarrow &
    \frac{\vev{\zeta\zeta \zeta}}{\vev{\zeta_g\zeta_g}^2} \simeq \frac{18}{5} f_\NL \lesssim \mathcal O(1\sim 10).
    \label{eq:f_nl_const}
\end{align}

If we take the total curvature perturbation to be dominated by $\zeta_{\chi}$ as $\zeta \simeq \zeta_{\chi}$, we get $\vev{\zeta_{\chi} \zeta_{\chi}} \sim 10^{-9}$.
In this case, Eqs.~(\ref{eq:zeta_a_2}) and (\ref{eq:zeta_a_3}) lead to 
\begin{align}
    \frac{\vev{\zeta\zeta \zeta}}{\vev{\zeta_g\zeta_g}^2} \sim \vev{\zeta_g\zeta_g}^{-1/2} \sim 10^{4}.
\end{align}
This is inconsistent with the Planck results, Eq.~(\ref{eq:f_nl_const}).
Therefore, let us consider the case where $\zeta_{\chi}$ is a subdominant component as $\zeta = \zeta_\phi + \zeta_{\chi}$ with $\zeta_\phi \gg \zeta_{\chi}$ and $\vev{\zeta_\phi\zeta_\phi} \simeq 10^{-9}$.
In this case, we obtain 
\begin{equation}
    \frac{\vev{\zeta\zeta \zeta}}{\vev{\zeta_g\zeta_g}^2} \sim \frac{\vev{\zeta_{\chi}\zeta_{\chi}}^{3/2}}{\vev{\zeta_\phi\zeta_\phi}^2} \sim \frac{\vev{\zeta_{\chi}\zeta_{\chi}}^{3/2}}{10^{-18}}.
\end{equation}
The Planck constraint in  Eq.~(\ref{eq:f_nl_const}) requires $\vev{\zeta_{\chi} \zeta_{\chi}} \sim \left(\frac{\vev{\delta \chi^2}}{\mpl^2}\right)^2 \lesssim 10^{-12}$.
Specifically, we obtain the following constraint on the long-wavelength modes of $\delta \chi$:
\begin{align}    
    \frac{\sqrt{\mathcal{P}_{\delta\chi} (k_{l})}}{\mpl}=
    \left.
    \begin{cases}
    2\sqrt{\cfrac{2}{3}} \sqrt{\mathcal{P}_{\delta_{S}}(k_{l})}  & (F_S \gg 1) \\    
    \sqrt{6} F_S^{1/2} \sqrt{\mathcal{P}_{\delta_{S}}(k_{l})} & (F_S \ll 1)
    \end{cases}
    \right\}
    \, \lesssim 10^{-3}.
    \label{eq:ng_const_d_chi}
\end{align}
This typically provides the strongest constraint on long-wavelength fluctuations of $S$, $\zeta_S (k_l)$.
In particular, it seems difficult to produce an observable amount of CMB B-mode polarization while satisfying the non-Gaussianity constraint. Note, however, that our estimation of the non-Gaussianity constraint and B-mode polarization is at the order-of-magnitude level. More precise estimation may reveal that B-mode polarization is observable in future datasets, which is beyond the scope of this paper.

\subsubsection{$k > 1/\eta_\ls$}
For the modes that enter the horizon before the last scattering ($k > 1/\eta_\ls$), 
the gravitational potential $\Phi_\chi$ also induces  temperature fluctuations through gravitational force.
Since $\Phi_\chi$ becomes maximal at its horizon entry and the order of the induced temperature fluctuations is the same as that of $\Phi_\chi$ at the horizon entry, the analysis in $k < 1/\eta_\ls$ can be applied to the modes of $k > 1/\eta_\ls$ and we find the same constraints, Eq.~(\ref{eq:ng_const_d_chi}).


\section{Detection prospects for the induced GWB and its anisotropy}
\label{sec:detectability}
In this section, we assess the strength of the GWB signal and its detectability with future GW experiments.

\begin{figure}[!t]
   \centering
   \subfloat[]{\includegraphics[width=0.84\linewidth, trim={2.5cm 3.5cm 2.5cm 4cm},clip]{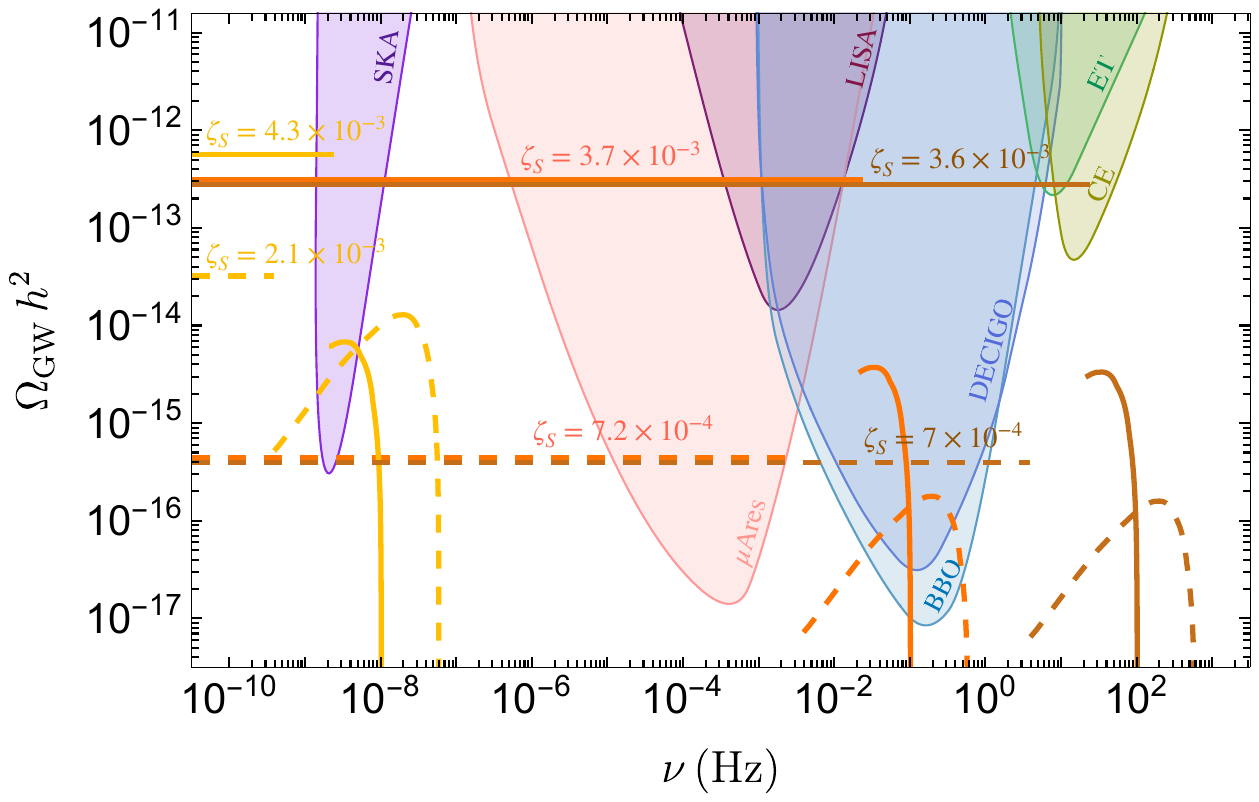}}\\[0.5em]
    \subfloat[]{\includegraphics[width=0.84\linewidth, trim={2.5cm 3.5cm 2.5cm 4cm},clip]{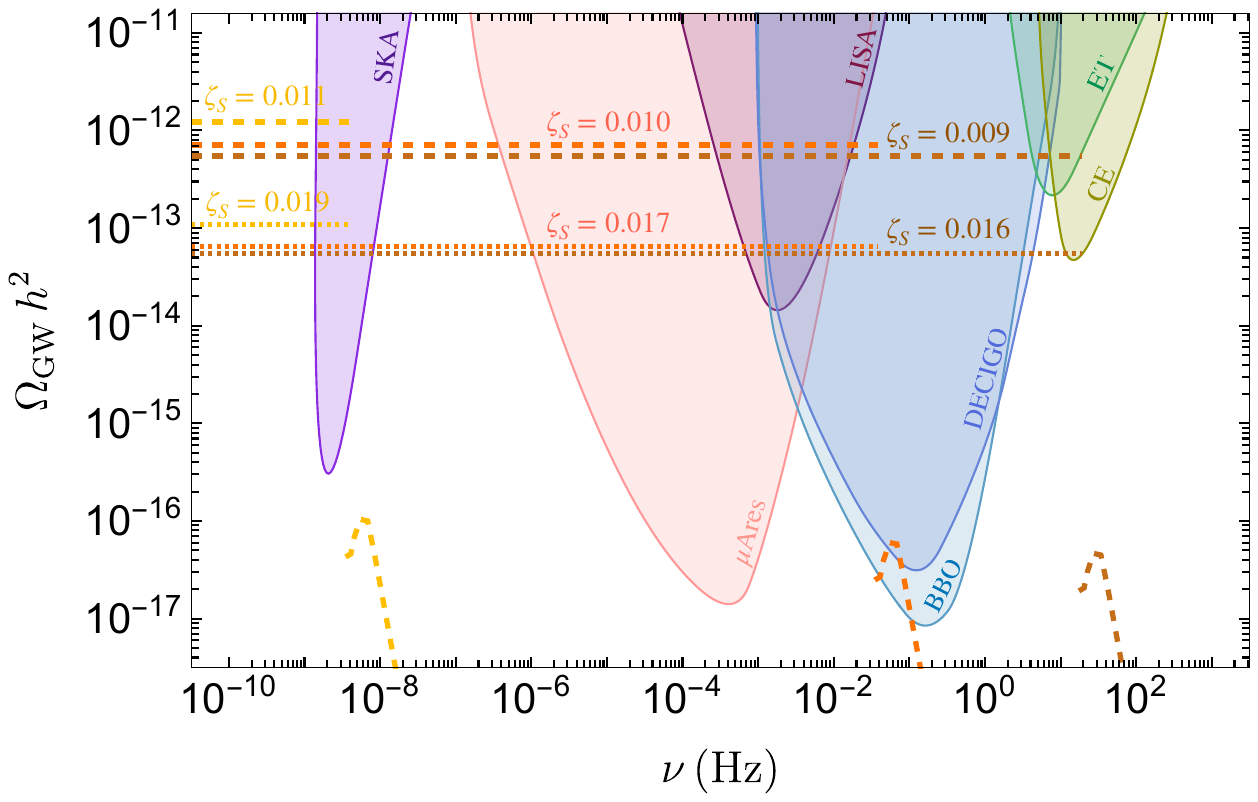}}
	\caption{
    Plots showing benchmarks for the cases of (a) rotation dominance with solid lines representing short eMD of $\frac{\afactor{MK}}{\afactor{RM}}=3$ or $\Omega_{S, \rm MK} = 0.75$, while dashed lines representing longer eMD with $\frac{\afactor{MK}}{\afactor{RM}}=100$ or $\Omega_{S, \rm MK} = 0.99$, and (b) rotation non-dominance with  short dashed lines corresponding to $\Omega_{S, \rm MK}=0.1$, while dotted lines representing $\Omega_{S, \rm MK}=0.01$.
    $T_{\rm MK}$ for yellow, orange, and brown curves are $100$ MeV, $10^3$ TeV, and $10^6$ TeV, respectively.
    The long-wavelength fluctuations $\zeta_S (k_l)$ are chosen to saturate the constraints in sections~\ref{sec:constraint_Tkr} -  \ref{sec:constraint_NG} with exact values given in the text.
    The amplitude of the short-wavelength fluctuation $\zeta_S (k_s)$, which is more relevant for the GWB production, is shown alongside each curve.
    }
	\label{fig:bechmarks_GW}
\end{figure}

Let us begin by considering a few benchmark scenarios at $T_{\rm MK} = 100$ MeV, $10^3$ TeV, and $10^{6}$ TeV.  
For each case, we choose the largest value of $\zeta_S$ that is consistent with the constraints discussed in section~\ref{sec:constraints}.
Constraints in sections~\ref{sec:constraint_Tkr} - \ref{sec:constraint_NG}, which only apply to long-wavelength modes that re-enter the horizon near or after recombination, are first saturated to fix $\zeta_S(k_l)$. 
It turns out that the non-Gaussianity constraint from \Sec{sec:constraint_NG} is the strongest in all cases.
$\zeta_S(k_l)$ is then used in the constraints of sections~\ref{sec:constraint_pbh} - \ref{sec:constraint_axions} to fix the short-wavelength fluctuations, $\zeta_S(k_s)$, which re-enter much earlier and  are responsible for the production of the GWB. 
The largest value of $\zeta_S (k_l)$ is typically in the range $\sim 10^{-4}-10^{-3}$, while that for  $\zeta_S (k_s)$ is in $\sim 10^{-3}-10^{-2}$.
In cases where the bound on $\zeta_S (k_l)$ is more stringent than that on $\zeta_S (k_s)$, we implicitly assume a mild blue tilt in the $\zeta_S$ spectrum so that both bounds can be saturated. 
Such a blue tilt can be easily accommodated into our model (see Appendix~\ref{app:susy_realization}).

\begin{figure}[h]
	\centering	
    \subfloat[\footnotesize{Duration of eMD $\frac{\afactor{MK}}{\afactor{RM}} =F_S =100$}]{\includegraphics[width=0.49\linewidth,trim={3cm 3.5cm 4.5cm 3.5cm},clip]{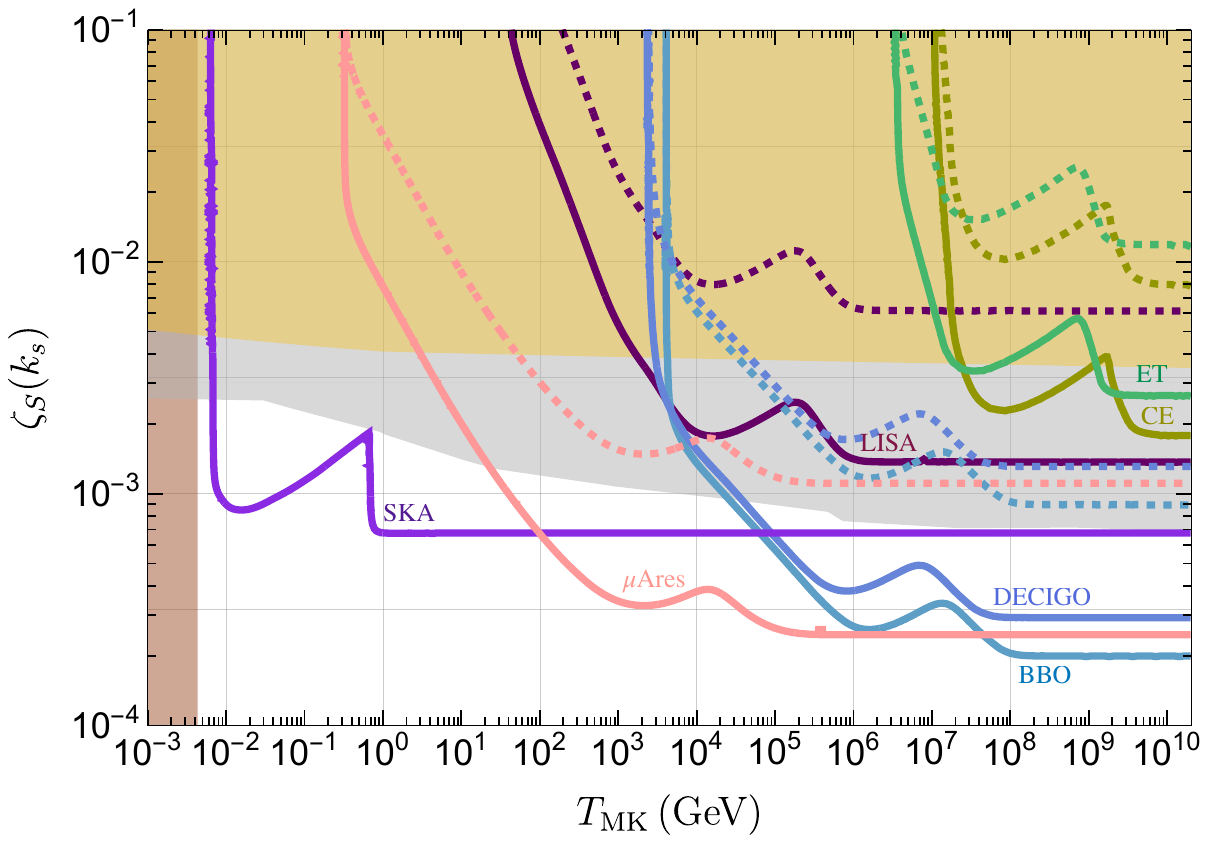}}
    \hspace{0.3em}
    \subfloat[\footnotesize{Duration of eMD $\frac{\afactor{MK}}{\afactor{RM}} =F_S =3$}]{\includegraphics[width=0.49\linewidth,trim={3.cm 3.5cm 4.5cm 3.5cm},clip]{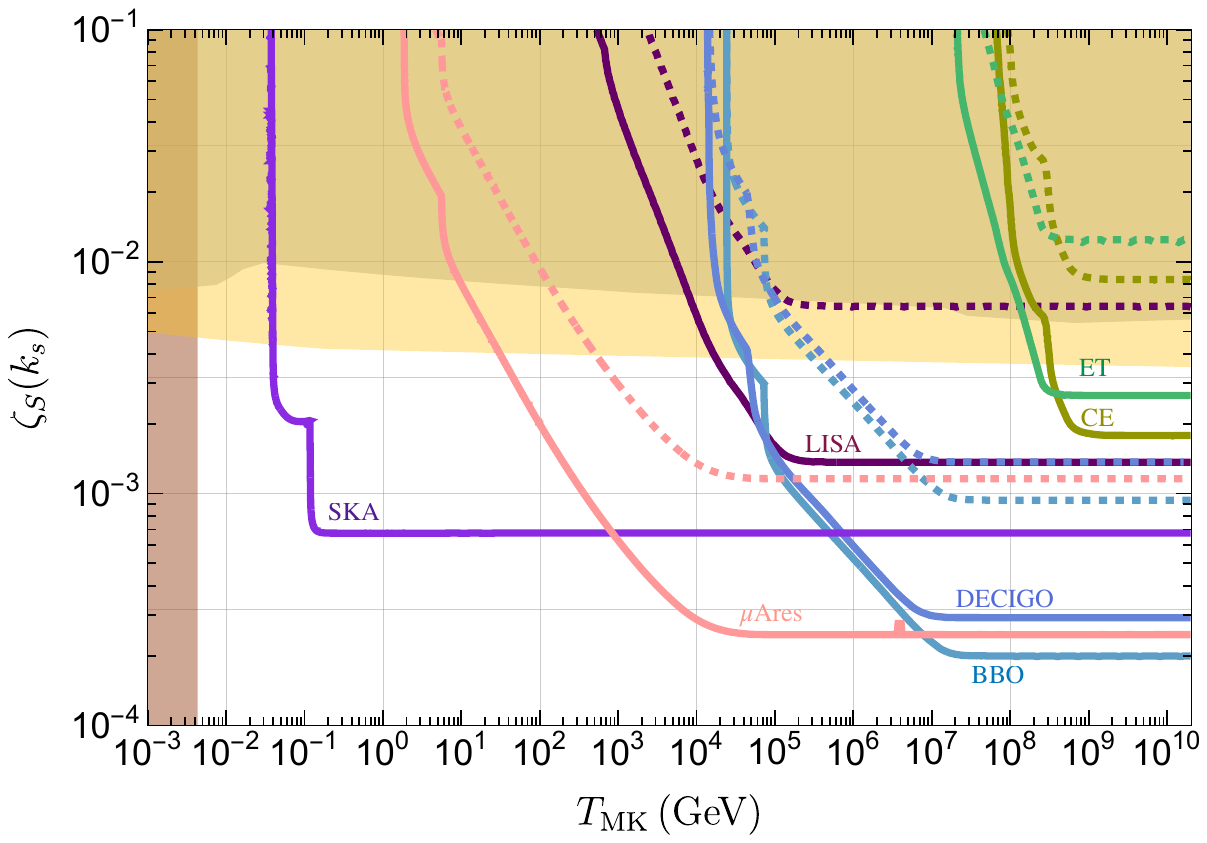}}\\[1.5em]   \subfloat[\footnotesize{$\Omega_{S,\rm MK} \approx F_S=0.1$}]{\includegraphics[width=0.49\linewidth,trim={3cm 3.5cm 4.5cm 3.5cm},clip]{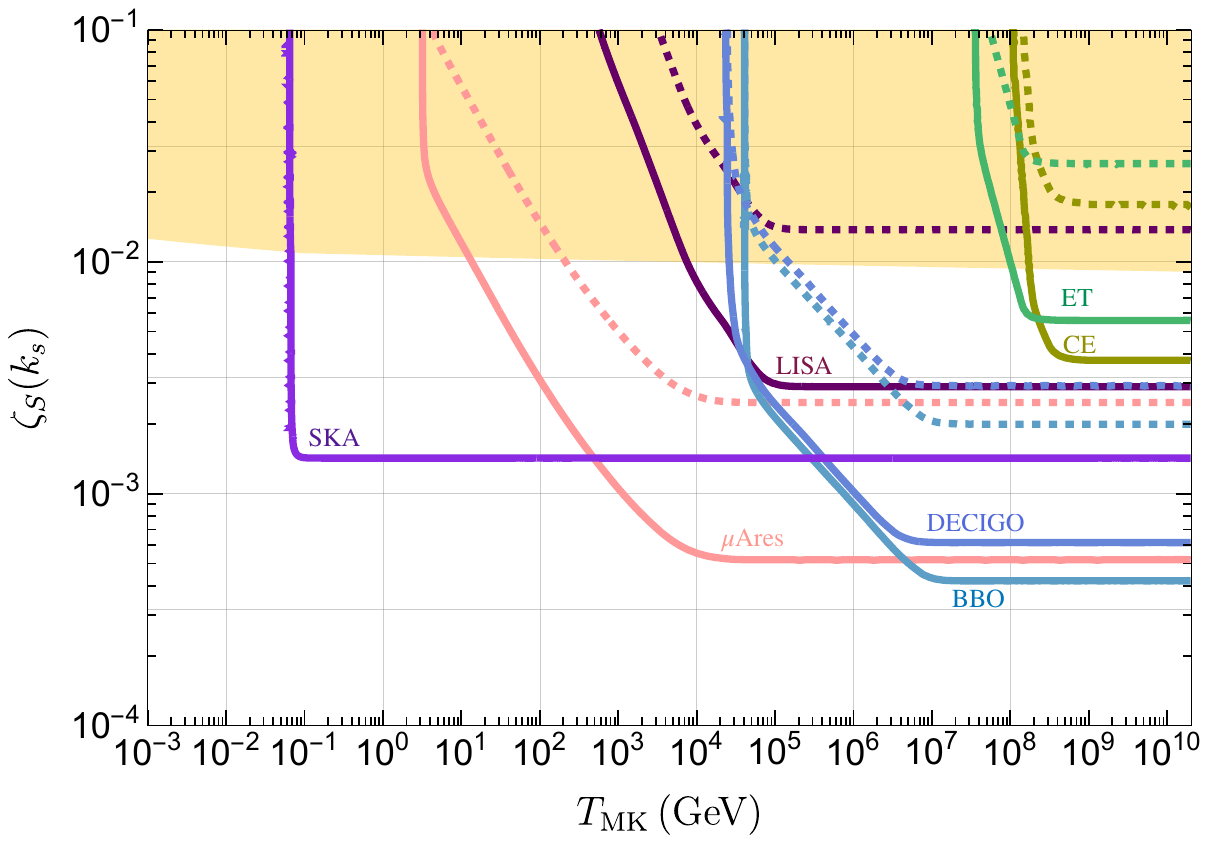}}
    \hspace{0.3em}   \subfloat[\footnotesize{$\Omega_{S,\rm MK}\approx F_S=0.01$}]{\includegraphics[width=0.49\linewidth,trim={3cm 3.5cm 4.5cm 3.5cm},clip]{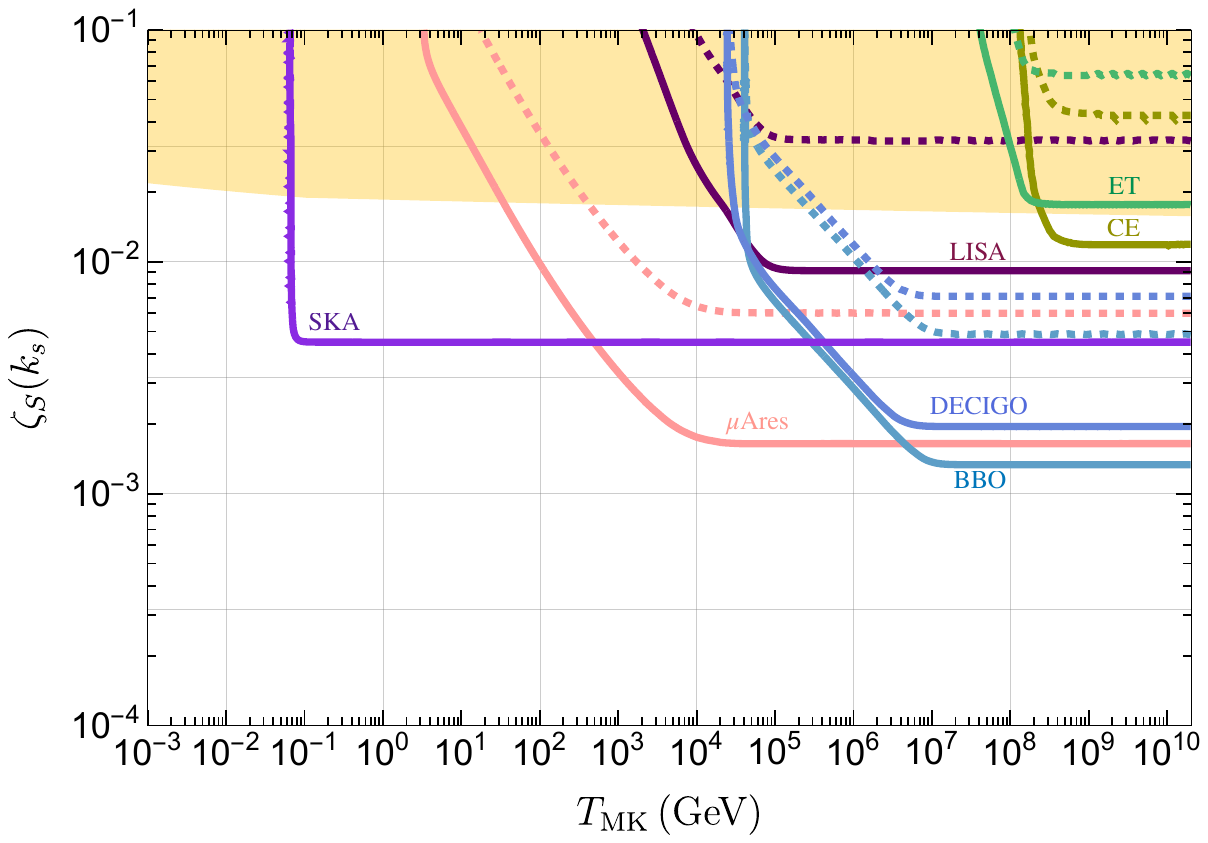}}
	\caption{Probed parameter space in the $(\zeta_S (k_s), T_{\rm MK})$ plane for four benchmark scenarios: (a, b) rotation-dominant cases with long ($a_{\rm MK}/a_{\rm RM} = 100$) and short ($a_{\rm MK}/a_{\rm RM} = 3$) eMD, and (c, d) rotation non-dominant cases with $\Omega_{S,\rm MK} = 0.1$ and $0.01$, respectively. Shaded regions show exclusions from PBH overproduction (gray), axion overproduction (yellow), and isocurvature constraints from late KR transition (brown). Solid (dashed) contours indicate regions where the GWB (its low-$\ell$ anisotropy) is detectable by future GW experiments.
    } 
	\label{fig:Delta_TKR}
\end{figure}

Figure~\ref{fig:bechmarks_GW} shows the GWB spectrum for our benchmarks
overlaid with the power-law integrated sensitivity curves of various GW experiments~\cite{Sesana:2019vho,Schmitz:2020syl}.
Panels (a) and (b) correspond to scenarios with and without rotation dominance, respectively.
In panel (a), solid lines correspond to a short eMD with $a_{\rm MK}/a_{\rm RM} = 3$ or $\Omega_{S, \rm MK} = 0.75$, and dashed lines to a long eMD with $a_{\rm MK}/a_{\rm RM} = 100$ or $\Omega_{S, \rm MK} = 0.99$.
In panel (b), short dashed lines correspond to $\Omega_{S, \rm MK} = 0.1$, and dotted lines to $\Omega_{S, \rm MK} = 0.01$.
In panel (a), the long-wavelength fluctuation is $\zeta_S (k_l) \approx 6 \times 10^{-4}$, while in panel (b), for $\Omega_{S, \rm MK} = 0.1$, $\zeta_S (k_l) \approx 10^{-3}$, and for $\Omega_{S, \rm MK} = 0.01$, $\zeta_S (k_l) \approx 4 \times 10^{-3}$.
The value of the maximum allowed $\zeta_S (k_s)$ is indicated for each benchmark alongside the corresponding curve.
The $\zeta_S (k_s)$ in the case of short eMD is primarily determined by the axion isocurvature constraint of \Sec{sec:constraint_axions}, while that for the long eMD is determined by PBH overproduction consideration from \Sec{sec:constraint_pbh}.
For the axion non-dominance case, axion isocurvature is the main consideration.
We see that the GW signal for a short eMD can be stronger than that for a longer one because the PBH constraints become stronger with the duration of the matter dominance. 
This effect is also visible in \fig{fig:Delta_TKR} panels (a) and (b), which will be discussed shortly.
In the case of rotation non-dominance, the constraint on $\zeta_S (k_s)$ weakens as $\Omega_{S,\rm MK}$ decreases.
However, the GWB signal weakens more since $\Omega_{\rm GW} \propto \Omega_{S, \rm MK}^2$.
As a result, GWB is stronger for $\Omega_{S, \rm MK} = 0.1$ compared to $\Omega_{S, \rm MK} = 0.01$.

The broken structure of the GW spectra in panels (a) and (b) arises from the two distinct contributions:
the peaked component is generated during the KD/ pre-MK era respectively, as discussed in \Sec{subsec:early_gw}, and a nearly scale-invariant component is sourced by the axion phase fluctuations, detailed in \Sec{subsec:late_gw}.
The apparent discontinuity is an artifact of our approximations in evaluating the GW production; a more refined calculation would yield a continuous transition between the two regimes. 

\begin{figure}[t]
   \centering
   \includegraphics[width=0.75\columnwidth, trim={5.2cm 4.7cm 5.9cm 5.5cm},clip]{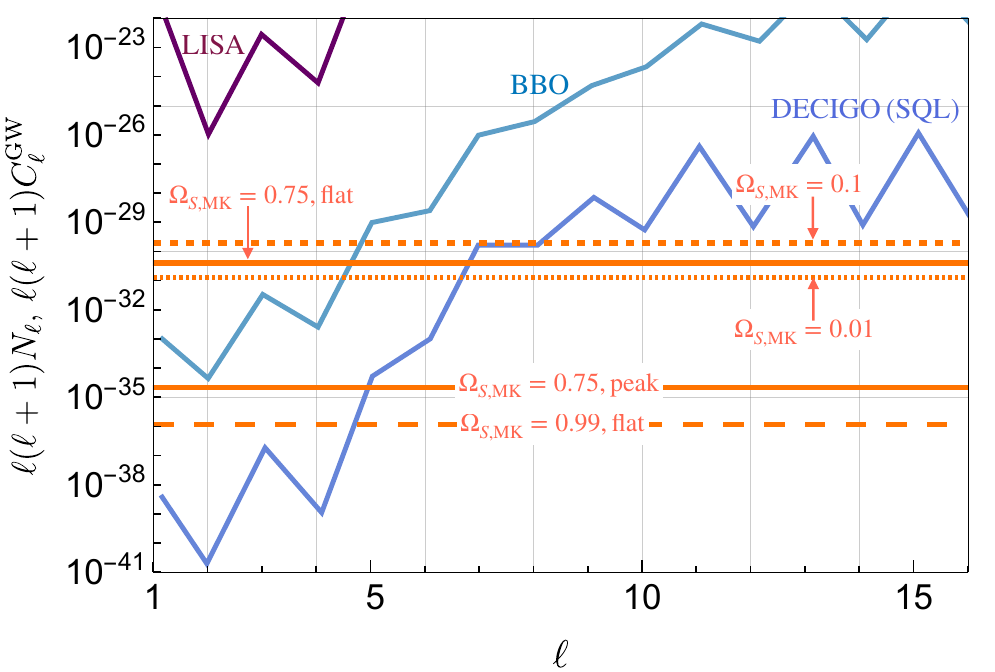}\\
	\caption{Anisotropies in relevant benchmarks ($C_{\ell}^{\rm GW}$ corresponding to $\langle \delta\Omega_{\rm GW}^2 \rangle$) compared with the angular noise power spectrum ($N_\ell$) for future experiments in mHz to deciHz range, such as LISA, BBO, and DECIGO (SQL-limited). 
    The long-wavelength fluctuation for the benchmark with short eMD ($\Omega_{S,\rm MK}=0.75$) and long eMD ($\Omega_{S,\rm MK}=0.99$) is taken to be $\zeta_S (k_l) = 6\times 10^{-4}$. In the cases of rotation non-dominance, for $\Omega_{S,\rm MK}=0.1$, $\zeta_S (k_l) \approx 10^{-3}$, and for $\Omega_{S,\rm MK}=0.01$, $\zeta_S (k_l) \approx 4 \times 10^{-3}$.
    }
	\label{fig:benchmarks_Anisotropy}
\end{figure}
An important feature of our scenario is the large anisotropy in the GWB, as discussed in \Sec{sec:anisotropy}. 
Detecting such anisotropies is generally challenging, as most GW detectors are optimized for measuring the isotropic (monopole) component of the signal. 
However, future space-based interferometers such as LISA, BBO, and DECIGO are expected to have sensitivity to low multipoles of the GWB.%
\footnote{Future PTAs, such as SKA, will also have some sensitivity to anisotropies \cite{Depta:2024ykq}. However, this sensitivity is typically limited by the number of pulsars, making it relatively weak. While the monopole component of the GWB could be detectable with SKA (see figures~\ref{fig:bechmarks_GW} and \ref{fig:Delta_TKR} for example), the anisotropies would likely remain out of reach, even if they are enhanced in our scenario.
}
In \fig{fig:benchmarks_Anisotropy}, we show the projected angular noise power spectra \( N_\ell \) for LISA, BBO, and DECIGO (assuming standard quantum limit (SQL) for noise\footnote{Note that this is a more optimistic noise limit than the one used to obtain the power-law integrated sensitivity curve in \fig{fig:bechmarks_GW}.}), based on calculations using the \texttt{schNell} code \cite{Alonso:2020rar,Braglia:2021fxn}. 
$N_\ell$ should be compared with the angular power spectrum $C_\ell^{\rm GW}$ corresponding to $\langle \delta\Omega_{\rm GW}^2 \rangle$ to assess detectability.
We have rescaled $N_\ell$ for each experiment by $\ell (\ell+1)$ such 
that the corresponding $\ell (\ell+1) C_{\ell}^{\rm GW}$ from an (approximately) scale-invariant spectrum appears flat in the plot.
We overlay these with the predicted anisotropies for benchmarks corresponding to $T_{\rm MK} = 10^3$ TeV from \fig{fig:bechmarks_GW} as their GWB spectra overlap with the target frequency bands of the detectors.
As discussed before, the large-scale fluctuation $\zeta_S (k_l)$ is chosen to saturate constraints in sections~\ref{sec:constraint_Tkr} - \ref{sec:constraint_NG} such that
for $\Omega_{S, \rm MK} = 0.75, \, 0.99$, $\zeta_S (k_l) \approx 6 \times 10^{-4}$, for $\Omega_{S, \rm MK} = 0.1$, $\zeta_S (k_l) \approx 10^{-3}$, and for $\Omega_{S, \rm MK} = 0.01$, $\zeta_S (k_l) \approx 4 \times 10^{-3}$.
The two different lines for $\Omega_{S,\rm MK} = 0.75$ correspond to the flat and the peaked parts of the spectrum.
As discussed in section~\ref{sec:anisotropy} (1), detecting both the anisotropy in the GW signal at the peak and the anisotropy in the peak frequency itself can provide evidence for the dynamics of a rotating axion.

We see that several of these benchmarks exhibit anisotropies large enough to be potentially observable with BBO and a more sensitive SQL-limited DECIGO setup, at least at low $\ell$s.
Therefore, if a stochastic GWB signal is detected in these experiments, its isocurvature origin could also be confirmed through its angular power spectrum, opening up a path to an altogether new cosmological map.
We hope this motivates the development of future GW detectors with enhanced angular resolution and anisotropic sensitivity.

The benchmark studies above indicate that the GWB signal in our model, and in some cases, its low-$\ell$ anisotropies, could be observable with proposed GW experiments. 
We now extend this analysis to explore the broader parameter space that can be probed.
In \fig{fig:Delta_TKR}, we show the $(\zeta_S(k_s), T_{\rm MK})$ parameter space for four scenarios: two with rotation dominance featuring different durations of eMD —(a) long, with $a_{\rm MK}/a_{\rm RM}= F_S = 100$ and (b)  short, with $a_{\rm MK}/a_{\rm RM} = F_S = 3$, and two in the rotation non-dominant regime — (c) $\Omega_{S,\rm MK} \approx F_S = 0.1$ and (d) $\Omega_{S,\rm MK} \approx F_S = 0.01$.
The shaded regions indicate exclusion bounds from the various constraints:
gray for PBH overproduction (section~\ref{sec:constraint_pbh}), 
yellow from axion radiation (section~\ref{sec:constraint_axions}), and brown for isocurvature limits arising from a late KR transition (section~\ref{sec:constraint_Tkr}). 
In the case of short eMD, the primary constraint on $\zeta_S (k_s)$ comes from axion radiation isocurvature, while PBH constraints dominate for long eMD. 
For the rotation non-dominant case,  PBH constraints are expected to be weaker as discussed at the end of \Sec{sec:constraint_pbh}, and therefore only axion isocurvature is relevant.

The solid (dashed) lines in \fig{fig:Delta_TKR}
outline regions where GWB (low-$\ell$ GWB anisotropy) is detectable with the corresponding GW experiment.\footnote{The dashed line corresponding to SKA is omitted, as its sensitivity to anisotropies is limited by the number of pulsars and is significantly weaker than the maximum allowed anisotropy in our scenario.}
Detectability is assessed using the signal-to-noise ratio (SNR) for the induced GWB, defined as~\cite{Schmitz:2020syl}
\begin{align}
    {\rm SNR}_{\rm GW} = \left[n_{\rm det} \, t_{\rm obs} \int_{\nu_{\rm min}}^{\nu_{\rm max}} \left(\frac{h^2 \Omega_{\rm signal}(\nu)}{h^2 \Omega_{\rm noise}(\nu)}\right)^{2} d\nu \right]^{1/2},
\end{align}
where $n_{\rm det}$ is the number of detectors in the experiment, $t_{\rm obs}$ is the observation time, and $\Omega_{\rm noise}(\nu)$ is the noise power spectrum of the detector. 
We adopt a detection threshold of ${\rm SNR}_{\rm GW} = 1$.
The noise spectra for various experiments are taken from Refs.~\cite{Sesana:2019vho,Schmitz:2020syl}.
The parameters assumed for each experiment are:
\begin{itemize}[itemsep=0pt, topsep=0pt, parsep=0pt, partopsep=0pt]
    \item LISA: $n_{\rm det} = 1$, $t_{\rm obs} = 4$ years
    \item $\mu$Ares / Cosmic Explorer (CE): $n_{\rm det} = 1$, $t_{\rm obs} = 5$ years
    \item BBO / DECIGO / Einstein Telescope (ET): $n_{\rm det} = 2$, $t_{\rm obs} = 5$ years
    \item PTA (e.g., SKA)\footnote{We take $n_{\rm det} = 2$ for PTAs because of their use of cross-correlation techniques.}
    : $n_{\rm det} = 2$, $t_{\rm obs} = 20$ years
\end{itemize}

\begin{figure}[t]
	\centering	
    \includegraphics[width=0.85\linewidth,trim={3.cm 3.5cm 3.cm 3.cm},clip]{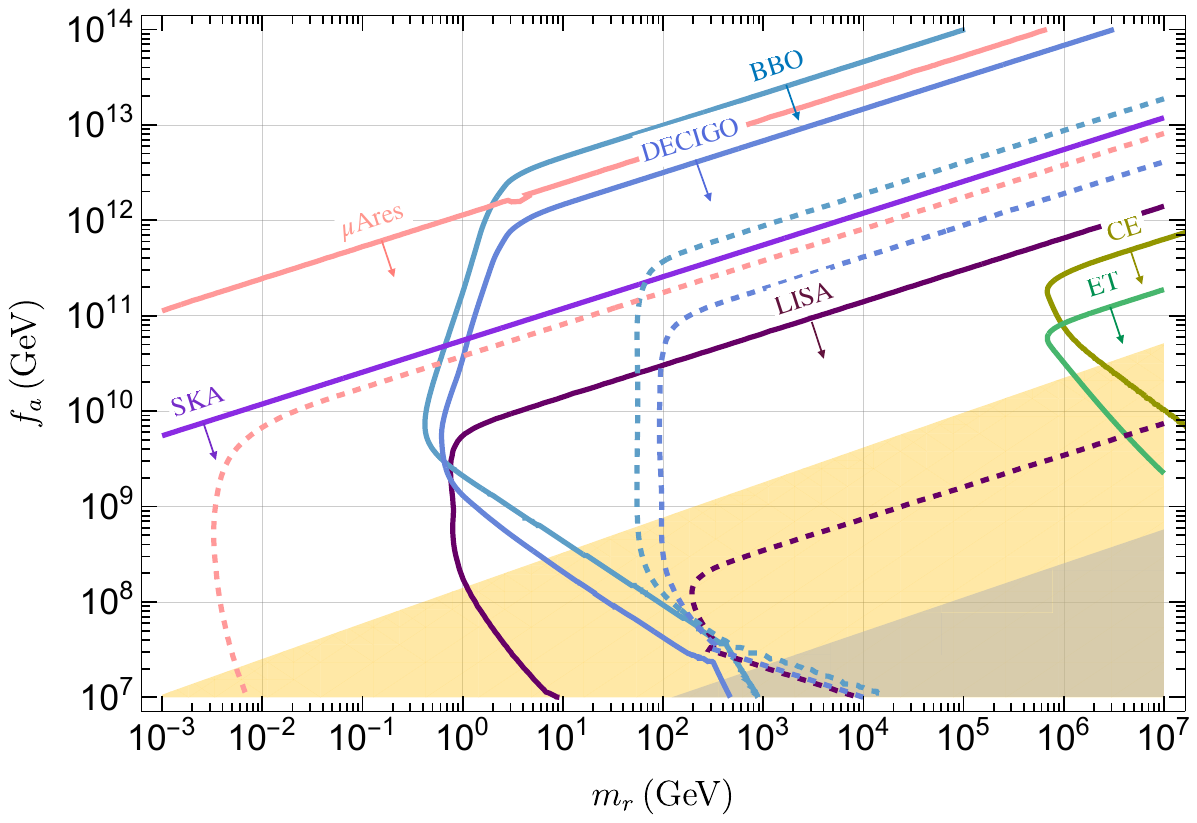}
	\caption{Parameter space of our model, $(f_a, m_r)$,  probed by various GW experiments for $\zeta_S(k_s) = 10^{-2}$.
    Other parameters are chosen such that the non-Gaussian contribution from the radiation component that is thermalized from the axion is within experiential bound (see \App{app:thermalization} for more details.)
    Shaded regions show exclusions, from PBH overproduction (gray) and axion overproduction (yellow). Solid (dashed) contours encircle regions where the GWB (its low-$\ell$ anisotropy) is detectable with future GW experiments.} 
	\label{fig:fa_mr}
\end{figure}

In \fig{fig:fa_mr}, we show the detectability of GW signals (and its low-$\ell$ anisotropy) in  the $(f_a, m_r)$ parameter space. 
The amplitude of the short-wavelength axion fluctuation is fixed to $\zeta_S(k_s) = 10^{-2}$, while the long-wavelength fluctuation saturates the constraint discussed in sections~\ref{sec:constraint_Tkr} - \ref{sec:constraint_NG}, as before. We take the maximal possible $Y_\theta$ allowed by the thermalization of the radial mode, which is taken to occur sufficiently early such that the non-Gaussianity induced by its resulting radiation remains within observational constraints (see \App{app:thermalization} for a more detailed discussion).
We see that future GW detectors can probe a substantial portion of the parameter space of our model.


\section{Conclusion}
\label{sec:conclusion}

In this work, we have studied GW production in a well-motivated model of a rotating axion field. 
We focused on the scenario where primordial fluctuations of the axion are larger than those of the inflaton that is taken to source the adiabatic perturbations observed in the CMB. 
This leads to several distinctive features in the resulting GWB:
\begin{itemize}
    \item The enhancement of curvature perturbations around matter-to-kination (MK) transition of  the axion and the superhorizon growth of axion phase fluctuations can lead to a detectably large induced GWB.

    \item The GWB thus produced inherits long-wavelength fluctuations from the axion, exhibiting large anisotropies on large scales.
    The anisotropies of such a GWB, therefore, offer a unique possibility to study the quantum fluctuations of the spectator axion field.

    \item The axion fluctuation on short-wavelengths can be as large as $\sim 10^{-2}$, constrained by PBH overproduction and isocurvature from axion radiation, while on long-wavelengths, it is constrained to be $\lesssim 10^{-3}$ mainly by the non-Gaussianity consideration. 
    
    \item 
    There is a distinctive change in the slope of the GW spectrum at the frequency corresponding to the MK transition,
    as seen in \figs{fig:omegw} and \ref{fig:gw_wo}. 
    The anisotropy in the amplitude of the GW signal at this point and its location in the frequency space can reveal the presence of axion rotation and its duration.
    This effect is potentially detectable when the axion dominates the energy density around MK (see \fig{fig:benchmarks_Anisotropy} for example).

    \item The GW spectrum is flat in the low-frequency region, potentially sourcing B-mode polarization that could be accessible to future CMB observations.

    \item The GW signal is large enough such that a large portion of the parameter space of our model can be probed with future GW experiments, as can be seen from figures~\ref{fig:Delta_TKR} and \ref{fig:fa_mr}. 
\end{itemize}

We would like to stress that GWBs have a unique ability to carry isocurvature anisotropies \cite{Geller:2018mwu}.
However, such an isocurvature GWB typically suffers from suppressed isotropic component, as elaborated in \Sec{sec:intro} and also in Ref~\cite{Bodas:2022urf}.
Our model adds to a small set of examples \cite{Bodas:2022urf} where this trade-off between anisotropy and amplitude is alleviated.
This is enabled by the matter-to-kination dynamics of the axion, which allows it to temporarily dominate the total curvature perturbation. 
As a result, the setup remains largely safe from stringent CMB isocurvature constraints, while providing a rare window into the inflationary dynamics of a light spectator axion field through its quantum fluctuations.

In this work, we focused on the long-wavelength axion fluctuations that are large as a distinctive signature of the mechanism.
An interesting variation would be to consider a blue-tilted axion power spectrum, such that the long-wavelength anisotropy is small (produced from the adiabatic perturbations), while only the power on short-wavelength fluctuations is enhanced.
In such a scenario, 
PBHs formed during the early matter era or residual axions (if they become non-relativistic before the matter-radiation equality) would not exhibit any isocurvature, and therefore, could constitute all of dark matter.
Such a dark matter would also be accompanied by a correlated gravitational wave signal, providing a distinctive signature for the model. 
The enhancement of the gravitational signals by kination domination and/or the poltergeist mechanism and the existence of a flat IR part can distinguish such a scenario from other scenarios of PBH formation.
We leave a detailed exploration of this possibility to future work.

\section*{Acknowledgments}
The authors thank Wayne Hu for helpful discussions.
AB and LTW are supported by the DOE under grant DE-SC-0013642, and KH under grant DE-SC0025242, both at the University of Chicago.
AB also acknowledges support from the Fermi Forward Discovery Group, LLC, under Contract No. 89243024CSC000002 with the DOE Office of Science, Office of High Energy Physics.
KH is further supported by the Grant-in-Aid for Scientific Research (No. 20H01895) from the Ministry of Education, Culture, Sports, Science, and Technology (MEXT), Japan, and by the World Premier International Research Center Initiative (WPI), MEXT, Japan, through Kavli IPMU.
KI acknowledges support from a JSPS Overseas Research Fellowship.
TT’s work was supported by the 34th (FY 2024) Academic Research Grant (Natural Science), No.~9284, from the DAIKO FOUNDATION.

\appendix

\section{Notation index}
\label{app:notation}
For ease of navigation, we provide here a list of notation frequently used in the paper, along with where each symbol first appears.
If the symbol does not have a separate equation, Eq.~$(\cdots)$* is used to indicate the equation around which it first appears.

\begin{center}
\renewcommand*{\arraystretch}{1.11}
\begin{longtable}{c p{10cm} c}
\toprule
\multicolumn{1}{c}{\textbf{Symbol}} &
\multicolumn{1}{l}{\textbf{Meaning}} &
\multicolumn{1}{c}{\textbf{Reference}} \\
\midrule
\endfirsthead
\multicolumn{3}{c}
{} \\
\toprule
\multicolumn{1}{c}{\textbf{Symbol}} &
\multicolumn{1}{l}{\textbf{Meaning}} &
\multicolumn{1}{c}{\textbf{Reference}} \\
\midrule
\endhead
\bottomrule
\endfoot
\bottomrule
\endlastfoot
        RM  & Radiation-to-matter transition & Figure~\ref{fig:S_class_dynamics}\\ 
        MK  & Matter-to-kination transition (when $r \rightarrow f_a$) & Section~\ref{sec:intro}\\ 
        KR  & Kination-to-radiation transition & Figure~\ref{fig:S_class_dynamics}\\ 
        $S = \frac{1}{\sqrt{2}} r e^{i\theta}$ &  A complex scalar hosting the axion within $\theta$ & \eq{eq:S_def} \\ 
        $\chi = r \, \theta$ &  The axion field & \eq{eq:theta_chi}\\ 
        $f_a$ &  The axion decay constant, the field value of $r$ at the potential minimum & \eq{eq:S_def}* \\         
        $m_r$ &  The mass of the radial field $r$ & \eq{eq:theta_chi}* \\ 
        $n_\theta = \dot{\theta} \, r^2$ &  The number density of angular charge & \eq{eq:Y_def}* \\ 
        $Y_{\theta}= n_\theta/s$  &  The yield of the angular charge & \eq{eq:Y_def}\\ 
        $\delta_S =\frac{\delta n_\theta}{n_\theta}=\frac{\delta Y_\theta}{Y_\theta}$  &  The conserved fluctuation of the charge density/yield & \eq{eq:delta_nTheta} \\ 
        $\zeta_S = \delta_S /3$ &  The gauge-invariant fluctuation of $S$ & \eq{eq:zeta_S} \\
        $\zeta_\chi (=\mathcal{O}(\zeta_S^2))$ & The curvature perturbations induced by the integrated Sachs-Wolfe effect from the time-dependence of the axion. & \eq{eq:Phi_in_terms_of_delta_chi}* \\
        $\rho_{i,\rm t} $ & The energy density in fluid $i$ at time $t$ & \eq{eq:rho_s_propto}*  \\ 
        $\Omega_{i,\rm t} = \frac{\rho_i}{\rho_{\rm tot}}$ & The fraction of energy density in fluid $i$ at time $t$ & \eq{eq:rho_s_propto}*  \\ 
        $\tilde{\Omega}_{i,\rm t}= \frac{\rho_i+P_i}{(\rho+P)_{\rm tot}}$ & The fraction of (energy density + pressure) in fluid $i$ at time $t$ & \eq{eq:zeta_tot}* \\
        $g_{\rho(s)}$ & The degrees of freedom for radiation (entropy) energy density & \eq{eq:omega_gw0} \\        
        $b_\rho = \rho_\rr/T^4$  & The normalized degrees of freedom for radiation energy density. ($b_\rho = \pi^2 g_\rho/30$) & \eq{eq:hubble_s_rho}* \\
        $b_s = s/T^3$  & The normalized degrees of freedom for entropy density. ($b_s= 2\pi^2 g_s/45)$ & \eq{eq:hubble_s_rho}*\\        
        $c = b_\rho b_s^{-4/3}$  & The combination of the normalized degrees of freedom for energy and entropy density. & \eq{eq:hubble_s_rho}*\\
        $F_S = \left. \frac{\rho_S}{\rho_\rr} \right|_{\rm MK}$  & The ratio of energy densities of $S$ and radiation at MK, $F_S >1$ corresponds to the rotation domination, while $F_S < 1$ to $S$ non-domination & \eq{eq:F_s_def}\\
        $T(k\eta)$ &  The transfer function of $\delta \chi_{\bfk}$ & \eq{eq:chi_trans} \\        
        $k_s$  & Short-wavelength or high-$k$ comoving modes & \eq{eq:GW_MK}*\\
        $k_l$ & Long-wavelength or low-$k$ comoving modes & \eq{eq:GW_MK}* \\
        $\delta_{i}(k_l ; k_s )$ & Long-wavelength ($k_l$) fluctuation of quantity $i$ sourced by the short-wavelength ($k_s$) fluctuation & \eq{eq:deltaGW_MK}\\
        $\nu_t$ & Frequency of the GW produced at time $t$ & \eq{eq:nu_KR}\\
\end{longtable}
\enlargethispage{.75cm}
\end{center}

\section{Thermalization}\label{app:thermalization}

For the existence of a kination phase, it is necessary for the rotation to become approximately circular. This occurs if $S$ can thermalize with the bath.
We consider an efficient thermalization scenario where the axion interacts with light fermions in the thermal bath through a Yukawa coupling: 
$y S \bar{\psi} \psi$ .
The rate of thermalization is given by~\cite{Mukaida:2012qn}:
\begin{align}
    \Gamma = b y^{2} T ,
\end{align}
where $b \simeq 0.1$. 
However, this thermalization channel is active only when the mass of the fermion is smaller than the plasma temperature, i.e, $m_{\psi} = y r < T$,
giving an upper bound on the rate of thermalization:
\begin{equation}
    \Gamma  = b y^2 T \leq b \frac{T^3}{r^2}.
\end{equation}

Assuming maximum rate from above, thermalization occurs when $\Gamma \sim H$, giving a relation between the radial field value and the temperature of the radiation bath at thermalization:
\begin{equation}\label{eq:r_th}
    r_{\rm th} = \paren{\frac{\sqrt{3}}{b_\rho^{1/2}} b \, T_{\rm th} \mpl}^{1/2}.
\end{equation}

During thermalization, the energy density of the radial mode $\rho_r$, which is comparable to or more than the energy density of the angular mode,%
\footnote{The ratio of the energy densities depends on the strength of the kick to the angular direction. In supersymmetric models discussed in appendix~\ref{app:susy_realization}, the ratio is ${\cal O}(1)$.}
goes to the thermal bath.
This component inherits the axion fluctuation and can generate a sizable non-Gaussianity as in the curvaron scenario~\cite{Lyth:2002my,Sasaki:2006kq} in the following way. 
The explicit $U(1)$-breaking potential not only gives the kick to the angular direction but also gives an initial radial energy in addition to that from the $U(1)$-symmetric potential. Because the magnitude of the explicit $U(1)$-breaking potential depends on $\theta$, the modulation of $\theta$ gives rise to that of $\rho_r$. Using the $\delta N$ formalism~\cite{Sasaki:1995aw,Wands:2000dp,Lyth:2004gb}, we find
\footnote{
Here we made the following approximation: The radial direction begins oscillations when $m_r \sim H$ and its energy density decreases in proportion to $a^{-3}$. However, when the explicit $U(1)$-breaking potential is comparable to the $U(1)$-conserving potential, the beginning of the oscillation is also modulated by $\theta$ and contributes to $\delta N$~\cite{Kawasaki:2011pd}. Also, the scaling of the energy density is not exactly $a^{-3}$ when the $U(1)$-symmetric potential deviates from a quadratic one, as in Eq.~\eqref{eq:pot_vs}. These effects can change $f_{\rm NL}$ by an $\mathcal{O}(1)$ factor.  
}
\begin{equation}
    \zeta = \zeta_{\phi} + \frac{1}{4}\tilde{\Omega}_{S, \rm th} \left( \frac{\partial{\rho_r}/\partial \theta}{\rho_r} \delta \theta + \frac{\partial^2{\rho_r}/\partial \theta^2}{2 \rho_r} \delta \theta^2 \right),
\end{equation}
where
$\zeta_{\phi}$ is the curvature perturbation from the inflaton and
we assume that the energy fraction of $S$ at the thremalization $\tilde{\Omega}_{S, \rm th} \ll 1$.
We see that while $\delta\theta$ is a Gaussian variable, $\zeta$ is inherently non-Gaussian,
The local non-Gaussianity parameter is then
\begin{equation}
    f_{\rm NL} = \frac{5}{27} \frac{\vev{\zeta^3}}{\vev{\zeta^2}^2}  = \frac{5}{384} \tilde{\Omega}_{S, \rm th}^3 \frac{\vev{\delta \theta^2}^2}{{\cal P}_\zeta^2} \frac{(\partial{\rho_r}/\partial \theta)^2 \partial^2{\rho_r}/\partial \theta^2 }{ \rho_r^3} 
    \sim \tilde{\Omega}_{S, \rm th}^3 \frac{{\cal P}_{\zeta_S}^2}{{\cal P}_\zeta^2},
\end{equation}
where in the second equality we assume that the explicit $U(1)$-breaking potential is comparable to the $U(1)$ symmetric potential when the rotation is initiated and that the initial value of $\theta$ is not at extrema or inflection points, so that $\rho_r \sim \partial \rho_r/\partial \theta \sim \partial^2 \rho_r / \partial \theta^2$.
The observational constraint $f_{\rm NL}\lesssim 10$~\cite{Planck:2019kim} gives an upper bound on $\tilde{\Omega}_{S, \rm th}$,
\begin{equation}\label{eq:tilde_Omega_S_therm}
   \tilde{\Omega}_{S, \rm th} \lesssim  \paren{10 \frac{{\cal P}_{\zeta}^2}{{\cal P}_{\zeta_S}^2}}^{1/3} \equiv \, \tilde{\Omega}_{S,\rm th}^{\rm max}.
\end{equation}

Now, $\tilde{\Omega}_{S,\rm th}$ can be written as
\begin{equation}
    \tilde{\Omega}_{S,\rm th} = \frac{3}{4} \frac{m_r^2 r_{\rm th}^2}{b_\rho T_{\rm th}^4}.
\end{equation}
Together with \eq{eq:tilde_Omega_S_therm}, this gives
\begin{align}
    T_{\rm th} > \paren{\frac{3 \sqrt{3}}{4} \frac{1}{b_{\rho}^{3/2} \tilde{\Omega}_{S,\rm th}^{\rm max}} b \,m_r^2 \, \mpl}^{1/3}.
\end{align}
Assuming maximum rate of thermalization and the resulting \eq{eq:r_th}, the yield is $Y_{\theta} = \frac{m_r r_{\rm th}^2}{b_s T_{\rm th}^3} \propto T_{\rm th}^{-2}$, and is maximized when $T_{\rm th}$ is the smallest.
Therefore, the largest yield that is consistent with the non-Gaussianity constraint above is 
\begin{align}
    Y_{\theta, \rm max} =\frac{ 4^{2/3} \, b_\rho^{1/2}}{3^{1/2} \, b_s} \paren{\frac{(\tilde{\Omega}_{S,\rm th}^{\rm max})^2 \,b \, \mpl}{m_r}}^{1/3}.
\end{align}
Using Eq.~(\ref{eq:F_s_def2}), the phenomenological parameters $(F_S, \, T_{\rm MK})$ can then be written in terms of $(m_r, \, f_a, \, Y_{\theta, \rm max})$ as
\begin{align}
    F_S &= \frac{1}{2}\frac{b_s^{4/3}}{b_\rho} \paren{\frac{m_r}{f_a}}^{2/3} Y_{\theta, \rm max}^{4/3}\,, \\ \nonumber
    T_{\rm MK} &= \paren{\frac{m_r f_a^2}{b_s Y_{\theta,\rm max}}}^{1/3}.
\end{align}
These relations are used in \fig{fig:fa_mr} in \Sec{sec:detectability}.


\section{A supersymmetric realization of axion rotation}
\label{app:susy_realization}

Let us now see a model that can realize the rotating axion dynamics described in \Sec{sec:scalar_rotations}.
In a supersymmetric model, the $U(1)$ symmetry can be explicitly broken by a superpotential~\cite{Dine:1995kz},
\begin{equation}
    W =  \frac{1}{n+1}\frac{S^{n+1}}{\Lambda^{n-2}},
\end{equation}
where $\Lambda$ is the UV cutoff scale.
This contributes to the potential as $V(S) \supset \left| \frac{\partial W}{\partial S}\right|^2 + \left(\frac{\alpha m_r}{2}\right) S \frac{\partial W}{\partial S} + $ h.c., where $\alpha \sim {\cal O}(1)$ is a dimensionless coefficient. The total potential is
\begin{align}\label{eq:V(S)}
    V(S) &= \frac{1}{2}(m_{r}^{2} + m_{H}^{2})r^2 + \frac{r^{2n}}{\Lambda^{2n-4}}  
    +\underbrace{\alpha\, m_{r} \frac{r^{n+1}}{\Lambda^{n-2}} \cos[(n+1)\theta]}_{V_{\cancel{U(1)}}}.
\end{align}
We consider a radial mass $m_r$ originating from soft supersymmetry breaking 
and $|m_{H}| \sim {\cal O}(H)$ is the Hubble-induced mass, which can be sourced by 1) coupling to the potential energy of the inflaton $\phi$, $\paren{ c_1 \frac{V(\phi)}{\mpl^2} r^2}$, 2) coupling to the kinetic energy of the inflaton, $\paren{ c_2 \frac{\dot{\phi}^2}{\mpl^2} r^2}$, 
and/or
3) coupling to the Ricci scalar $R$, $\paren{c_3  R r^2}$. 
We assume $m_H^2 <0$ during inflation and $m_{r} < H_{\inf}$, such that the radial mode is stabilized at a large field value, $r_{\rm inf}$, given by
\begin{equation}
    r_{\rm inf} = \Lambda \paren{\frac{m_H}{\sqrt{2 n} \Lambda}}^{1/(n-1)} \gg f_a .
\end{equation}
We later comment on the case with $|m_H^2| \ll H^2$ during inflation.
The mass of the radial mode at $r_{\rm inf}$ is $\sqrt{(2n-1)} m_H$. We will see that $n\geq 4$ is required for the rotation to occur. Then the radial direction can be heavier than $H_{\rm inf}$ if $m_H$ is not much smaller than $\hinf$, such that its quantum fluctuations can be neglected. On the other hand, the mass in the angular direction is $\ll \hinf$, resulting in a (approximately) scale-invariant spectrum.

\begin{figure}[tb!]
	\centering	\includegraphics[width=0.6\linewidth]{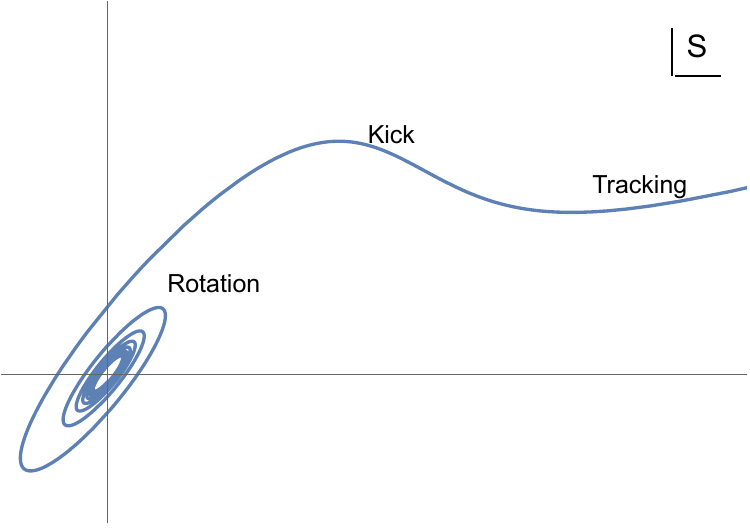}
	\caption{A schematic showing the initiation of axion rotations in supersymmetric theories.} 
	\label{fig:AD}
\end{figure}

The Hubble-induced mass may change after inflation if the coefficients $c_1$ and $c_2$ are different.
We consider two possibilities: (1) $m_H^2 <0$ and $|m_H^2| \gtrsim H^2 $, or (2) $m_H^2 >0$ and $m_H^2 \ll H^2 $. 
In the first case, there is a minimum at a non-zero field value, while in the second case, the Hubble friction dominates restricting the radial direction to a slow roll.
In both cases, the radial direction tracks the following expression~\cite{Dine:1995kz,Harigaya:2015hha}:
\begin{align}\label{eq:S_min}
    r_{\rm track} \sim \paren{\frac{H}{\sqrt{2n} \Lambda}}^{1/(n-1)} \Lambda.
\end{align}
This regime is shown as ``Tracking'' in \fig{fig:AD}.
When $H$ drops below $m_{r}$, the minimum shifts to $0$ in case (1) with a positive effective mass in the radial direction, while in case (2), the slow-roll ends.
The radial mode starts rolling down the potential and enters a matter-like phase.

For $n \geq 4\, (n \geq 3)$ in a radiation- (matter-)dominated universe, the initiation of rotation occurs concurrently with the start of the rapid rolling phase.
The strength of the gradient in the angular direction can be measured by $\epsilon \equiv \sqrt{|\partial V/\partial \theta | /r^2}/H$.
During the tracking regime, $\epsilon \propto r^{(n-1)/2}/H\propto H^{-1/2}$, so $\epsilon$ increases. After the radial mode starts rolling, $r \propto a^{-3/2}$, so $\epsilon \propto r^{(n-1)/2}/H\propto H^{3(n-1)/8-1}$ ($H^{(n-1)/2-1}$) during radiation (matter) domination. For $n\geq 4 (3)$, $\epsilon$ decreases.
Then the kick is strongest when the radial direction start rolling at $H \sim m_r$.
This is denoted by ``Kick'' in \fig{fig:AD}.
When $H\sim m_r$,
\begin{equation}
\epsilon \sim \sqrt{(\alpha (n+1)/\sqrt{2n})\,|\sin [(n+1)\theta]|}.
\end{equation}
For $\alpha \sim \mathcal O(1)$, the gradient to the angular direction is as large as that to the radial direction when the rotation begins. Then the resultant rotation has $\mathcal O(1)$ ellipticity.
After the kick, the field $S$ continues to rotate, shown as ``Rotation'' in \fig{fig:AD}, with a conserved angular momentum up to the dilution by cosmic expansion.

In the above discussion, we assume a negative Hubble-induced mass of ${\cal O}(H)$ during inflation, but this is not  necessary. As long as the Hubble induced mass is not a positive ${\cal O}(H)$ value, $r$ may take on a non-zero field value during inflation. (In this case, however, the Hubble induced mass after inflation must be negative and $|m_H|\gtrsim {\cal O}(H)$, so that $r$ may eventually take on a large field value and the kick by the $U(1)$-breaking higher dimensional operator is effective.) 
A small Hubble-induced mass may be naturally obtained by the RGE running of $m_H^2$~\cite{Elgamal:2025aol}. With a small Hubble-induced mass, it is possible that $r$ slowly decreases during inflation, so that the spectrum of $\delta \theta$ is blue-tilted, realizing the mild suppression of $\zeta_S(k_l)$ assumed in the main text.

The phenomenological parameters from \Sec{sec:scalar_rotations} can be mapped on to the parameters of this supersymmetric realization.
$S$ starts rolling and rotating at $H_{m_r} \sim m_r$. The field value at this point is $r_{m_r} \approx r_{\rm track}(H_{m_r})$.
In the case of the rotation dominance, the duration between the beginning of rotation and radiation-to-matter transition (RM) is
\begin{align}
    \frac{a_{m_r}}{a_{\rm RM}} \sim \paren{\frac{m_r}{\sqrt{2n} \Lambda}}^{2/(n-1)} \frac{\Lambda^2}{\mpl^2}.
\end{align}
Then the radial field value at RM is
\begin{align}
    r_{\rm RM} \approx r_{m_r} \paren{\frac{a_{m_r}}{a_{\rm RM}}}^{3/2} \sim \paren{\frac{m_r}{\sqrt{2 n} \Lambda}}^{4/(n-1)} \frac{\Lambda^4}{\mpl^3}.
\end{align}
Transition from matter to kination occurs when $r \sim f_a$. Assuming the transition from matter to kination is sufficiently fast, the duration of matter dominance is given by
\begin{align}
    \frac{a_{\rm MK}}{a_{\rm RM}} \approx \paren{\frac{f_a}{r_{\rm RM}}}^{-2/3} \sim \frac{\Lambda^{8/3}}{f_a^{2/3} \mpl^2} \paren{\frac{m_r}{\sqrt{2 n} \Lambda}}^{8/3(n-1)}.
\end{align}
Additionally, the temperature of the SM bath at MK is 
\begin{align}
    T_{\rm MK} &= T_{m_r} \paren{\frac{a_{m_r}}{a_{\rm MK}}} \approx   \frac{f_a^{2/3} m_r^{1/2} \mpl^{1/2}}{\Lambda^{2/3}} \paren{\frac{m_r}{\sqrt{2 n} \Lambda}}^{-2/3(n-1)}.
\end{align}

\section{Detailed calculation of GWs induced before kination phase}
\label{app:induced-GW}

Throughout this paper, we assume a gradual transition and study the strong GW production with large-amplitude isocurvature perturbations. 
In this appendix, we discuss the sources for GWs in $\eta < \eta_\MK$ other than those in section~\ref{subsubsec:gw_wo_ax_dom}, where we discuss the GW production with the assumption that $w = c_s = 0$ in $\eta < \eta_\MK$.
The main goal of this appendix is to study the GW production during a gradual transition from $w=0$ to $w = 1$ assuming a concrete model.
Specifically, we explain how to obtain the black dashed lines in \figs{fig:ph} and~\ref{fig:gw_wo}.

\subsection{Log potential model}

To be concrete, we consider the following Lagrangian in this appendix~\cite{Co:2021lkc}:
\begin{align}
	\mathcal L = -\frac{1}{2} (\partial r)^2 - \frac{r^2}{2} (\partial \theta)^2 - V(r),
\end{align}
where 
\begin{equation}
    \label{eq:pot_vs}
        V(r) = \frac{1}{2} m_r^2 r^2 \left( \log\left(\frac{r^2}{f_a^2} \right) - 1\right) + \frac{1}{2} m_r^2 f_a^2.
\end{equation}
At large radial field values, the potential is approximately quadratic, and gives a matter-like phase.
This potential has a minimum value at $r=f_a$ with $V(f_a)=0$ and leads to a gradual transition from matter to kination phase.

In the following, we focus on the evolution after thermalization has circularized the rotation orbit of the axion. 
From the circular mode condition, we obtain~\cite{Co:2021lkc}
\begin{align}
	{\dot {\bar\theta}}^2 = \frac{V'(\bar r)}{\bar r} = m_r^2 \ln \frac{\bar r^2}{f_a^2},
	\label{eq:dot_th}
\end{align}
where the bar denotes the background value, the dot denotes the physical time derivative, and $V'(r) \equiv \partial V(r)/\partial r$.
Then, we can express the background energy density and pressure of the circular mode as 
\begin{equation}
    \label{eq:rho_p_circular}
    \bar \rho = \frac{\bar r^2\dot {\bar \theta}^2}{2} + V(\bar r), \ \bar P = \frac{\bar r^2\dot {\bar \theta}^2}{2} - V(\bar r).
\end{equation}
From the conservation of the angular momentum, we obtain the scale factor dependence of $n_\theta$,
\begin{equation}
	n_\theta = \dot \theta \bar r^2 = m_r \bar r^2 \left(\ln \frac{\bar r^2}{f_a^2}\right)^{1/2} \propto a^{-3}.
\end{equation}
Using this, we can express the relation between $r$ and $a$ as 
\begin{align}
	\frac{\bar r^2}{r_i^2} \left(\frac{\ln \frac{r_i^2}{f_a^2} + \ln \frac{\bar r^2}{r_i^2}}{\ln \frac{r_i^2}{f_a^2}}\right)^{1/2} = \left(\frac{a_i}{a}\right)^3,
\end{align}
where the subscript $i$ represents the value at some initial time, much before $\eta_\MK$ but after the thermalization makes the orbit circular.
Solving this with respect to $r$, we obtain 
\begin{align}
	\bar r^2 &= f_a^2 \left(\exp\left[ W_0\left[ \frac{2 r_i^4 a_i^6 \log(r_i^2/f_a^2)}{f_a^4 a^6} \right] \right] \right)^{1/2}, 
	\label{eq:s_sq}
\end{align}
where $W_0$ is the principal branch of the Lambert $W$ function, satisfying $W_0(z) \ee^{W_0(z)} = z$.
Substituting 
Eq.~(\ref{eq:s_sq}) into Eq.~(\ref{eq:dot_th}), we obtain
\begin{align}
	\dot {\bar \theta}^2 = \frac{m_r^2}{2} W_0\left[ \frac{4 r_i^4 a_i^6 \log(r_i/f_a)}{f_a^4 a^6} \right].
\end{align}
With these, the equation-of-state parameter and the sound speed can be expressed as~\cite{Harigaya:2023mhl}
\begin{align}
    \label{eq:w}
	w &\simeq \frac{V'(\bar r) - 2 V(\bar r)/\bar r}{V'(\bar r) + 2 V(\bar r)/\bar r} = \frac{z^2 - 1}{1 - z^2(1- 4 \log z) }, \\
    \label{eq:cs_s}
	c_s^2 & \simeq \frac{V''(\bar r) - V'(\bar r)/\bar r}{V''(\bar r) + 3V'(\bar r)/\bar r} = \frac{1}{1 + 4 \log z},
\end{align}
where $z \equiv \bar r/f_a$.

\subsubsection{Perturbations}
\label{subsubsec:perturbations}

We here summarize some useful relations of perturbations in the log potential model (Eq.~(\ref{eq:pot_vs})), based on Ref.~\cite{Harigaya:2023mhl}.
The perturbations of the energy-momentum tensor of the circular mode are given by~\cite{Harigaya:2023mhl} 
\begin{align}
        \label{eq:rho_exa}
        -\delta T^0_{\ 0} &= \delta \rho = \frac{\bar r}{2} \left( -\frac{\nabla^2}{a^2} +  V'' +3 \frac{V'}{r} \right)\delta r, \\
        \label{eq:P_exa}       
        \frac{1}{3} \delta T^i_{\ i} &= \delta p = 
        \frac{\bar r}{2} \left( -\frac{\nabla^2}{a^2} + V'' - \frac{V'}{r} \right)\delta r, \\
       \label{eq:t_0i}
        \delta T^0_{\ i} &=  a^{-2} \left[ \bar r' \partial_i r + \bar r^2 \bar \theta' \partial_i \delta \theta \right] \simeq a^{-2} \bar r^2 \bar{\theta}' \partial_i \delta \theta, \\
      \label{eq:t_ij}        
        \delta T^i_{\ j(\neq i)} &=  a^{-2} \left[ \partial_i r \partial_j r + \bar r^2 \partial_i \delta \theta \partial_j \delta \theta \right] \simeq a^{-2} \bar r^2 \partial_i \delta \theta \partial_j \delta \theta,
\end{align}
where we have used $\delta \theta \gg |\delta r/\bar r|$ for Eqs.~(\ref{eq:t_0i}) and (\ref{eq:t_ij}) and the circular mode condition for Eqs.~(\ref{eq:rho_exa}) and (\ref{eq:P_exa}):
\begin{align}
    \label{eq:r_theta_rel}
    \left( a^2 V'' - ({\bar \theta}')^2\right) \frac{\delta r}{\bar r} \simeq 2 (\bar \theta' \delta \theta' - (\bar \theta')^2 \Phi).
\end{align}
Note that, in our case, $\delta \theta \gg 1$ is satisfied before the kination era begins, 
while $\delta r/\bar r < 1$ is always satisfied.
Specifically, we can find $(a^2 V'' - (\bar \theta')^2) \simeq a^2 V''$ around the beginning of kination era and $(\bar \theta'\delta \theta' - (\bar\theta')^2\Phi)/(a^2 V'')$ can be comparable to $\mathcal O(\delta r/\bar r)$ only when $\delta \theta \gg 1$.
This growth of $\delta \theta$ corresponds to $\delta(\Delta \theta)$, discussed in section~\ref{subsec:angle_fluc}.
In the fluid picture, $\delta T^0_{\ i}$ can be expressed with the velocity potential $\delta u$ as 
\begin{align}
    \delta T^{0}_{\ i} = \frac{1}{a} (\bar \rho + \bar P) \delta u_{,i}. 
\end{align}
Substituting Eq.~(\ref{eq:rho_p_circular}) into this, we obtain
\begin{align}
    \delta T^{0}_{\ i} = \frac{(\bar r \bar \theta')^2}{a^2} \frac{\delta u_{,i}}{a}.
\end{align}
Comparing this and Eq.~(\ref{eq:t_0i}), we can obtain 
\begin{align}
    \frac{\delta u}{a} = \frac{\delta \theta}{\bar \theta'}.
\end{align}
Then, we can express Eq.~(\ref{eq:t_ij}) as 
\begin{align}
    \delta T^i_{\ j(\neq i)} &\simeq a^{-4} (\bar r \bar \theta')^2 \delta u_{,i} \delta u_{,j} \nonumber \\
    &= a^{-2} (\bar \rho + \bar P) \delta u_{,i} \delta u_{,j}.
\end{align}
This expression is the same as the usual one fluid. 
This indicates that the axion rotation fluid can be regarded as the fluid with $w$ and $c_s^2$ given by Eqs.~(\ref{eq:w}) and (\ref{eq:cs_s}).
From this expression, we can also obtain the following late-time relation,
\begin{align}
    \sqrt{2\bar \rho} \delta u \simeq f_a \delta \theta \ \text{for } \eta \gg \eta_\text{MK}.
\end{align}

\begin{figure}  
\centering \includegraphics[width=0.8\columnwidth]{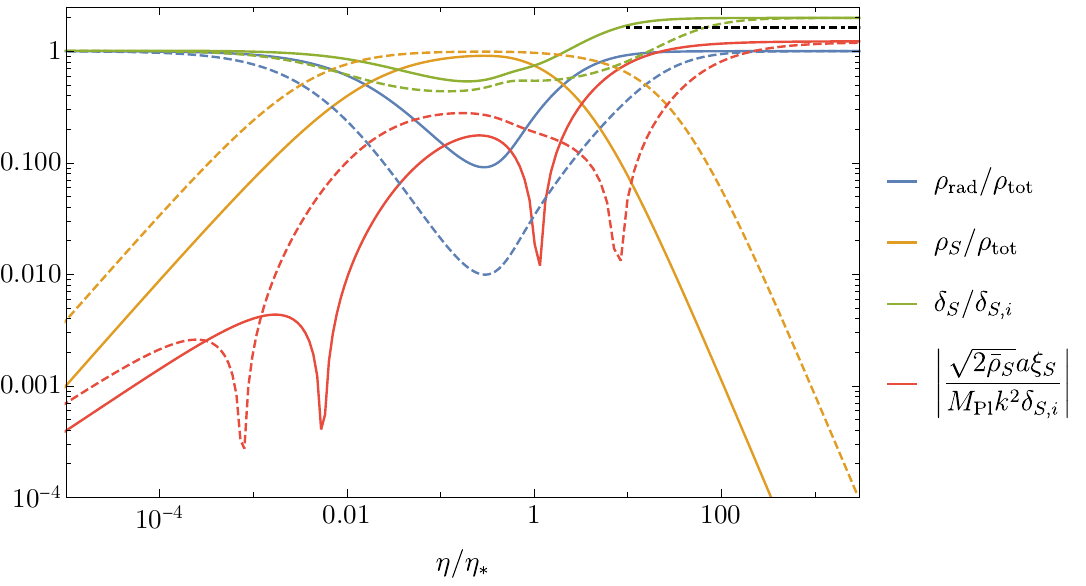}
\caption{
The background and superhorizon-limit perturbation evolution in the presence of the rotation domination. 
$\eta_*$ is defined as the conformal time at $c_s^2 = 0.95$.
The solid and dashed lines are for $F_S = 10$ and $=100$, respectively.
The black dot-dashed line is the analytic estimate of $\delta \chi/\mpl$ given by in Eq.~(\ref{eq:delta_chi_ini}), which should be compared to $|\sqrt{2\bar\rho_S}a\xi_S/(\mpl k^2 \delta_{S,i})|$ (red lines).
}
\label{fig:pertb_ax_dom}
\end{figure}

\begin{figure}  
\centering \includegraphics[width=0.8\columnwidth]{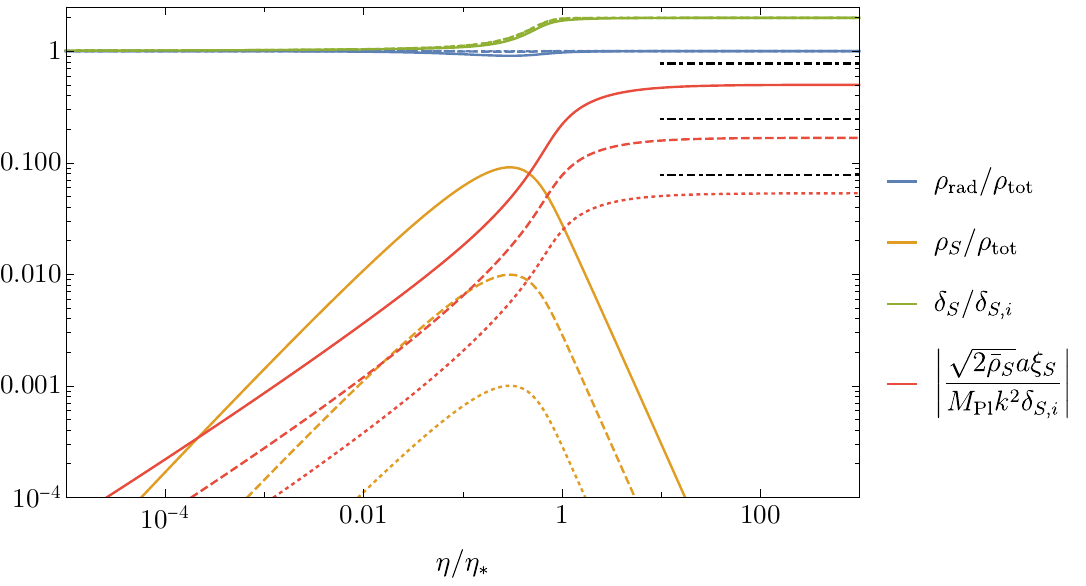}
\caption{
The evolution of the quantities without the rotation domination. 
The solid, dashed, and dotted lines are for $F_S = 0.1$, $=0.01$ and $=0.001$, respectively.
The black dot-dashed lines are the analytic estimate, Eq.~(\ref{eq:delta_chi_ini}).
}
\label{fig:pertb_wo_ax_dom}
\end{figure}

\begin{figure}  
\centering \includegraphics[width=0.9\columnwidth]{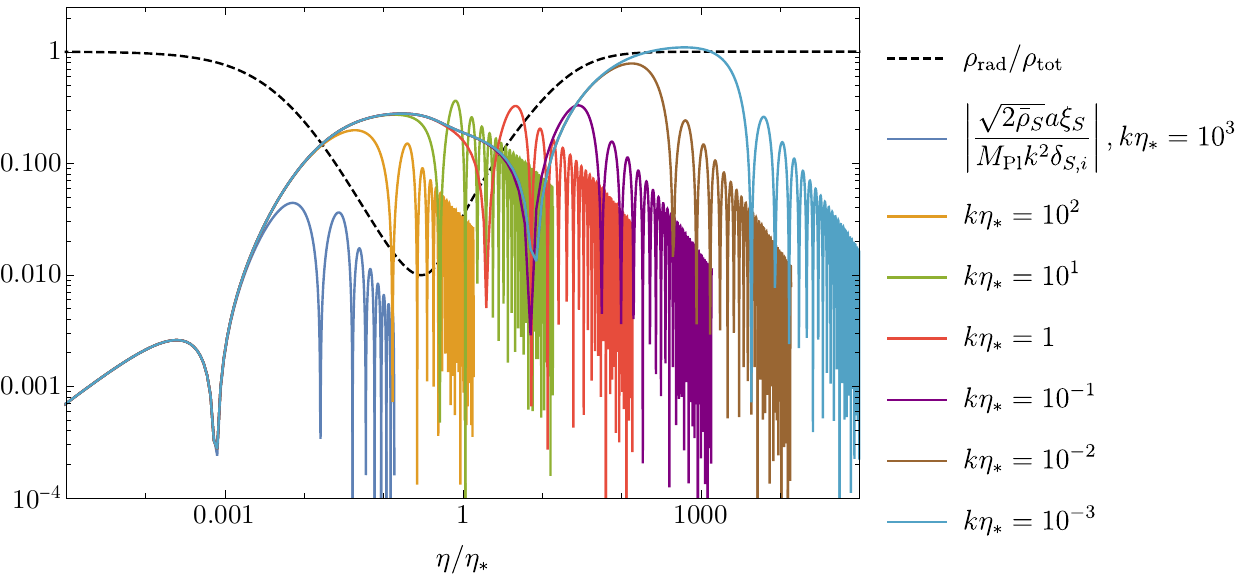}
\caption{
The evolution of $|\sqrt{2\bar\rho_S}a\xi_S/(\mpl k^2 \delta_{S,i})|$ from $k\eta_* = 10^{3}$ to $k\eta_* = 10^{-3}$ (solid lines).
$F_S = 100$ is taken for all lines and the black dashed line is $\rho_\rr/\rho_\tot$.
We show only the region in $\eta < 100/k$ for each line.
}
\label{fig:pertb_ax_dom_k}
\end{figure}

In our scenario, the Universe components consist of radiation and axion. 
To properly follow their evolution, we need to consider the axion rotation fluid and the radiation fluid, separately.
Specifically, the total energy-momentum tensor is given by 
\begin{align}
    T^\mu_{\ \nu} = T^\mu_{S\, \nu} + T^\mu_{\rr\, \nu},
\end{align}
where the subscript ``$S$'' and ``$\rr$'' mean the axion rotation and radiation components, respectively. 
Since we do not consider the energy or momentum transfer between axion and radiation after the thermalization, we can impose the energy-momentum conservation separately as
\begin{equation}
    T^\mu_{S\,\nu;\mu} = 0, \ T^\mu_{\rr\,\nu;\mu} = 0.
\end{equation}
From these conservation relations, we can obtain the following equation of motion:
\begin{align}
    &\bar \rho_S' + 3(1+w_S)\mathcal H \bar \rho_S =0, \\
    &\bar \rho_\rr' + 4\mathcal H \bar \rho_\rr = 0, \\
    &\delta_S' + (1+w_S) \xi_S + 3 \mathcal H (c_{s,S}^2-w_S) \delta_S  - 3\Phi' (1 + w_S) =0, \\
    &\xi_S' + \left[\frac{(\bar \rho_S(1+w_S))'}{\bar \rho_S(1+w_S)} + 4 \mathcal H\right] \xi_S - \frac{c_{s,S}^2}{1+w_S} k^2 \delta_S - k^2 \Phi = 0,\\
    &\delta_\rr' + \frac{4}{3} \xi_\rr - 4\Phi' =0,\\
    &\xi_\rr' - \frac{k^2}{4} \delta_\rr - k^2 \Phi = 0, \\
    &\Phi' = -\frac{(k^2 + 3 \mathcal H^2) \Phi + \frac{3}{2} \mathcal H^2 \left( \frac{\bar\rho_S \delta_S + \bar\rho_\rr \delta_\rr}{\bar \rho_S + \bar \rho_\rr} \right)}{3 \mathcal H},
\end{align}
where $\delta = \delta \rho/\bar \rho$ and $\xi \equiv \nabla^2 \delta u/a$.
We here consider the log potential model and $w_S$ and $c^2_{s,S}$ are given by Eqs.~(\ref{eq:w}) and (\ref{eq:cs_s}).
Similar to the main text, we assume that the adiabatic perturbation of radiation is negligibly small compared to the axion isocurvature perturbations. 
Then, we set the following initial conditions for the superhorizon perturbations:
\begin{align}
    \delta_S(\eta_i) = \delta_{S,i},\ \xi_S(\eta_i) = \delta_\rr(\eta_i) = \xi_\rr(\eta_i) = \Phi(\eta_i) = 0,
\end{align}
where $\eta_i$ is the initial time of the numerical calculation and $\eta_i \ll \eta_\text{MK}$.
We characterize the background with $F_S$, defined as the largest value of $\bar\rho_S/\bar \rho_\rr$ throughout the evolution.
With some fiducial parameter sets, we obtain \fig{fig:pertb_ax_dom} for the case with the rotation domination and \fig{fig:pertb_wo_ax_dom} for the case without the rotation domination.
These figures show that, if the rotation domination exists ($F_S > 1$), we find $|\sqrt{2\bar \rho_S}a \xi_S/(\mpl k^2)| \simeq |r \delta \theta/\mpl| \sim \mathcal O(1)\delta_{S,i}$ in the late time. 
Meanwhile, if the axion-rotatoin does not dominate the Universe, we find $|r \delta \theta/\mpl| \sim \mathcal O(1) F_S^{1/2} \delta_{S,i}$ in the late time.
These are consistent with the order estimates in Eq.~(\ref{eq:delta_chi_ini}).
Figure~\ref{fig:pertb_ax_dom_k} shows the evolution of $|\sqrt{2\bar \rho_S}a \xi_S/(\mpl k^2)|$ with different $k$ in the presence of the rotation domination.

Let us analytically derive the evolution of $\xi_S$ that enter the horizon in $\eta \gg \eta_\MK$.
In $\eta \gg \eta_\MK$, the energy fraction of axion decreases and the gravitational potential finally becomes negligible compared to $\delta_S$. 
Then, we can approximate the equation of motion for the axion kination fluid as
\begin{align}
    &\delta'_S + 2 \xi_S \simeq 0 \ \  (\eta \gg \eta_\MK), \\
    &\xi_S' - 2 \mathcal H \xi_S - \frac{k^2}{2} \delta_S \simeq 0 \ \  (\eta \gg \eta_\MK).
\end{align}
Taking the time derivative of the second equation, we obtain 
\begin{align}
    \xi_S'' - 2 \mathcal H' \xi_S - 2 \mathcal H \xi_S' + k^2\xi_S = 0.
\end{align}
With $\mathcal H = 1/\eta$ (RD era background), this becomes
\begin{align}
    &\xi_S'' - \frac{2}{\eta} \xi_S' + \left( k^2 + \frac{2}{\eta^2} \right)\xi_S = 0 \nonumber \\
    \Rightarrow & \ 
    \xi_S = D x \sin x + E x \cos x.
\end{align}
$D$ and $E$ are determined around the beginning of the RD era.
Figures~\ref{fig:pertb_ax_dom} and \ref{fig:pertb_wo_ax_dom} show that $\sqrt{2 \bar \rho_S} a \xi_S$ becomes constant in the late time on superhorizon scale.
Given $\sqrt{\bar \rho_S}a \propto a^{-2} \propto \eta^{-2}$ in the late time, we can find that $D x \sin x$ is the dominant solution in $\eta \gg \eta_\text{MK}$ and it is consistent with Eq.~(\ref{eq:trans_rd}).

\begin{figure}[h]
        \centering \includegraphics[width=0.8\columnwidth]{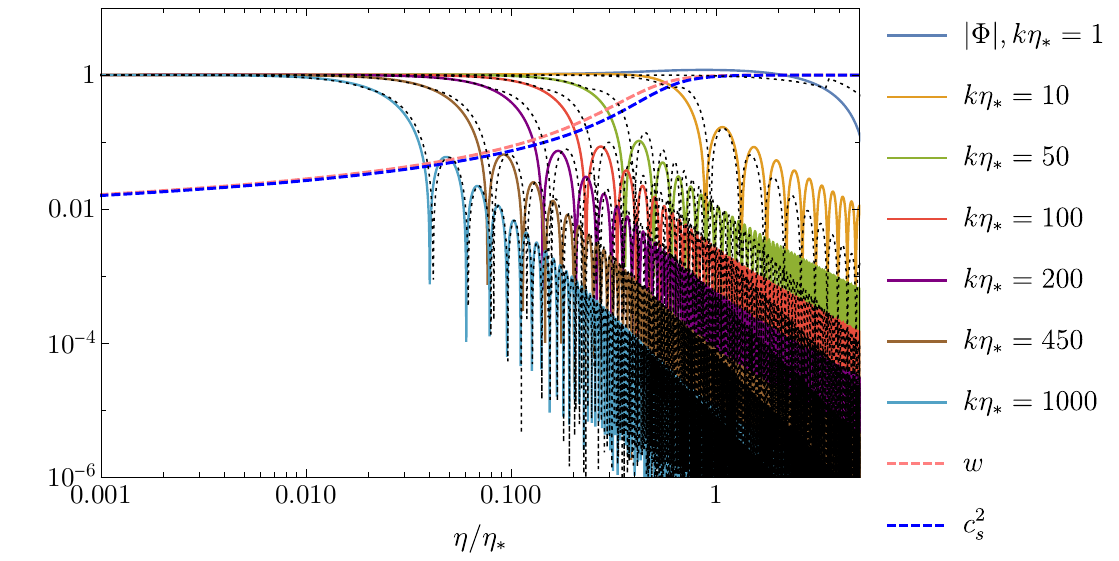}
        \caption{ 
        Evolution of $\Phi, w$, and $c_s^2$. 
        $\eta_*$ is defined as the conformal time at $c_s^2 = 0.95$.
        $\Phi$ is normalized so that $\Phi = 1$ much before the transition.
        The black dotted lines are the approximate formula Eq.~(\ref{eq:phi_approx}) with $A=2.5$, $B=-0.5$, and the parameters in \fig{fig:cs}.
		}
        \label{fig:phi2}
\end{figure}

\subsubsection{Analytic approximation of the gravitational potential}
For the calculation of the GW production during the rotation domination, the evolution of the gravitational potential is the most important.
We here discuss the analytical approximation of it.

\begin{figure}[t]
        \centering \includegraphics[width=0.7\columnwidth]{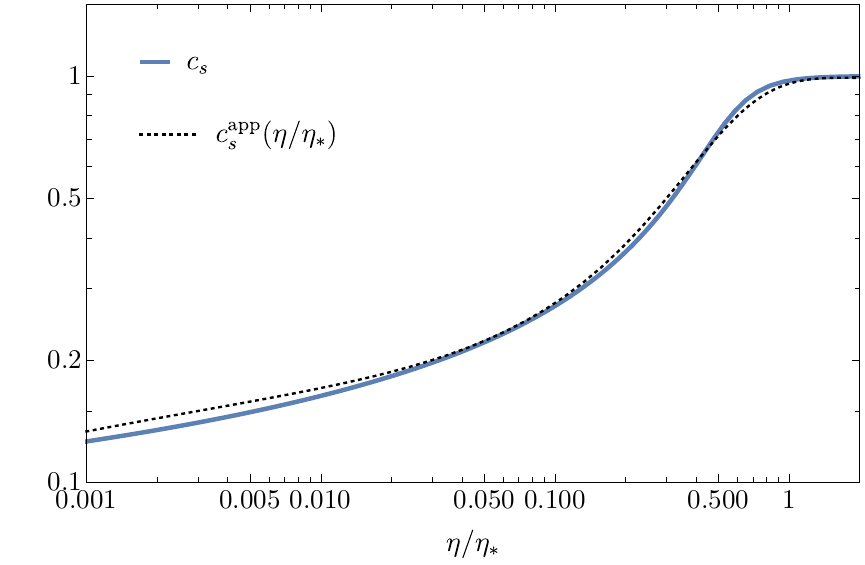}
        \caption{ 
        Evolution of $c_s$ and its approximate formula, given by Eq.~(\ref{eq:cs_app}), with $b_1 = 0.3$, $b_2 = 0.45$, and $b_3 = 0.03$.
		}
        \label{fig:cs}
\end{figure}

The equation of motion for the gravitational potential is given by~\cite{Mukhanov:991646}
\begin{align}
	\Phi'' + 3(1 + c_s^2) \mathcal H \Phi' + (c_s^2 k^2 + 3(c_s^2 - w)\mathcal H^2) \Phi = 0.
    \label{eq:phi_eom}
\end{align}
Figure~\ref{fig:phi2} shows the numerical results of the evolution of $\Phi, w$, and $c_s^2$ with the analytical approximation of $\Phi$, explained below.
From this figure, we can see that the gravitational potential, $\Phi$, starts to oscillate before $c_s^2 \simeq 1$ for $k \gg 1/\eta_*$ because of the nonzero $c_s$, which leads to the oscillations when $\eta \gtrsim 1/(k c_s)$.
To calculate the induced GWs within a reasonable computational cost, it is necessary to obtain the approximate formula for the gravitational potential.
If $w$ is constant and $c_s^2 = w$ is satisfied accordingly, we can solve Eq.~(\ref{eq:phi_eom}) as~\cite{Mukhanov:991646}
\begin{align}
    \Phi = C \eta^{-\nu} J_\nu(c_s k\eta), \ \nu \equiv \frac{1}{2} \left( \frac{5+3w}{1+3w} \right) \ \ (\text{if $w$ and $c_s$ are constant}).
\end{align}
However, the time dependence of the sound speed is non-negligible. 
Although we have obtained the $r$ dependence of $c_s$ in Eq.~(\ref{eq:cs_s}), what we need for the GW calculation is the analytical form with the explicit time dependence.
To this end, we first approximate the sound speed with\footnote{There is no physical meaning of this expression, which we found heuristically.}
\begin{equation}
	c_s^\app(y) = 
	\frac{1}{2} \left( 1+ \frac{\tanh\left[\frac{y-b_1}{b_2} \right]}{1+ b_3 \log y} \right),
	\label{eq:cs_app}
\end{equation}
where $y = \eta/\eta_*$. 
Figure~\ref{fig:cs} compares this approximate formula and the numerical result.
Using this, we approximate the gravitational potential as
\begin{align}
	\Phi^\text{app}(x) = \begin{cases}
	15 \sqrt{\pi/2} \left( \tilde c_s x \right)^{-5/2} J_{5/2}(\tilde c_s x) & (\eta_A < \eta) \\
	15 \sqrt{\pi/2} \left( \tilde c_{s,A} x \right)^{-5/2} \sqrt{\tilde c_s/\tilde c_{s,A}} J_{5/2}(\tilde c_s x) & (\eta_A > \eta)
	\end{cases},
	\label{eq:phi_approx}
\end{align}
where $x = k\eta$, $J_\nu$ is the Bessel function of the first kind, $\eta_A$ is the time when $\tilde c_s k \eta = A$, and 
\begin{align}
	\tilde c_s \equiv c_s^\app + B \frac{\dd c_s^\app}{\dd \log \eta}.
	\label{eq:til_cs_app}
\end{align}
In \fig{fig:phi2}, we compare the approximate formula of Eq.~(\ref{eq:phi_approx}) with the numerical results and can see that the approximate formula fits the numerical results to some extent, especially around the beginning of the oscillations.
Since the GWs are mainly produced around the beginning of the oscillation, we can use the approximate formula at least for a rough estimate of the induced GWs.

\subsection{GW production in the presence of rotation domination}

Let us calculate the GWs induced in $\eta < \eta_\MK$ in the presence of the rotation domination.
For simplicity, we approximate $w= 0$ until $\eta_*$ and $w= 1$ after that for the evolution of the induced GWs and background. 
On the other hand, we take into account the nonzero $c_s$ even when $c_s \ll 1$, which leads to the oscillations of the gravitational potential.
In the following, we calculate the induced GWs with the approximate formula of $\Phi$, obtained in the previous subsection.

The induced GWs can be calculated with Eq.~(\ref{eq:p_h_uv}). 
The modifications are incorporated in the kernel $I$:
\begin{equation}
  I(u,v,x) = \int^x_0 \dd \bar x\, k g_k(\eta; \bar \eta) f(u,v,\bar x),
  \label{eq:I_def}
\end{equation}
where Green's function satisfies 
\begin{align}
	\left[\frac{\partial^2}{\partial \eta^2} + 2 \mathcal H \frac{\partial}{\partial \eta} + k^2\right] g_k(\eta; \bar \eta) = \delta(\eta- \bar \eta).
\end{align}
The source function $f$ is given by 
\begin{align}
        f(u,v,\bar x) &= \frac{3}{25(1+w)} \left[ 2(5+3w) T_\Phi(u\bar{x})T_\Phi(v\bar{x}) +4 \mathcal H^{-1} \left(T_\Phi'(u\bar{x})T_\Phi(v\bar{x}) + T_\Phi(u\bar{x})T_\Phi'(v\bar{x})\right)\right. \nonumber \\
        &\left. 
        \qquad \qquad\qquad
         + 4 \mathcal H^{-2} T_\Phi'(u\bar{x})T_\Phi'(v\bar{x}) \right],
\label{eq:f_def}
\end{align}
where $T_\Phi(x)$ is the transfer function of the gravitational potential and defined as $\Phi_{\mathbf k}(\eta) = -(3/5) T_\Phi(k\eta) \zeta_{\mathbf k}$.
Note that $T_\Phi(x) = 1$ for the superhorizon-limit perturbations during the eMD era, followed by the kination era.
Note also that $T_\Phi'(x) = \partial T_\Phi(x)/\partial \eta = k \partial T_\Phi(x)/\partial x$.

For convenience, we separate the contributions into 
\begin{align}
  I(u,v,x) = I_{\text{eMD}}(u,v,x) + I_{\text{KD}}(u,v,x),
  \label{eq:I_def}
\end{align}
where
\begin{align}
	I_{\text{eMD}}(u,v,x) &\equiv \int^{x_*}_0 \dd \bar x\, k g^{\emd \to \kd}_k(\eta; \bar \eta) f(u,v,\bar x), \\
	I_{\text{KD}}(u,v,x) &\equiv \int^{x}_{x_*} \dd \bar x\, k g^\kd_k(\eta; \bar \eta) f(u,v,\bar x),
\end{align}
where $x_* = k\eta_*$.
In section~\ref{subsec:iGW_KD}, we focused on $I_\kd$ neglecting the perturbations that enter the horizon during the eMD era. 
In the following, we focus on the contribution of $I_\emd$:
\begin{align}
        \overline{\mathcal P^\emd_h(\eta, k)} = 4 \int^\infty_0 \dd v \int^{1+v}_{|1-v|} \dd u \left( \frac{4v^2 - (1+v^2-u^2)^2}{4 u v} \right)^2 \overline {I^2_\emd(u,v,x)} \mathcal P_\zeta(k u) \mathcal P_\zeta(k v).
        \label{eq:p_h_uv_emd}
\end{align}
First, we derive Green's function for $I_\emd$. 
During the eMD era, $g_k(\eta;\bar\eta)$ follows
\begin{align}
	\left[\frac{\partial^2}{\partial \eta^2} + \frac{4}{\eta}\frac{\partial}{\partial \eta} + k^2\right] g^\emd_k(\eta; \bar \eta) = \delta(\eta- \bar \eta) \quad (\text{for } \eta < \eta_*),
\end{align}
where we have set $w = 0$ during $\eta < \eta_*$.
This gives\footnote{
Note that the definition of Green's function in this paper is different from that in Ref.~\cite{Kohri:2018awv} by the factor of $a(\bar \eta)/a(\eta)$.
}
\begin{align}
	k g^\emd_k(\eta,\bar\eta) &= -\Theta(\eta-\bar \eta) \frac{\bar x^3}{x} (j_1(x) y_1(\bar x) - j_1(\bar x) y_1(x)) \quad (\text{for } x < x_*), 
	\label{eq:green_emd}
\end{align}
where $j_l$ and $y_l$ are the spherical Bessel functions of the first and the second kind, respectively. 
On the other hand, Green's function during the kination era follows
\begin{align}
	\left[\frac{\partial^2}{\partial \tilde \eta^2} + \frac{1}{\tilde \eta}\frac{\partial}{\partial \tilde \eta} + k^2\right] g_k(\eta; \bar \eta) = \delta(\eta- \bar \eta) \quad (\text{for } \eta > \eta_*),
\end{align}
where $\tilde \eta = \eta- (3/4)\eta_*$.
Since we are here interested in the GWs induced during the eMD era, we solve 
\begin{equation}
	\left[\frac{\partial^2}{\partial \tilde \eta^2} + \frac{1}{\tilde \eta}\frac{\partial}{\partial \tilde \eta} + k^2\right] g^{\emd \to \kd}_k(\eta; \bar \eta) = 0  \quad (\text{for } \eta > \eta_*),
\end{equation}
with the initial condition of $g_k$ and $g_k'$ at $\eta_*$, which are determined by Eq.~(\ref{eq:green_emd}) and their continuity at $\eta_*$.
Then, we obtain 
\begin{align}
	k g^{\emd \to \kd}_k(\eta; \bar \eta) = C(x_*, \bar x) J_0(x-(3/4)x_*) + D(x_*,\bar x) Y_0(x-(3/4)x_*),
\end{align}
where $Y_\nu$ is the Bessel function of the second kind. 
The coefficients $C$ and $D$ are given by 
\begin{align}
	C(x_*,\bar x) &= \frac{{k g^\emd_k}'(\eta_*;\bar \eta) Y_0(x_*/4) - {k g^\emd_k}(\eta_*;\bar \eta) Y_0'(x_*/4)}{J_0'(x_*/4) Y_0(x_*/4) - J_0(x_*/4) Y_0'(x_*/4)}, \\ 
	D(x_*,\bar x) &= \frac{{kg^\emd_k}(\eta_*;\bar \eta) J_0'(x_*/4) - {kg^\emd_k}'(\eta_*;\bar \eta) J_0(x_*/4)}{J_0'(x_*/4) Y_0(x_*/4) - J_0(x_*/4) Y_0'(x_*/4)},
\end{align}
where the primes here denote the derivative with respect to $x_*$ (e.g. $J_0'(x_*/4) = \dd J_0(x_*/4)/\dd x_*$).
Since we are interested in GWs in the late-time limit, we take the $x \gg 1$ limit and obtain
\begin{align}
	k g^{\emd \to \kd}_k(\eta; \bar \eta) \simeq & \ C(x_*, \bar x) \sqrt{\frac{2}{\pi x}} \cos\left( x - \frac{\pi}{4} \right) \nonumber \\
 &+ D(x_*,\bar x) \sqrt{\frac{2}{\pi x}} \sin\left( x - \frac{\pi}{4} \right) \quad \  (\text{for } x \gg x_* \text{ and } x \gg 1).
	\label{eq:g_emd_kd}
\end{align}
On the other hand, during the eMD era ($w=0$), $f$ is given by
\begin{align}
        f(u,v,\bar x) 
        &= \frac{3}{25} \left[ 10 T_\Phi(u\bar{x})T_\Phi(v\bar{x}) +2 \eta \left(T_\Phi'(u\bar{x})T_\Phi(v\bar{x}) + T_\Phi(u\bar{x})T_\Phi'(v\bar{x})\right) + \eta^2 T_\Phi'(u\bar{x})T_\Phi'(v\bar{x}) \right].
\label{eq:f_def_emd}
\end{align}
To take into account the oscillation of gravitational potential during the eMD era, we substitute $T_\Phi(x) = \Phi^\text{app}(x)$, given in Eq.~(\ref{eq:phi_approx}).

\begin{figure}[h]
        \centering \includegraphics[width=0.7\columnwidth]{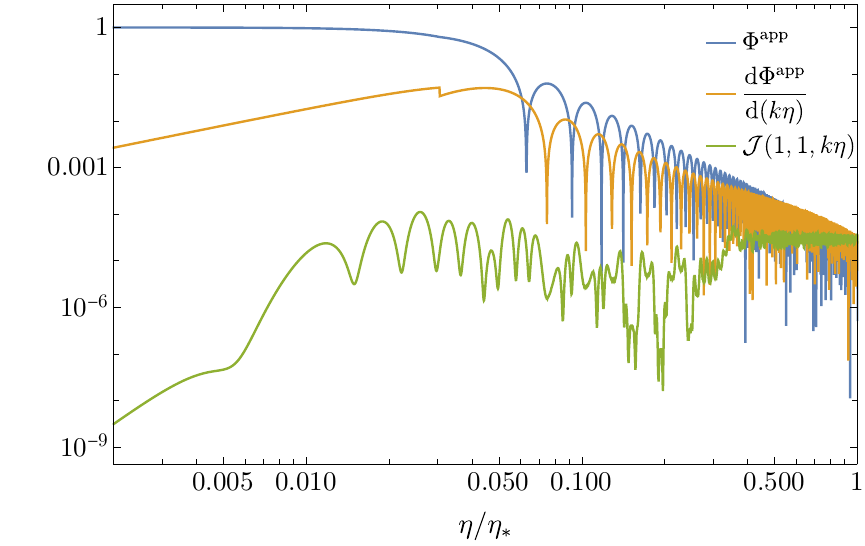}
        \caption{ 
        The evolution of $\Phi$, $\dd \Phi/\dd x$, and $\mathcal J$ with $u=v=1$ and $k \eta_* = 450$.
		}
        \label{fig:i_kernel}
\end{figure}

Combining these expressions, we can reexpress $I_\emd$ as 
\begin{align}
	I_\emd(u,v,x) &\equiv \int^{x_*}_0 \dd \bar x\, k g^{\emd \to \kd}_k(\eta; \bar \eta) f(u,v,\bar x) \nonumber \\
	&=  \sqrt{\frac{2}{x}} \left[\cos\left( x - \frac{\pi}{4} \right) \mathcal I_C(u,v,x_*) + \sin\left( x - \frac{\pi}{4} \right) \mathcal I_D(u,v,x_*) \right],
\end{align}
where 
\begin{align}
	\label{eq:ic_int}
	\mathcal I_C (u,v,x_*) &\equiv \int^{x_*}_0 \dd \bar x\, C(x_*, \bar x) \frac{1}{\sqrt{\pi}} f(u,v,\bar x), \\
	\label{eq:id_int}
	\mathcal I_D (u,v,x_*) &\equiv \int^{x_*}_0 \dd \bar x\, D(x_*, \bar x) \frac{1}{\sqrt{\pi}} f(u,v,\bar x).
\end{align}
Then, we can express the time average of $I_\emd^2$ as 
\begin{align}
	\overline{I_\emd(u,v,x)}^2 &= \frac{1}{x} \left( \mathcal I_C^2(u,v,x_*) + \mathcal I_D^2(u,v,x_*)\right).
\end{align}
Substituting this into Eq.~(\ref{eq:p_h_uv_emd}), we calculate the GWs induced during the eMD era. 
To see the time dependence of the integrals in Eqs.~(\ref{eq:ic_int}) and (\ref{eq:id_int}), we define the test function as 
\begin{align}
	\mathcal J(u,v,x) = \left(\int^{x}_0 \dd \bar x\, C(x_*, \bar x) \frac{1}{\sqrt{\pi}} f(u,v,\bar x)\right)^2 + \left(\int^{x}_0 \dd \bar x\, D(x_*, \bar x) \frac{1}{\sqrt{\pi}} f(u,v,\bar x)\right)^2.
\end{align}
Figure~\ref{fig:i_kernel} shows the evolution of this test function. 
From this, we can see that the integrals almost reach a constant before the kination era begins, which indicates that the tensor perturbations had decoupled from the source terms by then.

To focus on the contributions from the perturbations that enter the horizon during the eMD era, we consider the following power spectrum:
\begin{align}
	\mathcal P_\zeta = A_{\zeta_S} \Theta(k - k_{l,\text{cut}}) \Theta(k_{s,\text{cut}} - k), 
\end{align}
where $k_{l,\text{cut}}$ and $k_{s,\text{cut}}$ are the large- and small-scale cutoff scales.
The large-scale cutoff $k_{l,\text{cut}}$ is taken to be $k_{l,\text{cut}} \gg 1/\eta_*$ because the modes on $k \sim 1/\eta_*$ induce GWs when the approximation of $w = 0$ is not good ($\eta \simeq \eta_*$).
Also, we take a relatively large $k_{s,\text{cut}}$ (e.g. $k_{s,\text{cut}} = 1000/\eta_*$) as one fiducial parameter, which leads to nonlinear density perturbation if $w=c_s=0$ exactly holds. 
However, we note that, in our situation, the density perturbations stop growing and start oscillating much before $\eta_*$ because of the nonzero $c_s$.

\begin{figure}[h]
        \centering \includegraphics[width=0.7\columnwidth]{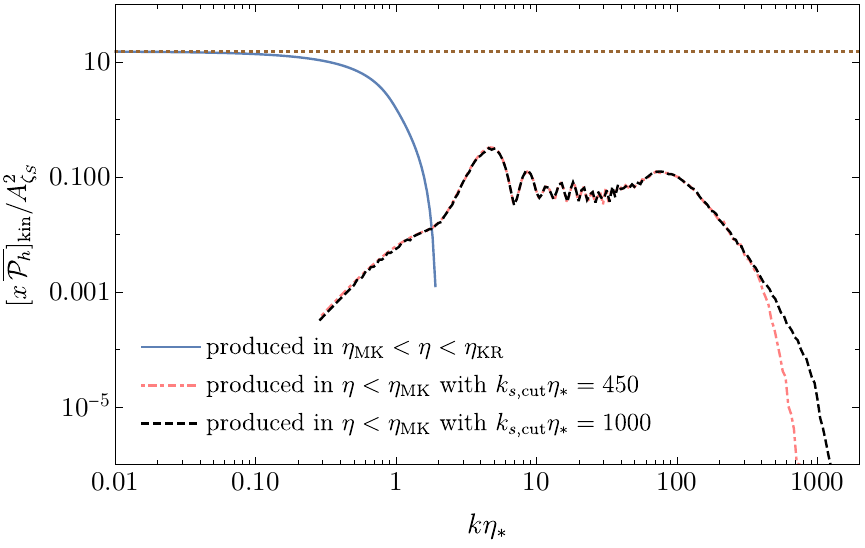}
        \caption{ 
        The power spectrum of GWs induced in $\eta < \eta_\KR$.
        Except for the pink line, all the lines are the same as \fig{fig:ph}, though we regard $\eta_*$ as $\eta_\MK$ there.
        We take $k_{l,\text{cut}} = 50/\eta_*$ for both the pink dot-dashed and the black dashed lines and take $k_{s,\text{cut}} = 450/\eta_*$ for the pink line and $k_{s,\text{cut}} = 1000/\eta_*$ for the black line. 
        The blue solid line shows the spectrum of GWs induced by the perturbations that enter the horizon during the kination era, which is discussed in section~\ref{sec:induced-GW}.
		}
        \label{fig:ph2}
\end{figure}

Figure~\ref{fig:ph2} shows the results. 
We here remark the caveats in our calculation. 
First, we stop the integral at $\eta_*$ in Eqs.~(\ref{eq:ic_int}) and (\ref{eq:id_int}). 
This may overlook further suppression of the induced GWs, similarly to the gradual transition case~\cite{Inomata:2019zqy}.
Second, we use the piecewise Green function, Eq.~(\ref{eq:g_emd_kd}), by assuming the sudden transition from the eMD era to the kination era. 
This may cause the unphysical oscillation of the power spectrum, as discussed in Ref.~\cite{Pearce:2023kxp}. 
Specifically, the oscillation of the GW spectrum around $\mathcal O(1) \lesssim k\eta_* \lesssim \mathcal O(10)$ would come from our choice of the Green function. 
Due to these caveats, our results for the GWs induced during the eMD era should be considered as a rough estimate of the upper bound of the GWs on the scales, $k > 1/\eta_*$.

\subsection{GW production in the absence of rotation domination}

Finally, let us roughly estimate the amplitude of GWs produced in $\eta < \eta_\MK$ in the absence of the rotation domination ($F_S < 1$).
Specifically, we derive the UV tail of the GW spectrum, produced by the oscillation of the perturbations due to non-zero $c_s$ in $\eta < \eta_\MK$ in the log potential model (Eq.~(\ref{eq:pot_vs})).
Since the gravitational potential is suppressed by $F_S$ in this case, we can safely focus only on the contributions from the fluid velocity $\delta u$ (or $\delta \chi$).
For simplicity, we focus on the GWs produced around the horizon entry of the source perturbations, which is expected to be the dominant contribution because the source perturbations decay on subhorizon scales. 
Then, we can roughly estimate the order of the induced tensor perturbation during its production as
\begin{align}
    \label{eq:h_ij_estimate}
    h_{ij,\bfk} \sim \frac{S_{ij,\bfk}}{k^2} \sim (\bar \rho_S + \bar P_S) (\delta u_\bfk)^2 
    \sim \bar \rho_S(1 + w_S) a^2 \left(\frac{\delta \theta_\bfk}{\bar \theta'}\right)^2 
    \sim \bar \rho_S(1 + w_S) \frac{1}{H^2} \left(\frac{\delta \theta'_\bfk}{\bar \theta'}\right)^2,
\end{align}
where we have used $\delta \theta \sim \delta \theta'/\mathcal H$. 
Using Eqs.~(\ref{eq:rho_exa}) and (\ref{eq:r_theta_rel}), we can obtain 
\begin{align}
    \frac{\delta \theta'}{\theta'} &\simeq \frac{V''-V'/r}{2V'/r} \frac{\delta r}{\bar r} \\
    &= \frac{c_{s,S}^2}{1+w_S} \delta_{S},
\end{align}
where we have neglected the gravitational potential in Eq.~(\ref{eq:r_theta_rel}). 
Using this, we can reexpress Eq.~(\ref{eq:h_ij_estimate}) as 
\begin{equation}
    h_{ij,\bfk} \sim \frac{\bar \rho_S}{\bar \rho_\tot} \frac{c_{s,S}^4}{1+w_S} \delta_{S,\bfk}^2.
\end{equation}
Roughly speaking, the GW production stops when $\delta_{S,\bfk}$ starts to oscillate, which occurs around $kc_s \eta \sim 1$ with the the same order of the initial amplitude of $\delta_{S}$ because $\delta_S$ in $\eta < \eta_\MK$ (during the RD era) cannot grow. 
Then, the final amount of GWs can be expressed as 
\begin{equation}
    \mathcal P_h(k) \sim \left(\frac{\bar \rho_S(\eta_\osc)}{\bar \rho_\tot(\eta_\osc)}\right)^2 \frac{c_{s,S}^8(\eta_\osc)}{(1+w_S(\eta_\osc))^2} \mathcal P_{\delta_S}(k),
\end{equation}
where $\eta_\osc \sim 1/(k c_s)$. 
Then, the energy density parameter of GWs is given by 
\begin{align}
    \label{eq:omega_gw_log_pot}
    \Omega_\GW(k) &\sim \left(\frac{k}{\mathcal H(\eta_\osc)}\right)^2 \left(\frac{\bar \rho_S(\eta_\osc)}{\bar \rho_\tot(\eta_\osc)}\right)^2 \frac{c_{s,S}^8(\eta_\osc)}{(1+w_S(\eta_\osc))^2} \mathcal P_{\delta_S}(k)\nonumber \\
    &\sim F_S^2 \left(k \eta_{\MK} \right)^{-2} \frac{c_{s,S}^4 (\eta_\osc)}{(1+w_S(\eta_\osc))^2} \mathcal P_{\delta_S}(k),
\end{align}
where we have used $\bar \rho_S(\eta_\osc)/\bar \rho_\tot(\eta_\osc) \sim F_S (c_s k\eta_\MK)^{-2}$.
This is used for the black dashed line in \fig{fig:gw_wo}.

\section{PBH formation during a short matter dominance}
\label{app:PBH_finiteMD}

In this appendix, we study PBH formation when the variance of the density perturbations does not become ${\cal O}(1)$ by the end of eMD due to either a short duration of eMD or a non-zero Jeans length (non-perfect matter fluid).
The derivation in this appendix is based on the method employed in Ref.~\cite{Harada:2016mhb}.  

Consider a sphere of comoving radius $R\sim k^{-1}$, or a physical radius $a(t) R$. Let us denote the average density across many such spheres to be $\bar{\rho}$.
The overdensity/underdensity in a particular volume $R^3$ can be captured by considering distortions of the sphere, which can then be considered an ellipsoid with three physical radii 
\begin{align}
    r_1 &= [a(t) - \alpha(k) b(t)] \, R, \nonumber\\
    r_2 &= [a(t) - \beta(k) b(t)] \, R, \nonumber\\
    r_3 &= [a(t) - \gamma(k) b(t)] \, R .   
\end{align}
The first term in above expressions gives homogeneous expansion, while the second term captures the squeezing/stretching of the sphere.
The deviation from the average density is then
\begin{align}\label{eq:overdensity}
    \rho \mathrm d^{3}r &= \bar{\rho} a^{3} \mathrm d^{3} R \nonumber \\
    \implies \, \rho &= \frac{a^{3}}{(a-\alpha b) (a-\beta b) (a-\gamma b)} \bar{\rho} \nonumber ,\\
    \xrightarrow{\rm linear \, regime} \, \delta &= \frac{\rho -\bar{\rho}}{\bar{\rho}} = (\alpha + \beta + \gamma) \frac{b(t)}{a(t)} + {\cal O}(\alpha^2, \beta^2, \gamma^2,\cdot \cdot).
\end{align}
During eMD, $\delta(t) \propto a(t) \implies b(t) \propto a^{2}(t)$. 
Therefore, $\alpha, \beta, \gamma \sim {\cal O}(\delta)$. 
Let us fix the constant of proportionality by defining 
\begin{align}
    \frac{b(t)}{a(t)} = \frac{a(t)}{a(t_i)},
\end{align}
where $t_i$ is the initial time when the comoving mode $k\sim R^{-1}$ re-enters the horizon.
Then the initial overdensity $\delta_i \approx \alpha+\beta+\gamma$, which means the standard deviation of the distributions of the parameters $(\alpha, \beta,\gamma)$ must be ${\cal O}(\delta_i)$.

Without loss of generality, let us take $\alpha \geq \beta \geq \gamma$. We also assume $\alpha > 0$ such that $r_1$ is the shortest direction. 
The collapse of the ellipsoid will start along this direction.
Let us use $t_f$ to denote the time at which the shortest direction is maximally expanded.
Therefore, $\dot{r}_1(t_f) =0$, giving
\begin{equation}
    \frac{b(t_f)}{a(t_f)} = \frac{1}{2 \alpha} \implies a(t_f) = \frac{1}{2\alpha} a(t_i).
\end{equation}
Let us denote the time of collapse along $r_1$ directions by $t_c$, which is fixed and given by the condition $r_{1}(t_c) =0$. Then
\begin{align}\label{eq:tc}
    \frac{b(t_c)}{a(t_c)} =  \frac{1}{\alpha} \implies a(t_c) = \frac{1}{\alpha} a(t_i).
\end{align}
At this time, the ellipsoid collapses into a disc with 
\begin{align}\label{eq:r2_r3_disc}
    r_2 (t_c) &= \paren{\frac{\alpha-\beta}{\alpha^2}}\, a(t_i) R,\nonumber \\ 
    r_3 (t_c) &= \paren{\frac{\alpha-\gamma}{\alpha^2}}\, a(t_i) R.
\end{align}

The hoop conjecture~\cite{1972mwm..book..231T}
gives a criterion for when a black hole can form.
It states that a black hole forms if and only if the mass $m$ is compactified into a region whose largest circumference is smaller than $2\pi r_g$.\footnote{If the hoop criterion is not satisfied, the ellipsoid can evolves in a complex way with the possibility of forming caustics in the middle, which can eject matter. This reduces $r_g$, further preventing PBH formation.}
Here, the mass within the ellipsoid and the corresponding gravitational (Schwarzschild) radius is given by
\begin{align}\label{eq:rg}
    m &= \frac{4 \pi}{3} (a(t_i) R)^{3} \bar{\rho}(t_i) = \frac{1}{2 G H(t_i)}, \nonumber \\
    \implies \, r_{g} &= 2 G m = \frac{1}{H(t_i)} = a(t_i)R,
\end{align}
where $G$ is the Newton's constant.
We consider a simpler criterion from hoop conjecture that a PBH forms if and only if the mass $m$ completely fits within a sphere of radius $r_g$. 
Then we get a simple conditions for PBH formation: $r_3 (t_c),\, r_2 (t_c) < r_g$. Using \eqs{eq:rg}{eq:r2_r3_disc}, these conditions translate to
\begin{align}\label{eq:hoop_criterion}
   (\alpha-\gamma) \lesssim \alpha^2 .
\end{align}

The probability distribution for $\alpha,\, \beta$, and $\gamma$ was derived in Ref.~\cite{doroshkevich1970spatial}, and is given by
\begin{align}
    p(\alpha, \beta, \gamma) \mathrm d\alpha \mathrm d\beta \mathrm d\gamma &= \frac{-27}{8 \sqrt{5} \pi \sigma^6} (\alpha-\beta)(\beta-\gamma) (\gamma-\alpha) \nonumber \\
    & \quad \times \exp\left[-\frac{3}{5 \sigma^2} \paren{(\alpha^2 +\beta^2 +\gamma^2)-\frac{1}{2}(\alpha\beta + \beta\gamma+\gamma\alpha)} \right] \mathrm d\alpha \mathrm d\beta \mathrm d\gamma,
\end{align}
where $\infty >\alpha \geq \beta \geq \gamma > -\infty$.
The factors in the first line above are responsible for the suppressed probability for spherically symmetric configurations.
The variance $\sigma$ is related to the density fluctuation as
\begin{equation}
    \sigma^2 = \frac{\bra \delta^{2}_{i} \ket}{5}.
\end{equation}
It is easier to work in redefined variables $x=\frac{\alpha+\beta+\gamma}{3}$, $y=\frac{(\alpha-\beta)-(\beta-\gamma)}{4}$, and $z=\frac{\alpha-\gamma}{2}$ such that the covariance matrix is diagonalized. Then
\begin{align}\label{eq:p_xyz}
    \tilde{p}(x,y,z) \mathrm dx \mathrm dy \mathrm dz 
    &= \frac{-27}{\sqrt{5} \pi \sigma^6} (4 y^2 z -z^3) \exp\left[-\frac{9}{10}\paren{\frac{x}{\sigma}}^{2} -2 \paren{\frac{y}{\sigma}}^{2} -\frac{3}{2} \paren{\frac{z}{\sigma}}^{2} \right] \mathrm dx \mathrm dy \mathrm dz,
\end{align}
where $\infty> x >-\infty$, $\infty> y >-\infty$, and $\infty> z \geq 2 |y|$.

Let us evaluate the PBH collapse fraction using \eqs{eq:hoop_criterion}{eq:p_xyz} in various cosmological eras.\\
1) \textbf{Long MD}: This is the scenario where MD is long enough such that all relevant modes become non-linear. As the $x$-integral is suppressed beyond $\sigma$, the dominant contribution to the integral comes from the region $ 0 < x \approx\frac{\alpha}{3} < \sigma$.
From the hoop criterion, the range of the other two integrals is constrained as $ 2|y|< z < \frac{\alpha^2}{2}$, and $ -\frac{\alpha^2}{4}< y < \frac{\alpha^2}{4}$.
Notice that $y$ and $z$ integrals have support on a much shorter range compared to standard deviation of the Gaussian, $\sigma (\sim \delta_i)$.  
The collapse fraction can therefore be estimated as
\begin{align}
    \beta_{\rm MD} \sim \frac{1}{\sigma^6} (\sigma^{2})^5 \sigma \sim \sigma^5 .
\end{align}
This matches the parametric form obtained in Ref.~\cite{Harada:2016mhb} for a period of long MD. \\
2) \textbf{non-MD}: For any era that is not MD (with an equation of state $w$), the PBH collapse predominantly occurs at horizon re-entry. 
Only volumes with overdensity larger than a critical overdensity $\delta_c = \frac{3(1+w)}{5+3w} \sin\paren{\frac{\sqrt{\pi w}}{1+3w}}= {\cal O}(1)$ form a black hole.
Since $\delta > \delta_c$ for PBH formation, 
$\alpha \gtrsim \delta_c$, i.e. $ x > \frac{\delta_c}{3}$ from \eq{eq:overdensity}. 
Since $\delta_c > \sigma$ in our case, the $x$-integral is exponentially suppressed with the maximum contribution coming from around $x\sim \frac{\delta_c}{3}$,
\begin{align}\label{eq:largex0_integral}
    \int_{x_0 \approx \delta_{c}/3}^{\infty} dx \exp\left[\frac{-9}{10}\paren{\frac{x}{\sigma}}^{2}\right] = \sqrt{\frac{5 \pi}{18}} \sigma \, {\rm Ercf}\left[\sqrt{\frac{9}{10}\paren{\frac{x_0}{\sigma}}^2}\right] \xrightarrow{x_0 \gg \sigma} \sqrt{\frac{10}{9}} \, \frac{\sigma^2}{\delta_c} \exp\left[-\frac{\delta_c^{2}}{2 \sigma^{2}}\right] ,
\end{align}
where ${\rm Erfc}(x)$ is the complementary error function.
The hoop criterion in this case is not very constraining, and the $y$ and $z$ integrals run over the full range of the Gaussian (i.e. $\sigma$).
This gives the form of $\beta_{\rm RD}$ as 
\begin{align}
    \beta_{\rm RD} \sim \frac{1}{\sigma^6} \, (\sigma)^5 \, \frac{\sigma^2}{\delta_c} \, e^{-\frac{\delta_c^{2}}{2 \sigma^{2}}} \sim \frac{\sigma}{\delta_c} \, e^{-\frac{\delta_{c}^{2}}{2 \sigma^2}}.
\end{align}
This matches the expected form of $\beta_{\rm RD}$~\cite{Carr:1975qj}.  \\
3) \textbf{Short MD}: 
Let $t_{\rm end}$ denote the time when the linear growth in density fluctuation stops. 
This may be because of the short duration of eMD or a finite Jeans length due to an imperfect matter fluid (as in our model).
We then require the collapse in the $r_1$ direction to occur before $t_{\rm end}$. 
That is, $\frac{1}{\alpha} = \frac{a(t_c)}{a(t_i)} < \frac{a(t_{\rm end})}{a(t_i)} \implies \alpha > \frac{a(t_i)}{a(t_{\rm end})}$.
Let us denote the standard deviation of the perturbation at $t_{\rm end}$ by $\delta_{\rm end}$. 
Then $\frac{a(t_i)}{a(t_{\rm end})} = \frac{\sigma}{\delta_{\rm end}}$. 
Therefore, $\alpha \geq \alpha_{\rm min} =\paren{\frac{\sigma}{\delta_{\rm end}}} \delta_c $.
Here $\delta_c$ corresponds to the critical density of the total fluid at $t_{\rm end}$.
We have included $\delta_c$ in $\alpha_{\rm min}$ since we expect the condition on $\alpha$ to become  $\alpha > \delta_c$ as $\delta_{\rm end} \rightarrow \sigma$ from the analysis of non-MD era above.
Since $\delta_{\rm end} <1$, $\alpha_{\rm min} > \sigma$, 
which means that the $x$-integral has maximum support around $\alpha_{\rm min}/3$.  
In this case, the other two integrals run over ranges $|y| < \frac{\alpha_{\rm min}^2}{4}$ and $2|y| < z < \frac{\alpha_{\rm min}^2}{2}$.
This gives rise to two distinct regimes:
\begin{itemize}
    \item $\alpha_{\rm min}^{2} > \sigma \rightarrow \sqrt{\sigma} > \delta_{\rm end}/\delta_c$: In this case, the $y$ and $z$ integrals run over the full range of the Gaussian, giving
    \begin{align}\label{eq:betaMD_smallDeltaEnd}
    \beta_{\rm MD}|_{\sqrt{\sigma} > \frac{\delta_{\rm end}}{\delta_c}} \sim \frac{1}{\sigma^6} \, \sigma^5 \, \frac{\sigma^2}{\alpha_{\rm min}} \exp\left[-\frac{\delta_{c}^{2}}{2 \delta_{\rm end}^{2}}\right] \sim \frac{\delta_{\rm end}}{\delta_c} \exp\left[-\frac{\delta_{c}^{2}}{2 \delta_{\rm end}^{2}}\right].
    \end{align}
    As expected, the form of this expression is similar to $\beta_{\rm RD}$, and it matches onto it when $\delta_{\rm end} \rightarrow \sigma$.

    \item $\alpha_{\rm min}^{2} < \sigma \rightarrow \sqrt{\sigma} < \delta_{\rm end}/\delta_c$: in this case, the $y$ and $z$ integrals do not run over the full range of the Gaussian, giving a smaller contribution,
    \begin{align}\label{eq:betaMD_largeDeltaEnd}
     \beta_{\rm MD}|_{\sqrt{\sigma} < \frac{\delta_{\rm end}}{\delta_c}} \sim \frac{1}{\sigma^6} \, (\alpha_{\rm min}^2)^5 \, \frac{\sigma^2}{\alpha_{\rm min}} \exp\left[-\frac{\delta_{c}^{2}}{2 \delta_{\rm end}^{2}}\right] \sim \frac{\sigma^5}{(\delta_{\rm end}^9 / \delta_c^9)} \exp\left[-\frac{\delta_{c}^{2}}{2 \delta_{\rm end}^{2}}\right].
    \end{align}
    This expression matches $\beta_{\rm MD}$ from long MD as $\delta_{\rm end} \rightarrow \delta_c$.
\end{itemize} 
The two forms in \eqs{eq:betaMD_smallDeltaEnd}{eq:betaMD_largeDeltaEnd} match at the transition $\delta_{\rm end}/\delta_c = \sqrt{\sigma}$.

For the modes that re-enter before eMD, the amplitude of the fluctuations remains constant. However, the gravitational radius reduces after horizon re-entry (in contrast to the constant $r_g$ for modes re-entering during eMD).
For these modes,  
\begin{align}
    \left.\frac{r_g}{r_1}\right|_{t_f} = 4 \alpha \paren{\frac{a_k}{a_{\rm RM}}}^2.
\end{align}
Comparing with $\frac{r_g}{r_{1,f}} = 4 \alpha$ for modes re-entering during eMD, we see that $r_g$ is much smaller for modes that re-enter before eMD.
The hoop criterion is therefore stronger for these modes,
\begin{align}\label{eq:hoopCriterion_eRD}
    (\alpha-\gamma) \lesssim \alpha^2 \paren{\frac{a_k}{a_{\rm RM}}}^2.
\end{align}
We can then follow the same procedure as above with the modified hoop criterion. 
For modes that re-enter sufficiently before eMD,
the collapse fraction is approximately given as 
$\beta(k< k_{\rm RM}) \approx \paren{\frac{a_k}{a_{\rm RM}}}^{10} \beta_{\rm MD}(k_{\rm RM})$.

\addcontentsline{toc}{section}{References}

\bibliographystyle{JHEP}
\bibliography{refs.bib}

\providecommand{\href}[2]{#2}\begingroup\raggedright\begin{thebibliography}{100}

\bibitem{LISA:2017pwj}
{\bf LISA} Collaboration, P.~Amaro-Seoane et~al., {\it {Laser Interferometer Space Antenna}},  \href{http://arxiv.org/abs/1702.00786}{{\tt arXiv:1702.00786}}.

\bibitem{Harry:2006fi}
G.~M. Harry, P.~Fritschel, D.~A. Shaddock, W.~Folkner, and E.~S. Phinney, {\it {Laser interferometry for the big bang observer}},  {\em Class. Quant. Grav.} {\bf 23} (2006) 4887--4894. [Erratum: Class.Quant.Grav. 23, 7361 (2006)].

\bibitem{Kawamura:2011zz}
S.~Kawamura et~al., {\it {The Japanese space gravitational wave antenna: DECIGO}},  {\em Class. Quant. Grav.} {\bf 28} (2011) 094011.

\bibitem{Sesana:2019vho}
A.~Sesana et~al., {\it {Unveiling the gravitational universe at $\mu$-Hz frequencies}},  {\em Exper. Astron.} {\bf 51} (2021), no.~3 1333--1383, [\href{http://arxiv.org/abs/1908.11391}{{\tt arXiv:1908.11391}}].

\bibitem{NANOGrav:2023gor}
{\bf NANOGrav} Collaboration, G.~Agazie et~al., {\it {The NANOGrav 15 yr Data Set: Evidence for a Gravitational-wave Background}},  {\em Astrophys. J. Lett.} {\bf 951} (2023), no.~1 L8, [\href{http://arxiv.org/abs/2306.16213}{{\tt arXiv:2306.16213}}].

\bibitem{Braun:2015zta}
R.~Braun, T.~Bourke, J.~A. Green, E.~Keane, and J.~Wagg, {\it {Advancing Astrophysics with the Square Kilometre Array}},  {\em PoS} {\bf AASKA14} (2015) 174.

\bibitem{Hall:2022dik}
E.~D. Hall, {\it {Cosmic Explorer: A Next-Generation Ground-Based Gravitational-Wave Observatory}},  {\em Galaxies} {\bf 10} (2022), no.~4 90.

\bibitem{Punturo:2010zz}
M.~Punturo et~al., {\it {The Einstein Telescope: A third-generation gravitational wave observatory}},  {\em Class. Quant. Grav.} {\bf 27} (2010) 194002.

\bibitem{Roshan:2024qnv}
R.~Roshan and G.~White, {\it {Using gravitational waves to see the first second of the Universe}},  {\em Rev. Mod. Phys.} {\bf 97} (2025), no.~1 015001, [\href{http://arxiv.org/abs/2401.04388}{{\tt arXiv:2401.04388}}].

\bibitem{Affleck:1984fy}
I.~Affleck and M.~Dine, {\it {A New Mechanism for Baryogenesis}},  {\em Nucl. Phys. B} {\bf 249} (1985) 361--380.

\bibitem{Co:2019wyp}
R.~T. Co and K.~Harigaya, {\it {Axiogenesis}},  {\em Phys. Rev. Lett.} {\bf 124} (2020), no.~11 111602, [\href{http://arxiv.org/abs/1910.02080}{{\tt arXiv:1910.02080}}].

\bibitem{Domcke:2020kcp}
V.~Domcke, Y.~Ema, K.~Mukaida, and M.~Yamada, {\it {Spontaneous Baryogenesis from Axions with Generic Couplings}},  {\em JHEP} {\bf 08} (2020) 096, [\href{http://arxiv.org/abs/2006.03148}{{\tt arXiv:2006.03148}}].

\bibitem{Co:2020xlh}
R.~T. Co, L.~J. Hall, and K.~Harigaya, {\it {Predictions for Axion Couplings from ALP Cogenesis}},  {\em JHEP} {\bf 01} (2021) 172, [\href{http://arxiv.org/abs/2006.04809}{{\tt arXiv:2006.04809}}].

\bibitem{Co:2020jtv}
R.~T. Co, N.~Fernandez, A.~Ghalsasi, L.~J. Hall, and K.~Harigaya, {\it {Lepto-Axiogenesis}},  {\em JHEP} {\bf 03} (2021) 017, [\href{http://arxiv.org/abs/2006.05687}{{\tt arXiv:2006.05687}}].

\bibitem{Co:2019jts}
R.~T. Co, L.~J. Hall, and K.~Harigaya, {\it {Axion Kinetic Misalignment Mechanism}},  {\em Phys. Rev. Lett.} {\bf 124} (2020), no.~25 251802, [\href{http://arxiv.org/abs/1910.14152}{{\tt arXiv:1910.14152}}].

\bibitem{Co:2020dya}
R.~T. Co, L.~J. Hall, K.~Harigaya, K.~A. Olive, and S.~Verner, {\it {Axion Kinetic Misalignment and Parametric Resonance from Inflation}},  {\em JCAP} {\bf 08} (2020) 036, [\href{http://arxiv.org/abs/2004.00629}{{\tt arXiv:2004.00629}}].

\bibitem{Eroncel:2022vjg}
C.~Er{\"o}ncel, R.~Sato, G.~Servant, and P.~S{\o}rensen, {\it {ALP dark matter from kinetic fragmentation: opening up the parameter window}},  {\em JCAP} {\bf 10} (2022) 053, [\href{http://arxiv.org/abs/2206.14259}{{\tt arXiv:2206.14259}}].

\bibitem{Eroncel:2022efc}
C.~Er{\"o}ncel and G.~Servant, {\it {ALP dark matter mini-clusters from kinetic fragmentation}},  {\em JCAP} {\bf 01} (2023) 009, [\href{http://arxiv.org/abs/2207.10111}{{\tt arXiv:2207.10111}}].

\bibitem{Eroncel:2025qlk}
C.~Er\"oncel, Y.~Gouttenoire, R.~Sato, G.~Servant, and P.~Simakachorn, {\it {A New Source for (QCD) Axion Dark Matter Production: Curvature-Induced}},  \href{http://arxiv.org/abs/2503.04880}{{\tt arXiv:2503.04880}}.

\bibitem{Bodas:2025eca}
A.~Bodas, R.~T. Co, A.~Ghalsasi, K.~Harigaya, and L.-T. Wang, {\it {Acoustic Misalignment Mechanism for Axion Dark Matter}},  \href{http://arxiv.org/abs/2503.04888}{{\tt arXiv:2503.04888}}.

\bibitem{Fasiello:2025ptb}
M.~Fasiello, J.~Lizarraga, A.~Papageorgiou, and A.~Urio, {\it {Kinetic Fragmentation of the QCD Axion on the Lattice}},  \href{http://arxiv.org/abs/2507.01822}{{\tt arXiv:2507.01822}}.

\bibitem{Co:2021lkc}
R.~T. Co, D.~Dunsky, N.~Fernandez, A.~Ghalsasi, L.~J. Hall, K.~Harigaya, and J.~Shelton, {\it {Gravitational wave and CMB probes of axion kination}},  {\em JHEP} {\bf 09} (2022) 116, [\href{http://arxiv.org/abs/2108.09299}{{\tt arXiv:2108.09299}}].

\bibitem{Gouttenoire:2021wzu}
Y.~Gouttenoire, G.~Servant, and P.~Simakachorn, {\it {Revealing the Primordial Irreducible Inflationary Gravitational-Wave Background with a Spinning Peccei-Quinn Axion}},  \href{http://arxiv.org/abs/2108.10328}{{\tt arXiv:2108.10328}}.

\bibitem{Tomita:1967wkp}
K.~Tomita, {\it {Non-Linear Theory of Gravitational Instability in the Expanding Universe}},  {\em Prog. Theor. Phys.} {\bf 37} (1967), no.~5 831--846.

\bibitem{Ananda:2006af}
K.~N. Ananda, C.~Clarkson, and D.~Wands, {\it {The Cosmological gravitational wave background from primordial density perturbations}},  {\em Phys. Rev.} {\bf D75} (2007) 123518, [\href{http://arxiv.org/abs/gr-qc/0612013}{{\tt gr-qc/0612013}}].

\bibitem{Baumann:2007zm}
D.~Baumann, P.~J. Steinhardt, K.~Takahashi, and K.~Ichiki, {\it {Gravitational Wave Spectrum Induced by Primordial Scalar Perturbations}},  {\em Phys. Rev.} {\bf D76} (2007) 084019, [\href{http://arxiv.org/abs/hep-th/0703290}{{\tt hep-th/0703290}}].

\bibitem{Kohri:2018awv}
K.~Kohri and T.~Terada, {\it {Semianalytic calculation of gravitational wave spectrum nonlinearly induced from primordial curvature perturbations}},  {\em Phys. Rev.} {\bf D97} (2018), no.~12 123532, [\href{http://arxiv.org/abs/1804.08577}{{\tt arXiv:1804.08577}}].

\bibitem{Inomata:2019ivs}
K.~Inomata, K.~Kohri, T.~Nakama, and T.~Terada, {\it {Enhancement of Gravitational Waves Induced by Scalar Perturbations due to a Sudden Transition from an Early Matter Era to the Radiation Era}},  {\em Phys. Rev.} {\bf D100} (2019), no.~4 043532, [\href{http://arxiv.org/abs/1904.12879}{{\tt arXiv:1904.12879}}].

\bibitem{Inomata:2020lmk}
K.~Inomata, M.~Kawasaki, K.~Mukaida, T.~Terada, and T.~T. Yanagida, {\it {Gravitational Wave Production right after a Primordial Black Hole Evaporation}},  {\em Phys. Rev. D} {\bf 101} (2020), no.~12 123533, [\href{http://arxiv.org/abs/2003.10455}{{\tt arXiv:2003.10455}}].

\bibitem{Harigaya:2023mhl}
K.~Harigaya, K.~Inomata, and T.~Terada, {\it {Gravitational wave production from axion rotations right after a transition to kination}},  {\em Phys. Rev. D} {\bf 108} (2023), no.~8 L081303, [\href{http://arxiv.org/abs/2305.14242}{{\tt arXiv:2305.14242}}].

\bibitem{Bartolo:2019oiq}
N.~Bartolo, D.~Bertacca, S.~Matarrese, M.~Peloso, A.~Ricciardone, A.~Riotto, and G.~Tasinato, {\it {Anisotropies and non-Gaussianity of the Cosmological Gravitational Wave Background}},  {\em Phys. Rev. D} {\bf 100} (2019), no.~12 121501, [\href{http://arxiv.org/abs/1908.00527}{{\tt arXiv:1908.00527}}].

\bibitem{Bartolo:2019yeu}
N.~Bartolo, D.~Bertacca, S.~Matarrese, M.~Peloso, A.~Ricciardone, A.~Riotto, and G.~Tasinato, {\it {Characterizing the cosmological gravitational wave background: Anisotropies and non-Gaussianity}},  {\em Phys. Rev. D} {\bf 102} (2020), no.~2 023527, [\href{http://arxiv.org/abs/1912.09433}{{\tt arXiv:1912.09433}}].

\bibitem{Dimastrogiovanni:2022eir}
E.~Dimastrogiovanni, M.~Fasiello, A.~Malhotra, and G.~Tasinato, {\it {Enhancing gravitational wave anisotropies with peaked scalar sources}},  {\em JCAP} {\bf 01} (2023) 018, [\href{http://arxiv.org/abs/2205.05644}{{\tt arXiv:2205.05644}}].

\bibitem{Chen:2022qec}
C.~Chen and A.~Ota, {\it {Induced gravitational waves from statistically anisotropic scalar perturbations}},  {\em Phys. Rev. D} {\bf 106} (2022), no.~6 063507, [\href{http://arxiv.org/abs/2205.07810}{{\tt arXiv:2205.07810}}].

\bibitem{Li:2023qua}
J.-P. Li, S.~Wang, Z.-C. Zhao, and K.~Kohri, {\it {Primordial non-Gaussianity f $_{NL}$ and anisotropies in scalar-induced gravitational waves}},  {\em JCAP} {\bf 10} (2023) 056, [\href{http://arxiv.org/abs/2305.19950}{{\tt arXiv:2305.19950}}].

\bibitem{Li:2023xtl}
J.-P. Li, S.~Wang, Z.-C. Zhao, and K.~Kohri, {\it {Complete analysis of the background and anisotropies of scalar-induced gravitational waves: primordial non-Gaussianity f $_{NL}$ and g $_{NL}$ considered}},  {\em JCAP} {\bf 06} (2024) 039, [\href{http://arxiv.org/abs/2309.07792}{{\tt arXiv:2309.07792}}].

\bibitem{Wang:2023ost}
S.~Wang, Z.-C. Zhao, J.-P. Li, and Q.-H. Zhu, {\it {Implications of pulsar timing array data for scalar-induced gravitational waves and primordial black holes: Primordial non-Gaussianity fNL considered}},  {\em Phys. Rev. Res.} {\bf 6} (2024), no.~1 L012060, [\href{http://arxiv.org/abs/2307.00572}{{\tt arXiv:2307.00572}}].

\bibitem{Yu:2023jrs}
Y.-H. Yu and S.~Wang, {\it {Anisotropies in scalar-induced gravitational-wave background from inflaton-curvaton mixed scenario with sound speed resonance}},  {\em Phys. Rev. D} {\bf 109} (2024), no.~8 083501, [\href{http://arxiv.org/abs/2310.14606}{{\tt arXiv:2310.14606}}].

\bibitem{Ruiz:2024weh}
J.~{\'A}. Ruiz and J.~Rey, {\it {Gravitational waves in ultra-slow-roll and their anisotropy at two loops}},  {\em JCAP} {\bf 04} (2025) 026, [\href{http://arxiv.org/abs/2410.09014}{{\tt arXiv:2410.09014}}].

\bibitem{Li:2025met}
J.-P. Li, S.~Wang, Z.-C. Zhao, and K.~Kohri, {\it {The isotropy, anisotropies and non-Gaussianity of the scalar-induced gravitational-wave background: diagrammatic approach for primordial non-Gaussianity up to any order}},  \href{http://arxiv.org/abs/2505.16820}{{\tt arXiv:2505.16820}}.

\bibitem{LISACosmologyWorkingGroup:2022kbp}
{\bf LISA Cosmology Working Group} Collaboration, N.~Bartolo et~al., {\it {Probing anisotropies of the Stochastic Gravitational Wave Background with LISA}},  {\em JCAP} {\bf 11} (2022) 009, [\href{http://arxiv.org/abs/2201.08782}{{\tt arXiv:2201.08782}}].

\bibitem{Kudoh:2005as}
H.~Kudoh, A.~Taruya, T.~Hiramatsu, and Y.~Himemoto, {\it {Detecting a gravitational-wave background with next-generation space interferometers}},  {\em Phys. Rev.} {\bf D73} (2006) 064006, [\href{http://arxiv.org/abs/gr-qc/0511145}{{\tt gr-qc/0511145}}].

\bibitem{Mentasti:2023gmg}
G.~Mentasti, C.~Contaldi, and M.~Peloso, {\it {Prospects for detecting anisotropies and polarization of the stochastic gravitational wave background with ground-based detectors}},  {\em JCAP} {\bf 08} (2023) 053, [\href{http://arxiv.org/abs/2304.06640}{{\tt arXiv:2304.06640}}].

\bibitem{Depta:2024ykq}
P.~F. Depta, V.~Domcke, G.~Franciolini, and M.~Pieroni, {\it {Pulsar timing array sensitivity to anisotropies in the gravitational wave background}},  {\em Phys. Rev. D} {\bf 111} (2025), no.~8 083039, [\href{http://arxiv.org/abs/2407.14460}{{\tt arXiv:2407.14460}}].

\bibitem{Cusin:2025xle}
G.~Cusin, C.~Pitrou, M.~Pijnenburg, and A.~Sesana, {\it {Measuring anisotropies in the PTA band with cross-correlations}},  \href{http://arxiv.org/abs/2502.17401}{{\tt arXiv:2502.17401}}.

\bibitem{Kumar:2021ffi}
S.~Kumar, R.~Sundrum, and Y.~Tsai, {\it {Non-Gaussian stochastic gravitational waves from phase transitions}},  {\em JHEP} {\bf 11} (2021) 107, [\href{http://arxiv.org/abs/2102.05665}{{\tt arXiv:2102.05665}}].

\bibitem{Bodas:2022zca}
A.~Bodas and R.~Sundrum, {\it {Primordial clocks within stochastic gravitational wave anisotropies}},  {\em JCAP} {\bf 10} (2022) 012, [\href{http://arxiv.org/abs/2205.04482}{{\tt arXiv:2205.04482}}].

\bibitem{Geller:2018mwu}
M.~Geller, A.~Hook, R.~Sundrum, and Y.~Tsai, {\it {Primordial Anisotropies in the Gravitational Wave Background from Cosmological Phase Transitions}},  {\em Phys. Rev. Lett.} {\bf 121} (2018), no.~20 201303, [\href{http://arxiv.org/abs/1803.10780}{{\tt arXiv:1803.10780}}].

\bibitem{Bodas:2022urf}
A.~Bodas and R.~Sundrum, {\it {Large primordial fluctuations in gravitational waves from phase transitions}},  {\em JHEP} {\bf 06} (2023) 029, [\href{http://arxiv.org/abs/2211.09301}{{\tt arXiv:2211.09301}}].

\bibitem{Domcke:2022wpb}
V.~Domcke, K.~Harigaya, and K.~Mukaida, {\it {Charge transfer between rotating complex scalar fields}},  {\em JHEP} {\bf 08} (2022) 234, [\href{http://arxiv.org/abs/2205.00942}{{\tt arXiv:2205.00942}}].

\bibitem{Shtanov:1994ce}
Y.~Shtanov, J.~H. Traschen, and R.~H. Brandenberger, {\it {Universe reheating after inflation}},  {\em Phys. Rev. D} {\bf 51} (1995) 5438--5455, [\href{http://arxiv.org/abs/hep-ph/9407247}{{\tt hep-ph/9407247}}].

\bibitem{Kofman:1994rk}
L.~Kofman, A.~D. Linde, and A.~A. Starobinsky, {\it {Reheating after inflation}},  {\em Phys. Rev. Lett.} {\bf 73} (1994) 3195--3198, [\href{http://arxiv.org/abs/hep-th/9405187}{{\tt hep-th/9405187}}].

\bibitem{Kofman:1997yn}
L.~Kofman, A.~D. Linde, and A.~A. Starobinsky, {\it {Towards the theory of reheating after inflation}},  {\em Phys. Rev. D} {\bf 56} (1997) 3258--3295, [\href{http://arxiv.org/abs/hep-ph/9704452}{{\tt hep-ph/9704452}}].

\bibitem{Fedderke:2025sic}
M.~A. Fedderke, J.~Huang, and N.~Siemonsen, {\it {Periodic Cosmic String Formation and Dynamics}},  \href{http://arxiv.org/abs/2503.03116}{{\tt arXiv:2503.03116}}.

\bibitem{Tkachev:1995md}
I.~Tkachev, {\it {Phase transitions at preheating}},  {\em Phys. Lett. B} {\bf 376} (1996) 35--40, [\href{http://arxiv.org/abs/hep-th/9510146}{{\tt hep-th/9510146}}].

\bibitem{Kasuya:1996ns}
S.~Kasuya, M.~Kawasaki, and T.~Yanagida, {\it {Cosmological axion problem in chaotic inflationary universe}},  {\em Phys. Lett. B} {\bf 409} (1997) 94--100, [\href{http://arxiv.org/abs/hep-ph/9608405}{{\tt hep-ph/9608405}}].

\bibitem{Kasuya:1997ha}
S.~Kasuya and M.~Kawasaki, {\it {Can topological defects be formed during preheating?}},  {\em Phys. Rev. D} {\bf 56} (1997) 7597--7607, [\href{http://arxiv.org/abs/hep-ph/9703354}{{\tt hep-ph/9703354}}].

\bibitem{Kasuya:1998td}
S.~Kasuya and M.~Kawasaki, {\it {Topological defects formation after inflation on lattice simulation}},  {\em Phys. Rev. D} {\bf 58} (1998) 083516, [\href{http://arxiv.org/abs/hep-ph/9804429}{{\tt hep-ph/9804429}}].

\bibitem{Tkachev:1998dc}
I.~Tkachev, S.~Khlebnikov, L.~Kofman, and A.~D. Linde, {\it {Cosmic strings from preheating}},  {\em Phys. Lett. B} {\bf 440} (1998) 262--268, [\href{http://arxiv.org/abs/hep-ph/9805209}{{\tt hep-ph/9805209}}].

\bibitem{Kasuya:1999hy}
S.~Kasuya and M.~Kawasaki, {\it {Comments on cosmic string formation during preheating on lattice simulations}},  {\em Phys. Rev. D} {\bf 61} (2000) 083510, [\href{http://arxiv.org/abs/hep-ph/9903324}{{\tt hep-ph/9903324}}].

\bibitem{Enqvist:1998pf}
K.~Enqvist and J.~McDonald, {\it {Observable isocurvature fluctuations from the Affleck-Dine condensate}},  {\em Phys. Rev. Lett.} {\bf 83} (1999) 2510--2513, [\href{http://arxiv.org/abs/hep-ph/9811412}{{\tt hep-ph/9811412}}].

\bibitem{Kasuya:2008xp}
S.~Kasuya, M.~Kawasaki, and F.~Takahashi, {\it {Isocurvature fluctuations in Affleck-Dine mechanism and constraints on inflation models}},  {\em JCAP} {\bf 10} (2008) 017, [\href{http://arxiv.org/abs/0805.4245}{{\tt arXiv:0805.4245}}].

\bibitem{Co:2022qpr}
R.~T. Co, K.~Harigaya, and A.~Pierce, {\it {Cosmic perturbations from a rotating field}},  {\em JCAP} {\bf 10} (2022) 037, [\href{http://arxiv.org/abs/2202.01785}{{\tt arXiv:2202.01785}}].

\bibitem{Inomata:2019zqy}
K.~Inomata, K.~Kohri, T.~Nakama, and T.~Terada, {\it {Gravitational Waves Induced by Scalar Perturbations during a Gradual Transition from an Early Matter Era to the Radiation Era}},  {\em JCAP} {\bf 10} (2019) 071, [\href{http://arxiv.org/abs/1904.12878}{{\tt arXiv:1904.12878}}]. [Erratum: JCAP 08, E01 (2023)].

\bibitem{Pearce:2023kxp}
M.~Pearce, L.~Pearce, G.~White, and C.~Bal\'azs, {\it {Gravitational Wave Signals From Early Matter Domination: Interpolating Between Fast and Slow Transitions}},  \href{http://arxiv.org/abs/2311.12340}{{\tt arXiv:2311.12340}}.

\bibitem{Domenech:2019quo}
G.~Dom\`enech, {\it {Induced gravitational waves in a general cosmological background}},  {\em Int. J. Mod. Phys. D} {\bf 29} (2020), no.~03 2050028, [\href{http://arxiv.org/abs/1912.05583}{{\tt arXiv:1912.05583}}].

\bibitem{Harigaya:2023pmw}
K.~Harigaya, K.~Inomata, and T.~Terada, {\it {Induced gravitational waves with kination era for recent pulsar timing array signals}},  {\em Phys. Rev. D} {\bf 108} (2023), no.~12 123538, [\href{http://arxiv.org/abs/2309.00228}{{\tt arXiv:2309.00228}}].

\bibitem{Domenech:2021and}
G.~Dom\`enech, S.~Passaglia, and S.~Renaux-Petel, {\it {Gravitational waves from dark matter isocurvature}},  {\em JCAP} {\bf 03} (2022), no.~03 023, [\href{http://arxiv.org/abs/2112.10163}{{\tt arXiv:2112.10163}}].

\bibitem{Domenech:2023jve}
G.~Dom\`enech, {\it {Cosmological gravitational waves from isocurvature fluctuations}},  {\em AAPPS Bull.} {\bf 34} (2024), no.~1 4, [\href{http://arxiv.org/abs/2311.02065}{{\tt arXiv:2311.02065}}].

\bibitem{Eroncel:2025bcb}
C.~Er\"oncel, Y.~Gouttenoire, R.~Sato, G.~Servant, and P.~Simakachorn, {\it {A universal bound on the duration of a kination era}},  \href{http://arxiv.org/abs/2501.17226}{{\tt arXiv:2501.17226}}.

\bibitem{Inomata:2021zel}
K.~Inomata, {\it {Bound on induced gravitational waves during inflation era}},  {\em Phys. Rev. D} {\bf 104} (2021), no.~12 123525, [\href{http://arxiv.org/abs/2109.06192}{{\tt arXiv:2109.06192}}].

\bibitem{Mukhanov:991646}
V.~Mukhanov, {\em {Physical Foundations of Cosmology}}.
\newblock Cambridge Univ. Press, Cambridge, 2005.

\bibitem{Planck:2018vyg}
{\bf Planck} Collaboration, N.~Aghanim et~al., {\it {Planck 2018 results. VI. Cosmological parameters}},  {\em Astron. Astrophys.} {\bf 641} (2020) A6, [\href{http://arxiv.org/abs/1807.06209}{{\tt arXiv:1807.06209}}]. [Erratum: Astron.Astrophys. 652, C4 (2021)].

\bibitem{Dodelson:1282338}
S.~Dodelson, {\em {Modern cosmology}}.
\newblock Academic Press, San Diego, CA, 2003.

\bibitem{Cai:2019cdl}
R.-G. Cai, S.~Pi, and M.~Sasaki, {\it {Universal infrared scaling of gravitational wave background spectra}},  {\em Phys. Rev. D} {\bf 102} (2020), no.~8 083528, [\href{http://arxiv.org/abs/1909.13728}{{\tt arXiv:1909.13728}}].

\bibitem{Rey:2024giu}
J.~Rey, {\it {A consistency relation for induced gravitational wave anisotropies}},  \href{http://arxiv.org/abs/2411.08873}{{\tt arXiv:2411.08873}}.

\bibitem{Saito:2008jc}
R.~Saito and J.~Yokoyama, {\it {Gravitational wave background as a probe of the primordial black hole abundance}},  {\em Phys. Rev. Lett.} {\bf 102} (2009) 161101, [\href{http://arxiv.org/abs/0812.4339}{{\tt arXiv:0812.4339}}]. [Erratum: Phys. Rev. Lett.107,069901(2011)].

\bibitem{Saito:2009jt}
R.~Saito and J.~Yokoyama, {\it {Gravitational-Wave Constraints on the Abundance of Primordial Black Holes}},  {\em Prog. Theor. Phys.} {\bf 123} (2010) 867--886, [\href{http://arxiv.org/abs/0912.5317}{{\tt arXiv:0912.5317}}]. [Erratum: Prog. Theor. Phys.126,351(2011)].

\bibitem{khlopov1980primordial}
M.~Y. Khlopov and A.~Polnarev, {\it Primordial black holes as a cosmological test of grand unification},  {\em Physics Letters B} {\bf 97} (1980), no.~3-4 383--387.

\bibitem{Polnarev:1985btg}
A.~G. Polnarev and M.~Y. Khlopov, {\it {COSMOLOGY, PRIMORDIAL BLACK HOLES, AND SUPERMASSIVE PARTICLES}},  {\em Sov. Phys. Usp.} {\bf 28} (1985) 213--232.

\bibitem{Harada:2016mhb}
T.~Harada, C.-M. Yoo, K.~Kohri, K.-i. Nakao, and S.~Jhingan, {\it {Primordial black hole formation in the matter-dominated phase of the Universe}},  {\em Astrophys. J.} {\bf 833} (2016), no.~1 61, [\href{http://arxiv.org/abs/1609.01588}{{\tt arXiv:1609.01588}}].

\bibitem{Harada:2017fjm}
T.~Harada, C.-M. Yoo, K.~Kohri, and K.-I. Nakao, {\it {Spins of primordial black holes formed in the matter-dominated phase of the Universe}},  {\em Phys. Rev.} {\bf D96} (2017), no.~8 083517, [\href{http://arxiv.org/abs/1707.03595}{{\tt arXiv:1707.03595}}].

\bibitem{Kokubu:2018fxy}
T.~Kokubu, K.~Kyutoku, K.~Kohri, and T.~Harada, {\it {Effect of Inhomogeneity on Primordial Black Hole Formation in the Matter Dominated Era}},  {\em Phys. Rev. D} {\bf 98} (2018), no.~12 123024, [\href{http://arxiv.org/abs/1810.03490}{{\tt arXiv:1810.03490}}].

\bibitem{deJong:2021bbo}
E.~de~Jong, J.~C. Aurrekoetxea, and E.~A. Lim, {\it {Primordial black hole formation with full numerical relativity}},  {\em JCAP} {\bf 03} (2022), no.~03 029, [\href{http://arxiv.org/abs/2109.04896}{{\tt arXiv:2109.04896}}].

\bibitem{Harada:2022xjp}
T.~Harada, K.~Kohri, M.~Sasaki, T.~Terada, and C.-M. Yoo, {\it {Threshold of primordial black hole formation against velocity dispersion in matter-dominated era}},  {\em JCAP} {\bf 02} (2023) 038, [\href{http://arxiv.org/abs/2211.13950}{{\tt arXiv:2211.13950}}].

\bibitem{Harada:2013epa}
T.~Harada, C.-M. Yoo, and K.~Kohri, {\it {Threshold of primordial black hole formation}},  {\em Phys. Rev.} {\bf D88} (2013), no.~8 084051, [\href{http://arxiv.org/abs/1309.4201}{{\tt arXiv:1309.4201}}]. [Erratum: Phys. Rev.D89,no.2,029903(2014)].

\bibitem{Carr:2017jsz}
B.~Carr, M.~Raidal, T.~Tenkanen, V.~Vaskonen, and H.~Veerm{\"a}e, {\it {Primordial black hole constraints for extended mass functions}},  {\em Phys. Rev.} {\bf D96} (2017), no.~2 023514, [\href{http://arxiv.org/abs/1705.05567}{{\tt arXiv:1705.05567}}].

\bibitem{carr2021constraints}
B.~Carr, K.~Kohri, Y.~Sendouda, and J.~Yokoyama, {\it Constraints on primordial black holes},  {\em Reports on Progress in Physics} {\bf 84} (2021), no.~11 116902.

\bibitem{Planck:2018jri}
{\bf Planck} Collaboration, Y.~Akrami et~al., {\it {Planck 2018 results. X. Constraints on inflation}},  {\em Astron. Astrophys.} {\bf 641} (2020) A10, [\href{http://arxiv.org/abs/1807.06211}{{\tt arXiv:1807.06211}}].

\bibitem{Akita:2022hlx}
K.~Akita and M.~Yamaguchi, {\it {A Review of Neutrino Decoupling from the Early Universe to the Current Universe}},  {\em Universe} {\bf 8} (2022), no.~11 552, [\href{http://arxiv.org/abs/2210.10307}{{\tt arXiv:2210.10307}}].

\bibitem{Komatsu:2001rj}
E.~Komatsu and D.~N. Spergel, {\it {Acoustic signatures in the primary microwave background bispectrum}},  {\em Phys. Rev. D} {\bf 63} (2001) 063002, [\href{http://arxiv.org/abs/astro-ph/0005036}{{\tt astro-ph/0005036}}].

\bibitem{Planck:2019kim}
{\bf Planck} Collaboration, Y.~Akrami et~al., {\it {Planck 2018 results. IX. Constraints on primordial non-Gaussianity}},  {\em Astron. Astrophys.} {\bf 641} (2020) A9, [\href{http://arxiv.org/abs/1905.05697}{{\tt arXiv:1905.05697}}].

\bibitem{Schmitz:2020syl}
K.~Schmitz, {\it {New Sensitivity Curves for Gravitational-Wave Signals from Cosmological Phase Transitions}},  {\em JHEP} {\bf 01} (2021) 097, [\href{http://arxiv.org/abs/2002.04615}{{\tt arXiv:2002.04615}}].

\bibitem{Alonso:2020rar}
D.~Alonso, C.~R. Contaldi, G.~Cusin, P.~G. Ferreira, and A.~I. Renzini, {\it {Noise angular power spectrum of gravitational wave background experiments}},  {\em Phys. Rev. D} {\bf 101} (2020), no.~12 124048, [\href{http://arxiv.org/abs/2005.03001}{{\tt arXiv:2005.03001}}].

\bibitem{Braglia:2021fxn}
M.~Braglia and S.~Kuroyanagi, {\it {Probing prerecombination physics by the cross-correlation of stochastic gravitational waves and CMB anisotropies}},  {\em Phys. Rev. D} {\bf 104} (2021), no.~12 123547, [\href{http://arxiv.org/abs/2106.03786}{{\tt arXiv:2106.03786}}].

\bibitem{Mukaida:2012qn}
K.~Mukaida and K.~Nakayama, {\it {Dynamics of oscillating scalar field in thermal environment}},  {\em JCAP} {\bf 01} (2013) 017, [\href{http://arxiv.org/abs/1208.3399}{{\tt arXiv:1208.3399}}].

\bibitem{Lyth:2002my}
D.~H. Lyth, C.~Ungarelli, and D.~Wands, {\it {The Primordial density perturbation in the curvaton scenario}},  {\em Phys. Rev. D} {\bf 67} (2003) 023503, [\href{http://arxiv.org/abs/astro-ph/0208055}{{\tt astro-ph/0208055}}].

\bibitem{Sasaki:2006kq}
M.~Sasaki, J.~Valiviita, and D.~Wands, {\it {Non-Gaussianity of the primordial perturbation in the curvaton model}},  {\em Phys. Rev. D} {\bf 74} (2006) 103003, [\href{http://arxiv.org/abs/astro-ph/0607627}{{\tt astro-ph/0607627}}].

\bibitem{Sasaki:1995aw}
M.~Sasaki and E.~D. Stewart, {\it {A General analytic formula for the spectral index of the density perturbations produced during inflation}},  {\em Prog. Theor. Phys.} {\bf 95} (1996) 71--78, [\href{http://arxiv.org/abs/astro-ph/9507001}{{\tt astro-ph/9507001}}].

\bibitem{Wands:2000dp}
D.~Wands, K.~A. Malik, D.~H. Lyth, and A.~R. Liddle, {\it {A New approach to the evolution of cosmological perturbations on large scales}},  {\em Phys. Rev. D} {\bf 62} (2000) 043527, [\href{http://arxiv.org/abs/astro-ph/0003278}{{\tt astro-ph/0003278}}].

\bibitem{Lyth:2004gb}
D.~H. Lyth, K.~A. Malik, and M.~Sasaki, {\it {A General proof of the conservation of the curvature perturbation}},  {\em JCAP} {\bf 05} (2005) 004, [\href{http://arxiv.org/abs/astro-ph/0411220}{{\tt astro-ph/0411220}}].

\bibitem{Kawasaki:2011pd}
M.~Kawasaki, T.~Kobayashi, and F.~Takahashi, {\it {Non-Gaussianity from Curvatons Revisited}},  {\em Phys. Rev. D} {\bf 84} (2011) 123506, [\href{http://arxiv.org/abs/1107.6011}{{\tt arXiv:1107.6011}}].

\bibitem{Dine:1995kz}
M.~Dine, L.~Randall, and S.~D. Thomas, {\it {Baryogenesis from flat directions of the supersymmetric standard model}},  {\em Nucl. Phys. B} {\bf 458} (1996) 291--326, [\href{http://arxiv.org/abs/hep-ph/9507453}{{\tt hep-ph/9507453}}].

\bibitem{Harigaya:2015hha}
K.~Harigaya, M.~Ibe, M.~Kawasaki, and T.~T. Yanagida, {\it {Dynamics of Peccei-Quinn Breaking Field after Inflation and Axion Isocurvature Perturbations}},  {\em JCAP} {\bf 1511} (2015), no.~11 003, [\href{http://arxiv.org/abs/1507.00119}{{\tt arXiv:1507.00119}}].

\bibitem{Elgamal:2025aol}
S.~Elgamal and K.~Harigaya, {\it {Dynamical Solution to the Eta Problem in Spectator Field Models}},  \href{http://arxiv.org/abs/2502.01898}{{\tt arXiv:2502.01898}}.

\bibitem{1972mwm..book..231T}
K.~S. {Thorne}, {\it {Nonspherical Gravitational Collapse--A Short Review}},  in {\em Magic Without Magic: John Archibald Wheeler} (J.~R. {Klauder}, ed.), p.~231.
\newblock 1972.

\bibitem{doroshkevich1970spatial}
A.~Doroshkevich, {\it Spatial structure of perturbations and origin of galactic rotation in fluctuation theory},  {\em Astrophysics} {\bf 6} (1970), no.~4 320--330.

\bibitem{Carr:1975qj}
B.~J. Carr, {\it {The Primordial black hole mass spectrum}},  {\em Astrophys. J.} {\bf 201} (1975) 1--19.

\end{thebibliography}\endgroup
\end{document}